# Photonic Crystals Engineering For Light Manipulation:
## low symmetry,
## graded index and parity time symmetry

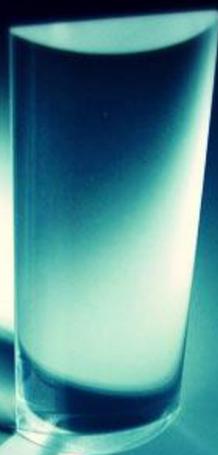

### Mirbek Turduev

# Photonic Crystals Engineering For Light Manipulation: Low Symmetry, Graded Index and Parity Time Symmetry

A dissertation submitted to the Graduate School of Science and Technology
of TOBB University of Economics and Technology

by

## Mirbek Turduev

In partial fulfillment of the requirements for the degree of
Doctor of Philosophy in
the Department of Electrical and Electronics Engineering

Under the supervision of Prof. Dr. Hamza Kurt

Ankara, 2015

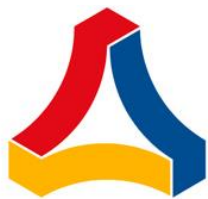

TOBB UNIVERSITY OF ECONOMICS AND TECHNOLOGY

# PHOTONIC CRYSTAL ENGINEERING FOR LIGHT MANIPULATION: LOW SYMMETRY, INDEX GRADIENT AND PARITY-TIME SYMMETRY

## ABSTRACT

The great interest to the two and three dimensionally periodic structures, called photonic crystals (PCs), has begun with the pioneer works of Yablonovitch and John as one can efficiently control the propagation of the electromagnetic (EM) waves in the same manner with semiconductors that affect the electron conduction. One of the main peculiar properties of PCs is that they have photonic band gap in the transmission spectrum similar to electronic band gap and hence, they are able to prevent the light to propagate in certain frequency regions irrespective of the propagation direction in space. Inside the band gaps, neither optical modes nor spontaneous emissions exist. Breaking the rotational and mirror symmetries of PC unit cells provides rich dispersive features such as tilted self-collimation, and wavelength de-multiplexing effects. Another important issue in PC designs is that it is feasible to design graded index medium if the parameters of the two dimensional PCs is intentionally rearranged. That type of configuration is known as Graded index photonic crystals (GRIN PCs). The implementations of GRIN via periodic structures provide great flexibilities in terms of designing different index gradient and photonic integrated circuit components such as couplers, lenses, and mode order converters. It is crucial to deliver optical signal without any loss for the long distances where light diffraction plays an important role. Hence, dealing with the alternative solution to light diffraction phenomena using 2D axicon shape annular type photonic structure is another topic of this thesis. In addition to conventional photonic all dielectric structures, we have proposed gain-loss modulated parity-time (PT-) symmetric photonic structures to obtain strong asymmetric light transmission close to the crystallographic resonances or, equivalently, close to high-symmetry points.

# ACKNOWLEDGMENTS

My long journey with TOBB ETÜ began in 2008 and now ends after 7 years. First I was a Master student at the Electrical and Electronics Department and after a PhD student in the same department. Hence, this gratitude is not only for PhD but also for all of these 7 years. In this respect, it is a pleasure for me to express my gratitude to all members of the Electrical and Electronics Engineering Department.

I would like to express my deepest gratitude and respect to my PhD and thesis advisor, Assoc. Prof. Dr. Hamza Kurt, whose tolerance, insight and patience helped me throughout my PhD journey. It was a great privilege to be his student.

I would like to thank to the members of my thesis committee, Asst. Prof. Dr. Rohat Melik, Asst. Prof. Dr. Israfil Bahçeci, Asst. Prof. Dr. Hümeyra Çağlayan, and Asst. Prof. Dr. İlyas Evrim ÇOLAK for their guidance. Also financial support of the TÜBİTAK via project number 110T306 is greatly acknowledged.

I would like to express my pleasure to have collaborated with Spanish UPC research team: Prof. Dr. Kestutis Staliunas, Assoc. Prof. Dr. Muriel Botey, and Assoc. Prof. Dr. Ramon Herrero. It was a great privilege to know and to work with them.

I wish to thank to my lab colleagues, Bilgehan B. Öner, Melih G. Can, Zeki Hayran, İbrahim H. Giden, Khalil Dadashi, Kadir Üstün, Ali Özmen, Eyüp M. Gayur, Ali B. Parım, İbrahim İ. Taşkıran, Zeynep Özmen, Sümeyye Gökçe, Sinan E. Tandoğan, Yusuf A. Yılmaz, Muhammed F. Ballı for their support and fruitfull discussions. I would never forget all the chats and beautiful moments I shared with them. I should also thank to my friends, Alisher Suyundukov, Altynbek Raimbekov, Zhyrgal Toktoshev, Ernist Tilenbaev, Tashtan Satiev, and Nurbek Tabyshev for their sincere friendship, their endless and unconditional support.

Finally, I would like to thank my wife, Aizhan Zhamanakova she was always there cheering me up and stood by me through the good and bad times.

*This thesis is dedicated to all the people who never stop believing in me. Especcially to my mother and father, Atyrgül and Kerezbek, who first taught me the value of education and critical thought, to my brilliant and outrageously loving and supportive wife, Aizhan, our exuberant, sweet, and kind-hearted little girl, Zhibek and to my always encouraging, ever faithful brothers Erkinbek, Suimonkul, Yryskul and Syimyk. And last but not least to my lovely Grandmother Burul, who taught me to get up after a fall and start again.*

# CONTENTS

# ABBREVIATIONS

Abbreviations used in this thesis are presented below with corresponding explanations.

| Abbreviations | Explanations |
| --- | --- |
| **PC** | Photonic crystal |
| **PWM** | Plane wave expansion method |
| **FDTD** | Finite difference time domain method |
| **PML** | Perfectly matched layer |
| **GRIN** | Graded index |
| **FWHM** | Full width at half maximum |
| **TE** | Transverse electric |
| **TM** | Transverse magnetic |
| **IFC** | Iso-frequency contour |
| **PBG** | Photonic band gap |
| **CPBG** | Complete photonic band gap |
| **HS** | Hyperbolic secant |
| **PT** | Parity-time |

# LIST OF SYMBOLS

Symbols used in this thesis are presented below with corresponding explanations.

| Symbols | Explanations |
|---|---|
| $k$ | Wave vector |
| $c$ | Light speed in air |
| $n$ | Refractive index of dielectric material |
| $a$ | Lattice constant |
| $\omega(k)$ | Angular frequency |
| $n_g$ | Group refractive index of dielectric material |
| $v_g$ | Group velocity of electromagnetic wave |
| $r$ | Position vector |
| $S$ | Light trajectory/path |
| $F$ | Filling factor |
| $\varepsilon_r$ | Dielectric permittivity of the photonic crystal rods |
| $d$ | Diffraction coefficient |
| $k_L$ | Spatial dispersion |
| $\theta_n$ | Beam shifting angle |
| $\Delta F$ | Focal length |
| $J_0$ | Zero order Bessel function |
| $\theta$ | Open angle |
| $L_x$ | Width of the structure |
| $L_y$ | Height of the structure |

# 1. INTRODUCTION

Nature has always been an invaluable source of inspiration for technological progress in the history of human development. In most of the mankind history, we are improving on how to control and utilize the mechanical properties of nature materials. At the very beginning age of civilization, human being learnt how to use and manipulate with the mechanical property of stone to make stereotype tools for everyday life. Later on, people studied how to get metal and alloy from ore. Even more recent invention of steam locomotive was based on how to improve the mechanical movement which made modern civilization possible.

Although the electrical and optical properties were noticed by us long time ago, the theoretical study of fundamental electrical and optical phenomena is not done until around two hundred years ago due to the tiny size of electron, photon and atomic structure. The Scottish physicist James Clerk Maxwell demonstrates that electric and magnetic fields travel through space in the form of waves at the constant speed of light and that electricity, magnetism and even light are all manifestations of electromagnetism. In 1864 he collected together the laws originally derived by Carl Friedrich Gauss, Michael Faraday and André-Marie Ampère into a his famous equations known as Maxwell's Equations. Later in 1898 Sir John Joseph Thomson first discovered electrons the very small, negatively charged, sub-atomic particles. In 1926, Schrodinger published his quantum mechanics paper to describe how electrons behave. In the middle of 20th century, with the efforts of both theoretical and experimental physicists, we can control the motion of electrons by introducing defects into pure crystals or semiconductors. After we had the ability to control the electrical properties, the electrical engineering industry development is possible and it has profound impact on our daily life.

In 21 century the science efforts are going day by day towards miniaturization and high speed at today's information and communications technology. Here an important role plays electronic circuits that provide us the ability to control the transport and storage data information by means of electrons. However, the performance of electronic integrated circuits is now becoming rather limited when digital information needs to be processed and sent from one point to another. According to Moore's Law originated in 1965 states that overall data processing power for computing technologies will double every two years. It means that every two years the density of electronic integrated circuits/transistors will be doubling. Consequently, the size of transistors should be decreased in a same rate. Today, transistors on integrated circuits have reached a size so small that it would take more than 2,000 of them stacked next to each other to equal the thickness of a human hair. The transistors on personal computers' latest chips are only 45 nanometers wide: the average human hair is about 100,000 nanometers thick. In the characterization and fabricating of nanoscale electronic devices the classical physics is no more enough and we should utilize quantum mechanics. The rules of physics in the quantum world are very different from the way things work on the macro scale. For example, quantum particles like electrons can pass through extremely thin walls even if they don't have the kinetic energy necessary to break through the barriers. Quantum physicists call this phenomenon quantum tunneling. Because electronics depend upon controlling the flow of electrons to work, issues like quantum tunneling create serious problems.

Replacement of electrons by photons can be an effective solution in nanoscale technology miniaturization problems. In addition, photons can travel faster than electrons in the medium, it can carry larger information than electrons and since photons are not strongly interacting particles as electrons, this helps reduce energy losses. The next question is how to use light as the information carrier instead of electrons. Photonic crystals are ideally suited for this task.

The history of photonic crystals starts with the early idea for electromagnetic wave propagation in a periodic medium, which was studied by Lord Rayleigh in 1887, that corresponds to 1D photonic crystals [1]. That study showed that it is possible to find a photonic band gap (PBG) in one-dimensional periodic structures. After 100 years, in 1987, two independent works appeared that are considered as the starting point of the research field. One was the paper was by Yablonovitch, titled "Inhibition of spontaneous emission of electromagnetic radiation using a three dimensionally periodic structure" [2]. Yablonovitch's idea was to understand controlling the spontaneous emission by modifying the photonic density of states of the medium using periodic dielectrics. The second paper by Sajeev John was titled "Strong localization of photons in certain disordered dielectric super-lattices" [3]. John's aim was to understand how a random refractive-index variation affects photon localization. Both scientists claimed that the interaction of photons with the dielectric structures can create unique properties in the electromagnetic spectrum. This can be possible if the structure has wavelength-scale geometrical features along with a high-contrast refractive index variation instead of uniform/homogenous medium. Such materials with unique electromagnetic properties such as electromagnetic band gap are later named as "photonic crystals" (PCs). Periodic photonic crystals have periodic 'potentials' due to lattices of macroscopic dielectric media in place of atoms. If the dielectric constant of the materials in the crystal are different enough, then, the absorption of light is minimal, then scattering at the interfaces can produce many of the same phenomena for photons (i.e. light modes) as the atomic potential does for electrons. One solution to the issue of optical control and manipulation is thus the photonic crystal, a low-loss periodic dielectric medium.

Photonic crystals are very tempting for use in a new generation of integrated circuit design because of their unique ability to confine light within certain regions of space. One of the basic characteristics of PCs is that the refractive index variation may appear in one, two, or even three-dimensions.

The PC concept has been extensively studied in the photonics field since 1987 because of its ability to control the flow of light. One of the most attractive aspects of PCs is that the light-matter interaction in PCs enables unique optical conditions that cannot be observed in standard optical waveguides, e.g., slow-light, graded-index PC design, optical cavities with high Q-factor, super-prisms, self-collimators, sensitive bio-chemical PC based sensors, specific light sources, and lasers [4-15]. Furthermore, light motion inside PCs can be analyzed by scale-invariant Maxwell's equations, so that structural PC unit-cell parameters can be easily tuned either to millimeter or micron-scale [16-19]. One of the important applications of photonic crystals is the possibility to design compact integrated optical devices [20, 21] that operate entirely with light. Replacing relatively slow electrons with photons as the carriers of information can dramatically increase the speed and the bandwidth of advanced communication systems, thus revolutionizing the telecommunication industry.

## 1.1. Photonic Crystals: Properties of the Periodic Photonic Structures

Photonic crystals are periodic arrangements of materials with different refractive index. This spatial distribution gives rise to a periodic dielectric function that, as for the periodic potential generated by regular arrays of atoms and molecules, produces an energy band structure in which band gaps may occur. The presence of photonic band gaps forbids the propagation for specific frequencies and in certain directions. This feature makes the photonic crystals an excellent framework to engineer the materials for the optical control and manipulation.

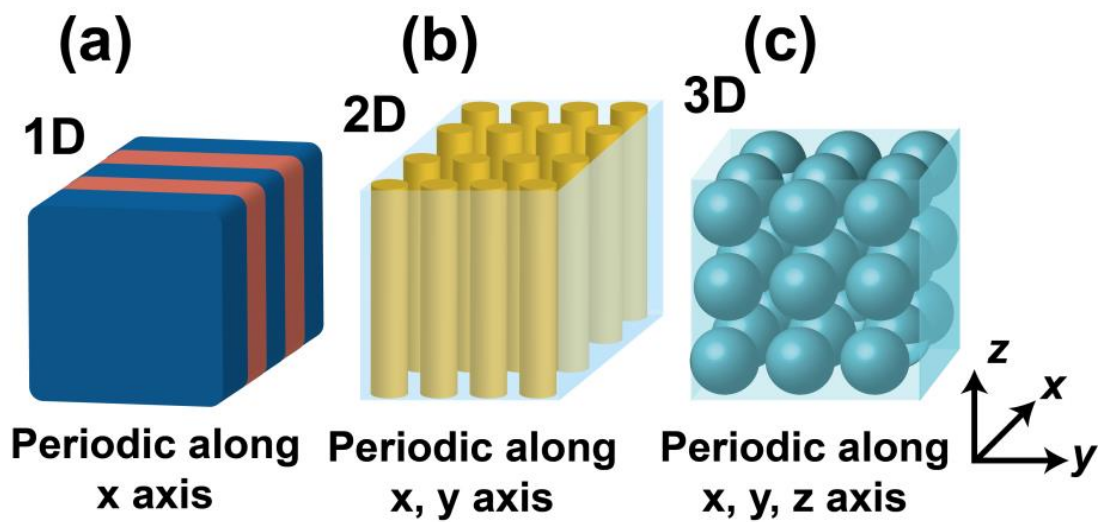

Figure 1.1. Examples of (a) 1D, (b) 2D and (c) 3D PCs.

Fig. 1.1 shows three different examples of realization of photonic crystal structure in (a) 1D, (b) 2D and (c) 3D periodicity. The spatial period can be named as lattice constant and it can be chosen on the order of the wavelength of the incident light involved in the optical process. The discrete translational symmetry of a photonic crystal makes possible to classify the electromagnetic modes with respect to their wavevectors **k**. The modes can be expanded in Bloch form consisting of a plane wave modulated by a periodical function that takes into account the periodicity of the crystal [4]. Therefore, for example, the magnetic field into a PC can be written as

$$\mathbf{H_k}(\mathbf{r}) = e^{i\mathbf{kr}}\mathbf{u}(\mathbf{r}) = e^{i\mathbf{kr}}\mathbf{u}(\mathbf{r}+\mathbf{R}) \tag{1.1}$$

where **R** is the spatial vector that accounts for the lattice periodicity and it is named lattice vector. Defining the reciprocal lattice vector **G** as the vector that satisfies the relationship $exp(i\mathbf{G}\cdot\mathbf{R}) = 1$, from Eq. (1.1) it follows that a mode with wavevector **k** and a mode with wavevector **k** + **G** are the same mode. This means that it is convenient to restrict the attention to a finite zone in reciprocal space (space of **k**) in which it is not possible to get from one part to another of the lattice by adding any **G**; this zone is known as the Brillouin zone. Fig 1.2 shows an example of (a) square lattice, of its (b) reciprocal space, and of the corresponding (c) Brillouin zone. The periodicity is obtained by using materials with different dielectric constants.

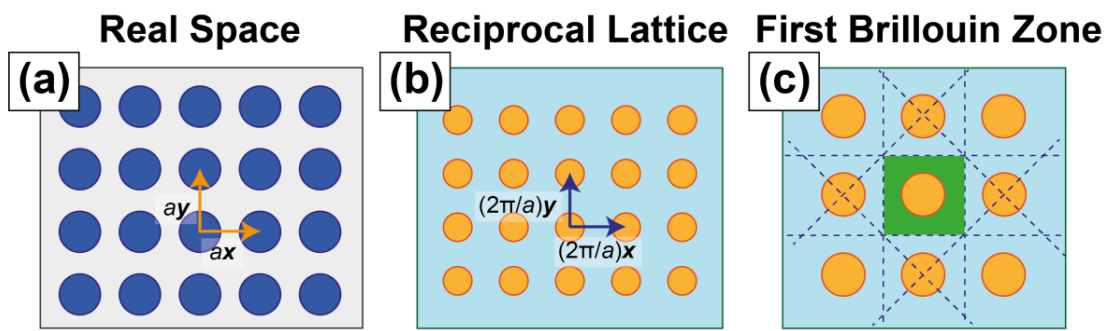

Figure 1.2. (a) The Real space, (b) the corresponding reciprocal space, and (c) the Brillouin zone of a square lattice.

## 1.2. Analysis of PC Structures in Frequency and Time Domains

The characterization and detailed analysis of PCs in both frequency and time domain can be performed by exploiting various numerical approaches. The plane wave method (PWM) and finite-difference time-domain (FDTD) method are widely used techniques for the investigation of PCs.

In order to design photonic crystals to take advantage of their unique properties, a calculation method is necessary to determine how light will propagate through a particular crystal structure. Specifically, given any periodic dielectric structure, we must find the allowable frequencies for light propagation in all crystal directions and be able to calculate the field distributions in the crystal for any frequency of light. There are several capable techniques, but one of the most studied and reliable method is the PWM method. The method based on the time-harmonic decomposition of the eigenmodes and the dielectric constant. It was used in some of the earliest studies of photonic crystals [22-25] and is simple enough to be easily implemented.

On the other hand, FDTD provides EM field fluctuation in spatial space with respect to time. The transmission and reflection efficiencies of finite structures can be evaluated easily and the wave propagation trough the medium can be observed in time. As a result, this method maybe more favorable to direct comparison with experiments. In addition, frequency dependence and loss can be included in this method [26, 27]. Using FDTD method one can also calculate dispersion relation of the PCs. However, utilizing FDTD method is a very tedious process as the selection of the initial excitation field is important to excite all possible modes. Similarly, the detection points shouldn't be placed at a high symmetry point. Moreover, if the structure has very sharp edges then uniform meshing may not predict the characteristics well enough. It may be needed to use non-uniform meshing and as a result, the size of the computational domain increases. For this reason, the combination of these two methods depending on the case is beneficial for the study of periodic dielectric structures. As a result, I performed the designs and analysis of PC structures employing these two methods.

**1.3. Outline of the Thesis**

The thesis is organized as follows. The general information and historical background of the photonic crystals are briefly introduced in the introductory section, chapter 1. Also in chapter 1, the necessary numerical tools for the study of the PCs are presented: the plane wave expansion method and the finite-difference time-domain technique are touched in that chapter.

The chapter 2 deals with the benefits of symmetry reduction in highly symmetric periodic photonic media. The main idea of the chapter 2 is to search for anomalous optical characteristics so that these types of PCs can be used in the design of novel optical devices by intentionally introducing reduced-symmetry to the PCs. Breaking either translational or rotational symmetries of PCs provides enhanced and additional optical characteristics such as creation of a complete photonic bandgap, wavelength demultiplexing, super-collimation, tilted self-collimation, and beam deflecting/routing properties. Utilizing these characteristics allows the design of several types of photonic devices such as beam deflectors, splitters, routers and wavelength demultiplexers.

In section 2.1 we present optical properties of crescent-shaped dielectric nano-rods that comprise a square lattice periodic structure named as crescent-shaped photonic crystals (CPC). The circular symmetry of individual cells of periodic dielectric structures is broken by replacing each unit cell with a reduced symmetry crescent shaped structure. The created configuration is assumed to be formed by the intersection of circular dielectric and air rods. The degree of freedom to manipulate the light propagation arises due to the rotational sensitivity of the CPC. The interesting dispersion property of designed CPC occurs due to the anisotropic nature of the iso-frequency contours that yield tilted self-collimated wave guiding. Furthermore, this feature allows focusing, routing, splitting and deflecting light beams along certain routes which are independent of the lattice symmetry directions of regular PCs. The propagation direction of light can be tuned by means of the opening angle of the crescent shape.

In section 2.2, we have investigated the optical properties of a new type of PC named star-shaped PC (STAR-PC) with anomalous iso-frequency contours. Intentionally introducing low-symmetry in the primitive unit cell gives rise to progressively tilting flat contours, which are observed in the fifth band of the transverse magnetic mode. Due to the intrinsic dispersive feature of the proposed PCs, *i.e.* tilted self-collimation, the incident signal with different wavelengths can be successfully separated in a spatial domain without introducing any corrugations or complexities inside the structure. We show numerical investigations of wavelength selective characteristic of the proposed PC structure in both time and frequency domains. The STAR-PC approach can be considered a good candidate for the wavelength division applications in the design of compact photonic integrated circuits. For the purpose of wavelength separation implementations, the proposed structure may operate within the wavelength interval of 1484.5*nm* - 1621.5*nm* with a broad bandwidth of 8.82%. The corresponding inter-channel crosstalk value is as low as -19 dB and the calculated transmission efficiency is above 97%.

In chapter 3, the detailed investigation of light propagation within the inhomogeneous, *i.e.* "Graded index (GRIN) medium", medium using ray and wave theory is reported. Consequently, in section 3.1 we proposed the design of an inhomogeneous artificially created graded index medium to enrich the optical device functionalities of light by using periodic all-dielectric materials. Continuous GRIN profile with hyperbolic secant index distribution is approximated using 2D PC dielectric rods with a fixed refractive index. The locations of each individual cell that contain dielectric rods of certain radii are determined based on the results of the frequency domain analysis. The desired index distribution is attained at long wavelengths using dispersion engineering approach. The frequency response of the transmission spectrum exhibits high transmission windows appearing at both larger and smaller wavelengths regions. Two regions are separated by a local band gap that blocks the incident light for a certain frequency interval. Light manipulation characteristics such as focusing, de-focusing and collimation are systematically and quantitatively compared for artificially designed GRIN medium within low and high frequency regimes. We show different field manipulation capabilities and focal point movement dynamics of the GRIN medium by special adjustment of the length of the structure. In addition, an analytical formulation based on ray theory is derived to investigate the focusing, de-focusing and collimation properties of proposed GRIN medium. The analytical approach utilizes Ray theory and computational tools are based on plane wave expansion and finite-difference time-domain methods. Implementing the GRIN medium by periodic optical materials provides frequency selectivity and strong focusing effects at higher frequency region. The designed structure can be used in integrated nanophotonics as a compact optical element with flat surfaces.

In section 3.2, we have worked on a novel mode conversion method using asymmetric GRIN (A-GRIN) PC structure. Proposed optical configuration enables transformation of the propagating mode from fundamental to higher order modes by utilizing A-GRIN structures. Refractive index variations of two different asymmetric gradient profiles, *i.e.* Exponential and Luneburg lens profiles have been approximated by two-dimensional photonic crystals. The basic structure is composed of constant radii with different lattice sizes. The designed GRIN mode converters provide relatively high transmission efficiency in the spectral region of interest and achieve the transformation in compact configuration. Numerical approaches utilizing the finite-difference time domain and plane wave expansion methods are used to analyze the mode conversion phenomenon of proposed GRIN PC media. Analytical formulation based on Ray theory is outlined to explore both ray trajectories and physical concept of wavefront retardation mechanism.

In chapter 4, we presented the design of a photonic structure for the generation of in-plane 2D limited diffraction beam. We have numerically investigated the characteristics of the light propagation passing through a two-dimensional square lattice annular type photonic crystal shaped in an axicon configuration. Careful selection of the operating frequency as well as the optimization of the apex rod position creates less diffracted beam whose transverse intensity profile closely resembles zero-order Bessel function. The created beam dramatically resists against the spatial spreading over a propagation distance of 50 micro-meters, after focusing with a spot size of ~0.23 micro-meters. The self-healing capability of the generated limited diffraction beam is demonstrated by placing obstacles with different sizes and shapes along the optical axis. The two features that accompany with such beams, i.e., diffraction-limited propagation and re-construction ability after encountering obstructions may strengthen its usage in manipulation of light propagation in various environments.

In chapter 5, we proposed a simple realistic two-dimensional complex parity-time-symmetric photonic structure that is described by a non-Hermitian potential but possesses real-valued eigenvalues. The concept is developed from basic physical considerations to provide asymmetric coupling between harmonic wave components of the electromagnetic field. The structure results in a nonreciprocal chirality and asymmetric transmission between in- and out-coupling channels into the structure. The analytical results are supported by a numerical study of the Bloch-like mode formations and calculations of a realistic planar semiconductor structure.

Finally, chapter 6 summarizes the achievements in the thesis.

## 2. REDUCED SYMMETRY IN PERIODIC PHOTONIC CRYSTAL CONFIGURATIONS

Meanwhile, research into non-periodic and disordered PC structures has attracted the much attention [28]. The interaction of photons with these types of structures allows exciting optical phenomena to be obtained; light scattering in disordered media may provide strong photon localization [29]. Disordered structures have potential in some applications such as random lasing, Anderson localization, sub-wavelength imaging, and novel light-source designs [30]. In a recent work, a compact spectrometer with high resolution was designed by intentionally introducing disorder into the photonic medium [31]. Utilizing random gain medium for lasing action is another research topic that exploits light scattering and amplification in disordered materials [32-35].

Periodic structures may be disadvantageous in some cases because of their high-symmetry. For example, high-symmetric structures are very sensitive to structural deformation. Moreover, the operating bandwidth may be quite small for the high-symmetric PC case. Besides, structural degradation during the fabrication process can be considered as another possible problem, since it causes deviation from the ideal cases. Lastly, unusual optical characteristics may be expected while reducing the symmetry of PC structures.

In addition to periodic and disordered PC configurations, quasi-crystals are a topic of much interest and have been intensively studied. Translational symmetry is broken in quasi-periodic structures, whereas rotational symmetry is kept intact [36]. Although random and disordered PCs do not have any spatial symmetry property, quasi-periodic structures possess a reduced symmetry characteristic; these types of periodic structures have high rotational symmetry and, therefore, anomalous characteristics may arise, especially in transmission spectra and photonic band gaps [37-40]. Due to the high rotational symmetries of quasi-crystals, their forbidden band gaps and light transport properties are superior to regular PCs [41-47]. Furthermore, using these types of PC designs, unique properties appear in transmission, reflection, refraction, localization, radiation of photons, symmetry in Fourier space, nonlinear optical, and diffraction characteristics. For example, enhancement of the light radiation in polymer light-emitting diodes has been achieved by using quasi-periodic PC structures [48].

Band diagram engineering has been performed, paying special attention to enlarging the gap opening and maximizing the gaps' overlap. On the other hand, both band movements (slope and form change of the dispersion curves) and degeneracy point splitting, at the symmetry points of the irreducible Brillouin zone, occur depending on symmetry-reduction in PC unit-cell. Iso-frequency contour engineering is an additional mechanism to inspect photonic periodic structures. For the low symmetry unit-cells, iso-frequency contours may indicate unique optical properties for photons. Symmetry reduced photonic media have great potential for several important concepts such as light propagation, reflection, refraction, slow-light, diffraction-free beam propagation, and wavelength de-multiplexing. In the present review, we discuss recent progress in the field by referencing papers, mainly by the current authors, and speculate on feasible, future research directions.

Square lattice PCs have translational symmetry with respect to lattice vectors, $a_1$ and $a_2$ so that the dielectric permittivity of the periodic structure can be defined by, $\varepsilon(r)=\varepsilon(r+la_1+ma_2)$, in which $l$ and $m$ are integers. As shown in Fig. 2.1(a), for the square lattice PC cylinders, the corresponding lattice unit vectors are, $a_1 = a\hat{x}$ and $a_2 = a\hat{y}$, where $\hat{x}$ and $\hat{y}$ are the base vectors in the spatial domain and a is the lattice constant. In addition to translational symmetry, two-dimensional PCs may have other types of symmetries such as mirror and rotational symmetries. If the PC design is invariant under the mirror reflection along the $x$-axis by the operation, $\sigma_x$ then the corresponding dielectric constant function does not change depending on sign of $x$, i.e., $\varepsilon(x,y)= \varepsilon(-x,y)$ Similarly, if the PC structure has a mirror symmetry under an operation, $\sigma_y$, then the dielectric permittivity function is invariant to the change of sign y, $\varepsilon(x,y)= \varepsilon(x,-y)$ The mirror symmetry operators, $\sigma_x$ and $\sigma_y$ are given as insets in Fig. 2.1(a). The rotational symmetry operation is another symmetry operation to be considered. It is denoted by $C_n$ which means the PC structure can be rotated by $2\pi/n$ radian in a counterclockwise direction about the origin without altering its geometry.

Figure 2.1(a) shows schematic representations of two examples of rotational symmetry operations, namely $C_2$ and $C_4$.

The band structure of a crystal provides significant information about its optical properties. When the PC lattice has rotational or mirror symmetry, then the band structures also have that symmetry. In such a case, we do not need to consider every k point in the Brillouin zone. The smallest region within the Brillouin zone is called the Irreducible Brillouin zone, where the symmetries in frequency bands cannot be taken into account. Figure 2.1(b) shows a schematic diagram of the first Brillouin zone of the square lattice PC, in which the Irreducible Brillouin zone is represented by the shaded region. On the other hand, when either the mirror or rotational symmetry of the structure is broken at a unit-cell scale, by reducing the symmetry of PC rods, the photonic band calculations in the Irreducible Brillouin zone are not sufficient anymore. Instead, every *k*-point at the edges of first Brillouin zone should be considered, and, thus, the band structure of low-symmetric PCs should be calculated along the [Γ-X-M- X1- M1- X2- M2- X3- M3- X- Γ] path, which is shown by the arrows in Fig. 2.1(b). In such a low-symmetric PC case, maxima and minima of photonic bands at high-symmetry points in the Brillouin zone may shift accordingly, which results in the variation of the band gap boundaries.

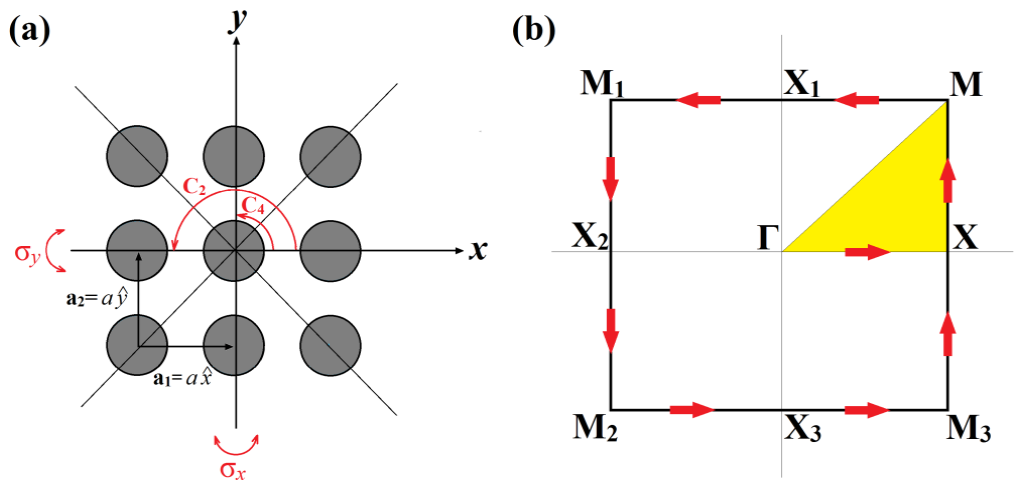

Figure 2.1. (a) Symmetry operations for the square-lattice PCs. (b) The corresponding Brillouin zone.

## 2.1. Crescent Shaped Dielectric Periodic Structure For Light Manipulation*

### 2.1.1. Introduction

The research in the field of photonic crystals was emerged in 1987 [4]. Since that time, highly symmetric periodic dielectric structures with a large refractive index contrast have been heavily investigated with an ultimate aim of achieving complete photonic band gap (PBG) that may appear in the dispersion diagram [4, 23]. The PC acts as a mirror reflecting the entire incident light wave whose wavelength falls inside the PBG region. The band gap features of pure periodic 3D and 2D PCs were soon demonstrated [25, 49]. Perturbing the periodicity of the PC may host artificially created optical modes that are surrounded by the upper and lower boundaries of PBG. Waveguides, sharp corners and cavities have become the ingredients of photonics research [50-52].

Meanwhile, it has been realized that the unperturbed structure also possesses rich spectral characteristics such as self-collimation, negative refraction, super-prism and super-lens [8, 53-55]. All these listed peculiar dispersion properties may not need structural defects. The two cases (unperturbed periodicity vs. broken periodicity) comprise various device applications frequently demanded for photonic devices.

Considering all of the previously investigated common PC configurations, we can conclude that PCs are highly symmetric structures and do possess fixed structural patterns. The two mostly explored and utilized PCs are square- and hexagonal- (also known as triangular) lattice photonic structures. The ingredient element of PC is usually circularly shaped unit cell although there are other types of unit cell shapes such as rectangular, elliptical or annular ones [56-60]. The ultimate aim of these studies is to achieve complete PBG for all polarizations (TE and TM) and design polarization insensitive optical devices [61-63].

In a rather different perspective, the re-oriented unit cells of the PCs may give rise to the implementation of graded index mediums [64, 65]. The implementations of GRIN via periodic structures provide great flexibilities in terms of designing different index gradient and photonic integrated circuit components such as couplers, lenses and super-bending device [66-71].

In this section, we propose a novel type of PC structure named as crescent-shaped photonic crystals. To the best of our knowledge, this structure has not been studied as a periodic dielectric structure, yet. In this study, the designed CPC enables us to arbitrarily route light beams by exploiting the engineered dispersion diagrams. There is no need to infiltrate any type of anisotropic material and the approach does not possess asymmetric PC patterns.

In the CPC structure, the geometrical adjustments are implemented at the level of unit cells not that of structural lattice arrangements. This brings extensive parametric tunabilities in realization of ultra-compact photonic integrated devices. Moreover, although CPCs are formed by isotropic materials, designed structure exhibits anisotropic optical properties similar to optical birefringence. The other unique feature of the CPC structure is due to the fact that the operating frequency of the structure can be easily shifted to any spectral region due to the scalability of the Maxwell's equations and availability of different lossless dielectric materials.

There have been various mechanisms that may induce optical anisotropy for light propagation in PCs. The anisotropy introduced into the periodic medium can be either in terms of selecting specific materials (dielectric parameter) or structural configuration (unit cell's shape or type) [72-79, 56, 58, 59, 61]. The optical properties of the former approaches can be dynamically tuned by an external applied electric field. The infiltrations of liquid crystals in 2D PCs involving anisotropic media were studied for tuning their photonic band structures [71, 76]. The lower symmetry periodic structures have been investigated for different applications. The rectangular lattice PC was used in the study of angular super-prism effect in Ref. 80 and broad angle self-collimation characteristic was explored in Ref. 81. PCs made of parallelogram lattice structure were investigated for light focusing device that utilize self-collimation phenomenon [82]. The self-collimated waveguide bends with different angles have also been implemented [83]. Special attention should be given while joining the rotated blocks of parallelogram lattice PC because the junction planes with complex geometries may be induced [82]. Similarly, the interface at the bend region should be carefully handled for the self-collimated waveguide bends [83]. As a result, these approaches may provide limited capabilities for beam deflecting and routing applications. However, the proposed structure in the present work enables us to easily integrate different blocks made of square-lattice crescent PC. The effects of symmetry reduction in PC were heavily explored with a goal of obtaining larger band gaps. We should emphasize that lower symmetry structures with complex configurations such as crescent-shape have not been investigated for the dispersion contours engineering and light manipulation applications. Instead of altering lattice type or introducing material anisotropy into periodic medium, we preferred to modify the circular shape of dielectric cylinders. The engineering of the iso-frequency contours (IFCs) can be performed at a level of unit cell and composite structures can be realized in such a way that the interfaces are free from complex geometries. It is possible to use other complex shape unit elements such as modified version of the crescent shape, U or V shapes instead of crescent one. However, it is expected that the degree of rotation of IFCs and focusing power may become different in each case. That aspect of the interpretation needs additional work which is kept for a future study.

### 2.1.2. 2D Crescent Shaped Photonic Crystals and Dispersion Analysis

In this work, we purposely break the circular (rotational, four-fold) symmetry of the unit cell by replacing it with a crescent shaped structure. The expectation is to enhance light manipulation capability inside the photonic structure without depending on artificially introduced structural defects. The geometrical shape of the individual cell provides the construction of complex photonic structures that may yield distinct spectral features as we show in the present work. It is versatile to tune the focal point locations and deflection angle of a beam via rotationally manipulating the structure. We show that the photon manipulation (propagation direction and focusing point) is greatly tailored due to the anisotropic nature of the IFCs. Introducing certain amount of rotational degree to each individual cells yields shifting of focal points along both *x*- and *y*-axes. It is worth noting that while rotational symmetry is lifted, we keep the translational symmetry intact. The beam flows along the direction which is dictated by the IFCs according to the following relation [84],

$$\vec{v}_g(x,y) = \nabla_k \omega\big(\vec{k}=(k_x,k_y)\big) = \frac{\partial \omega}{\partial k_x}\hat{x} + \frac{\partial \omega}{\partial k_y}\hat{y},  \quad (2.1.1)$$

where $k_x$ and $k_y$ represent wave-vector components along *x* and *y* directions, respectively.

The orientation of the CPCs strongly influences the direction of light propagation inside the medium. Fig. 2.1.1(a) shows the geometry of corresponding unit cell for two-dimensional (2D) CPCs. When it is spatially distributed in a square-lattice pattern, Fig. 2.1.1(b) appears as the schematic of the structure. The combination of two circular rods (one is made of dielectric and the other is air) in an overlapped form gives rise to a crescent shape. The regarding opto-geometric parameters describing the structure are denoted in the same figure as well. The refractive index of the CPCs is taken to be *n*=3.13 and the radii of the dielectric/air rods are denoted by $R_1$ and $R_2$. Their values are $R_1=R_2=0.30a$, where *a* is the lattice constant. The related unit cell filling factor, *f* is defined by the formula $fa^2 = \frac{2\pi}{3}R_1^2 - \frac{\pi}{3}R_2^2 + \frac{\sqrt{3}}{2}R_2^2$. In the case of $R_1=R_2=0.30a$ the value of the filling factor becomes 0.1722. The opening angle of the crescent shape is defined by $\theta$ and is altered by rotating the composite cell in clock-wise (CW) and counter-clock-wise (CCW) directions as shown in Fig. 2.1.1(a). The center to center distance of each circle is represented by *D* and this parameter is set to $D=R_1=R_2=0.30a$. The dimensions of the complete structure is denoted by $(L_i)$x$(W_i)$.

PWE method is performed in order to extract the dispersion characteristics of the CPC structure [85]. In our case, the photonic band structure calculations are traced along the Brillouin zone edges starting at the Γ point as can be seen in Fig. 2.1.1(c). In the spatial beam routing applications, we may not need any type of structural defects. In such a case, the shapes of the IFCs become crucial. Traditional PCs composed of cylindrical rods or holes provide symmetric IFCs with respect to *x*- and *y*-directions. On the other hand, lifting the symmetry of the predefined structure by radially shifting the location of inner air-rod brings anisotropic shape to the IFCs. Hence, the light propagation direction can be arranged by solely controlling crescent open-angle *θ*. The first band of the IFCs is isotropic due to validity of the effective medium theory [86]. The anisotropy occurs with respect to *θ* for the second and higher bands. For these higher order bands, there are three basic spectral characteristics. They are self-collimation, super-prism and focusing properties. The capability of the adjusting self-collimation direction (in this case it occurs not only along *x*- or *y*-directions but also along a certain angle) and the focal point of the light beam are the additional benefits of low-symmetry unit cell implementation. As a result, there is no need to alter the structure orientation or the incidence angle to adjust the focusing location.

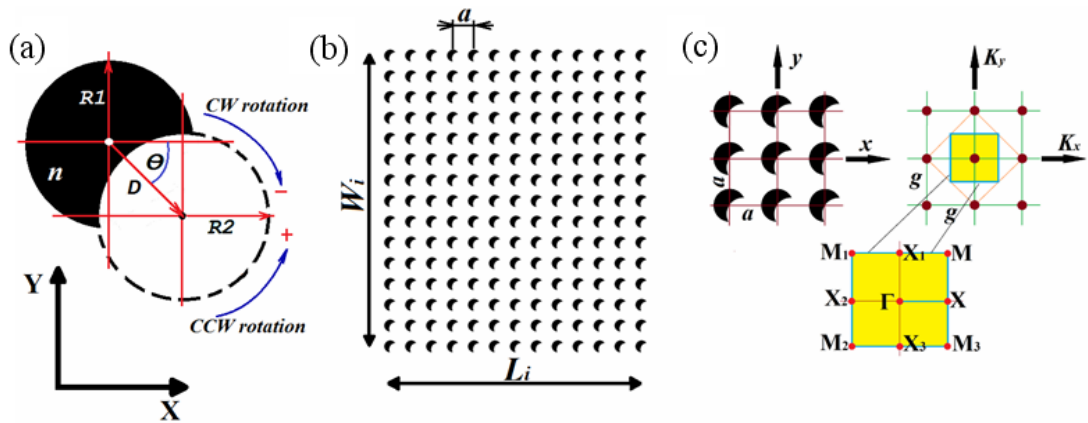

Figure 2.1.1. (a) The designed crescent-shaped PC (CPC). The refractive index of the dielectric rod is $n=13.13$ and the radii of two rods are $R_1=R_2=0.30a$. The distance between the circles is $D=R_1=R_2$ (b) Finite size square lattice CPC structure is created. (c) Brillouin zone of the structure.

Fig. 2.1.2 is a collection of dispersion curves when $\theta$ traces values from $0°$ to $90°$ in CCW direction. Frequency domain approach is employed while calculating IFCs. As we change the crescent open-angle $\theta$, different spectral regions appear in the dispersion plots. Figs. 2.1.2(a)-(c) correspond to different frequency contours chosen at fixed operating frequencies for each region: (1) $a/\lambda=0.416$, $\theta=(0°, 5°, 10°, 20°, 30°)$, (2) $a/\lambda=0.39$, $\theta=(40°, 50°, 60°)$, and (3) $a/\lambda=0.412$, $\theta=(70°, 80°, 90°)$. The three different frequencies are selected based on their IFC shapes so that strong focusing behavior is promoted. It can be clearly observed from Figs. 2.1.2(a)- 2.1.2(c) that CPC has tilted IFCs for the second band and the tilt amount can be regulated by only adjusting the crescent open-angle. It can be deduced from these figures that the orientation of the tilting amount directly depends on the steering of CPC opening. As the direction of light propagation is perpendicular to IFCs, the shift of focal point tracks an opposite path with respect to the crescent opening angle. These curves are selected to be representative cases of anisotropic IFCs that provide manipulation of focal point. In the next part of the paper, we present time-domain outcomes of the numerical studies.

### 2.1.3. FDTD Analysis of the Crescent Shape PC

The computational analysis of this section is based on time domain methods by employing two-dimensional FDTD [27]. In order to eliminate the back reflections coming from the ends of the finite computational window, the boundaries are surrounded by the perfectly matched layers [87]. We launched a source with a Gaussian distribution in the time-domain. For the numerical studies, transverse magnetic (TM) guided mode is used and the concerned non-zero electric and magnetic field components are $E_z$, $H_x$, and $H_y$. Then, the operational frequencies are chosen according to the different regimes of anisotropic IFCs as demonstrated in Figs. 2.1.2(a) - 2.1.2(c).

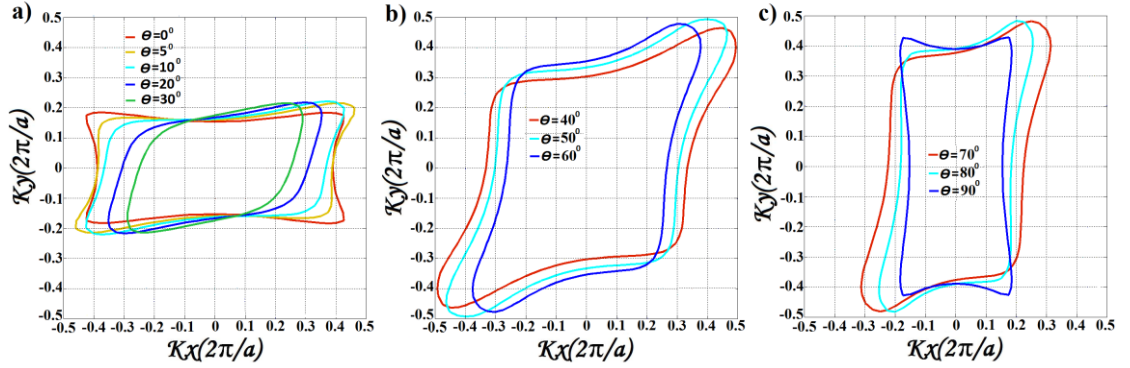

Figure 2.1.2. The operating frequency contours of the different crescent open-angle, $\theta$. The operating frequencies are $a/\lambda$=0.416 in (a) $a/\lambda$=0.39 in (b) and $a/\lambda$=0.412 in (c). The media file presents IFCs of different crescent open angle that varies from 0° to 90°.

The steady state electric field ($E_z$) intensity distributions of the CPC structures for the TM mode are shown in Figs. 2.1.3(a) - 2.1.3(c). The $\alpha$ parameter denotes the angle between the optical axis and focal point. The location of the focus is represented by $F$. We noticed that the $\alpha$ value can be changed by altering $\theta$. To exemplify, the crescent open angle is -30° in Fig. 2.1.3(a) and focal point occurs at above the optical axis. On the other hand, $\theta$ is 30° in Fig. 2.1.3(b) and then $\alpha$ becomes negative. The crescent open-angle $\theta$ parameter can be set as an input control parameter that is scanned between -90° to 90°. When the crescent open-angle $\theta$ is at 0°, the focal point location in y-direction is not changed and centered at the optical axis. An oscillation occurs in the structure and a strong focusing is observed at the end face of CPC (point $F_1$). The position of focal point is close to end face of CPC, as shown in Fig. 2.1.3(c). Due to interference of the side lobes, there occurs another secondary focal point which is represented by $F_2$. Three important remarks can be inferred from Figs. 2.1.3. First, the closeness of the focal point to CPC's end face is an indication of strong curvature (focusing power) due to special form of IFC. Second, the degree of anisotropy determines the amount of focal shift along y-direction, i.e. the values of $\alpha$. Finally, the output angle $\alpha$ depends on the input angle $\theta$ in a rather different manner. The functional dependency between the two parameters can be summarized in three sections as follows: first case is $\theta$=0°→$\alpha$=0°, the second case is 0° <$\theta$≤ 90°→ $f(\theta)$=-$\alpha$, and finally the third case is -90° ≤$\theta$<0° → $f(\theta)$=+$\alpha$. This dependency is summarized in Fig. 2.1.3(d). The maximum shift of focal point occurs when $\theta=-20°$. If one desires to obtain focal point residing on the optical axis, then $\theta=-0°, \pm 90°$.

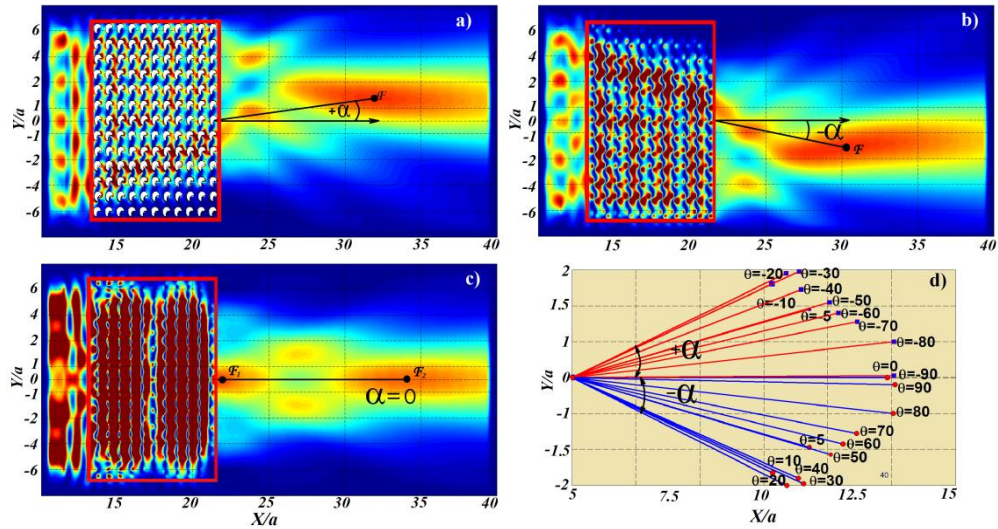

Figure 2.1.3. The steady state e-field intensity distribution of CPC structure is shown. (a) $\theta=-30°$ (b) $\theta=30°$ and (c) $\theta=0°$ (d) The schematic view of the locations of focal points and the output angle $\alpha$ variations for different $\theta$ values.

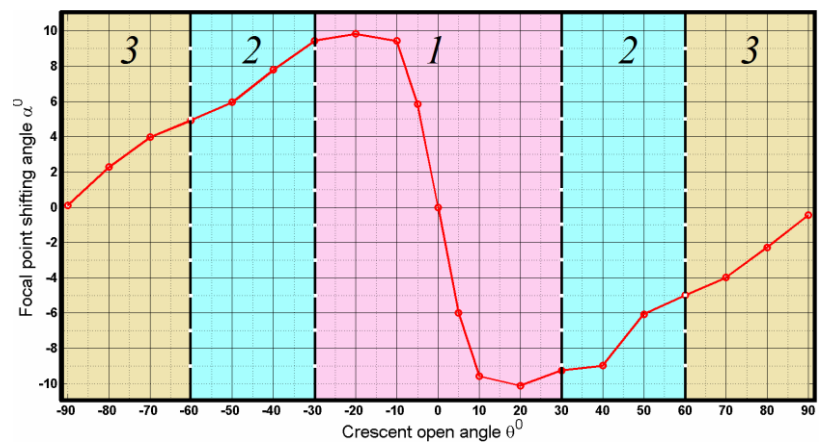

Figure 2.1.4. The dependency of $\alpha$ to $\theta$ parameter is sketched. There are three operational frequencies used for each region. The different colors designate the three regions.

In Fig. 2.1.4, the different operational frequency regions are displayed by different colors. The center frequencies of the input pulse for each region are set to $\omega_1=0.416(2\pi c/a)$, $\omega_2=0.390(2\pi c/a)$ and $\omega_3=0.412(2\pi c/a)$, respectively. We can see that variance of $\alpha$ with respect to $\theta$ resembles a sinusoidal pattern. The maximum lateral shift of focal point occurs at $\theta=\pm20°$ for a selected operating frequency. It can be seen from the figure that $\alpha$ initially increases quickly and then starts to decrease slowly as we increase $\theta$. When we consider the employed discretization process in FDTD small discrepancy occurs while reading the locations of focus points. With a finer spatial resolution, odd-symmetric version of the ($\theta$-$\alpha$) graph can be obtained.

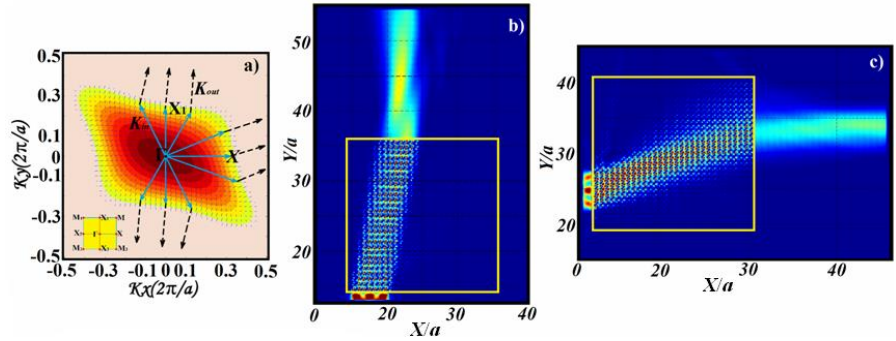

Figure 2.1.5. (a) Iso-frequency contours corresponding to the second band of the CPC with a crescent open angle θ=-30° The observed tilted self collimation characteristics along different propagation directions are presented in (b) and (c). The yellow boxes in (b) and (c) show the location of CPC. The normalized frequency is taken to be 0.416.

In addition to manipulating the focusing location, there is also a self-collimated behavior of the CPC. Similar to the previous results, we analyzed the IFCs of TM mode for the second band as shown in Fig. 2.1.5(a). The black dashed arrows in Fig. 2.1.5(a) represent the directions of the group velocities and blue arrows represent the wave vectors. The asymmetric characteristic of the CPC has direct impaction on IFCs. As a result, the calculated IFCs for the second band are deformed from a square shape to a tilted square as shown in Fig. 2.1.5(a). The steady state e-field is extracted by using 2D FDTD method to observe the tilted self-collimation properties of the CPCs when $\theta=30°$ and the results are presented in Figs. 2.1.5(b) and 2.1.5(c). To show this effect along tilted direction, the source wave is allowed to propagate along $\Gamma X_1$ and $\Gamma X$ directions, in Figs. 2.1.5(b) and 2.1.5(c), respectively. The flat contours can be used to laterally confine light in the CPC structure. In fact, for a range of incident wave-vectors, the propagation will be normal to the IFC. The flat portion of the contour allows input source to propagate inside periodic CPC without diffraction.

## 2.1.4. Discussions and Selected Applications

In the current work, we propose a novel type of photonic structure, called "CPC", and by means of this structure, we are able to design miniaturized optical medium that control both the propagation direction and focusing behavior of the electromagnetic fields. The great capability of CPC to adjust the focusing and deflection of light beams is due to lowering the symmetry of the proposed structure. The two fundamental light manipulation schemes were investigated: focusing and self-collimation effects. In addition to these features, one can implement beam splitters, routers and deflectors as well. The design methodologies are briefly depicted in Figs. 2.1.6(a) and 2.1.6(b). This can be achieved by advisedly combining differently-positioned CPCs with various crescent open-angles. For instance, suppose that the upper half of the structure lying above the optical axis has negative value for $\theta$ and the lower part has a positive value for $\theta$, as shown in Fig. 2.1.6(a). Then, the composite structure can act as a beam splitter. Half of the beam can be directed upwards and the other part is molded in the reverse direction. On the other hand, if the $\theta$ value is adjusted as gradually varying along the propagation direction (sweeping from $0°$ to $90°$), then beam routers can be implemented, which is schematically demonstrated in Fig. 2.1.6(b). The details of these proposals are kept the outside of the current study. However, we present an example that shows a two-step tilted light collimation process. To achieve this, a composite version of the structure can be obtained by cascading two pieces of CPC as shown in Fig. 2.1.7. While the first part has negative $\theta$, the second part may have positive $\theta$. The consequence of this combination yields self-collimated beam propagation having both positive and negative tilt angles. The source is placed at the left-side of the structure (the position is indicated by an arrow). The central part of the light beam follows the path that is highlighted with white arrows. When light travels inside the first part of the composite CPC, it bends upward. The second part of the structure routes the light wave downward. Due to the equal values of $\theta$ for both sections, the incidence and reflectance angles of the beam at the interface are equal to each other.

One of the observations that can be deduced from Fig. 2.1.7 is that e-field concentrates strongly at the sharp edges of each crescent shaped cells.

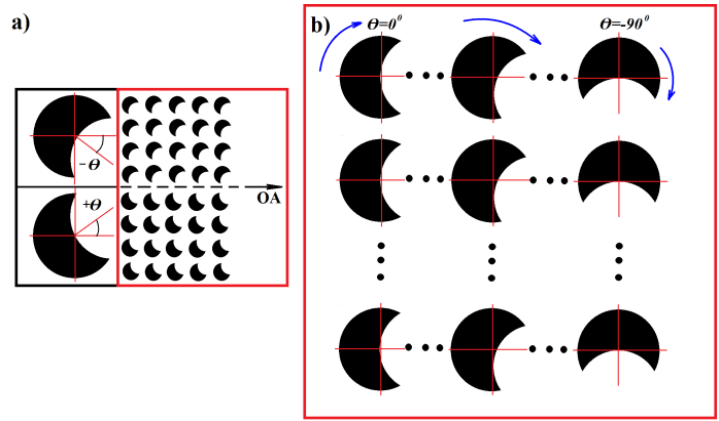

Figure 2.1.6. The representation of the construction methods of CPCs for various application areas: (a) the design of beam-splitting and (b) beam-deflectors and routers.

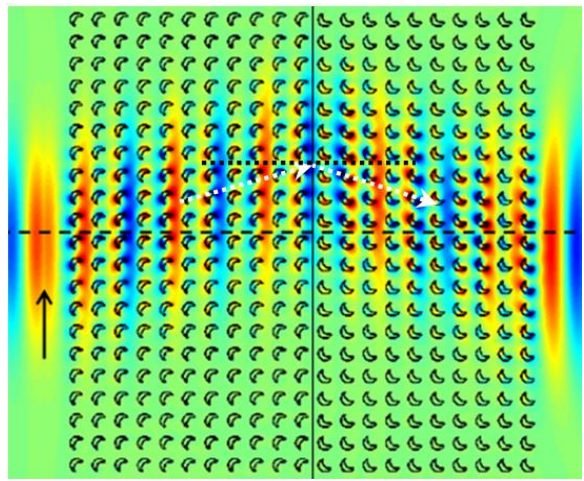

Figure 2.1.7. A composite CPC set up and steady-state electric field distribution. The cascade structure is obtained by combining two-block of CPC, one is negative $\theta$ and the second part has positive $\theta$. The blue and red colors correspond to minimum and maximum values of e-field's amplitude. Black arrow shows the location of source and the dashed-white one demonstrates the path of the propagation

The idea of splitting input power equally into two branches can be achieved by the help of the lower symmetry of CPC. The numerical investigation of Fig. 2.1.6(a) was performed and the result is shown in Fig. 2.1.8(a). The source is placed in the middle of the structure at the left side. The normalized operating frequency is selected to be $\omega_1=0.421(2\pi c/a)$. The light is divided into two self-collimated branches as can be observed from the plot. The amount of spatial separation between the two lobes at the end of the structure can be adjusted by means of CPC's length. The transverse e-field profile is represented in Fig. 2.1.8(b). Almost identical peaks show the successful splitting of light beam by using the designed CPC. By adjusting the location of input source, light splitting with variable intensity ratio can be achieved as well. In addition to that, splitting angle can be controlled by altering the opening angle of the crescent shape cells. The media file in Fig. 2.1.8(a) designates the propagation of the input light throughout the splitting structure.

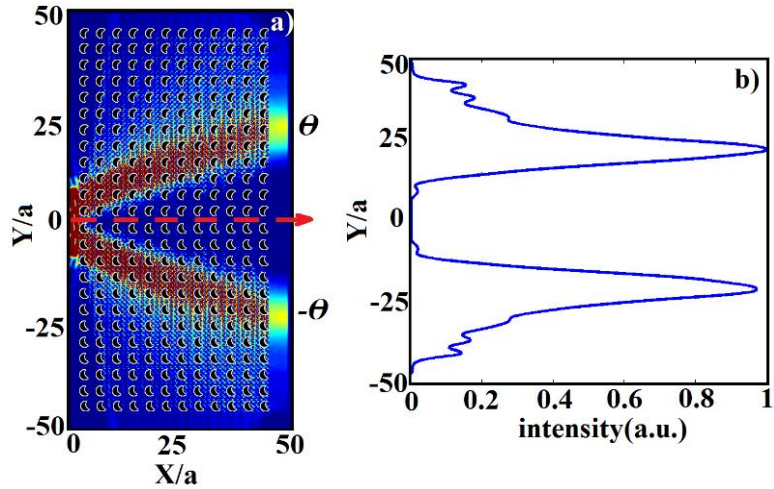

Figure 2.1.8. Beam splitter configuration. The upper and lower parts of the CPC have opposite angle $\theta=\pm20°$. (a) The steady-state intensity distribution of electric field throughout the structure. (b) The transverse intensity profile at the end of the structure.

One of the interesting properties of the CPCs is that although the material of the structure is itself isotropic, the formed structure may exhibit anisotropic characteristics due to its asymmetric shape of IFCs. For normalized frequencies above 0.40, the anisotropy ratio $a_r$, defined as $a_r=n_g(\Gamma X)/n_g(\Gamma M)$ is higher than 1.50 and $a_r=n_g(\Gamma X)\neq n_g(\Gamma X_1)$ [26]. This implies that CPCs can display different optical properties for different propagation directions of the same polarized light wave and can be approximated as an anisotropic media. Usually, the anisotropic feature of materials belongs to certain type of crystals that the nonlinear optics applications heavily use them [88]. The proposed structure may offer an alternative way to create similar optical effect (form birefringence) that can be realized by structural manipulation of pure transparent periodic dielectric materials. The response of the structure should be investigated for both polarizations. The scope of the present work is intended not to cover this property of the CPC.

Thanks to the recent development in the fields of applied physics and photonics, the difficulties on the fabrication of complex shaped PCs can be surmounted [89-92]. Thus, it is expected that the fabrication of theoretically designed CPCs can be realized by the state-of-the-art fabrication methods featured in semiconductor devices. E-beam lithography, focused ion beam lithography and atomic layer deposition technique can be among the choices. Besides, one can always introduce symmetry reduced configurations that may demand less difficulty for fabrication steps but still show similar effects for light beams.

Converting the normalized values such as lattice constant, structure dimensions *etc.*, in terms of measurable physical quantities gives us following results. When we tune frequency at 1550*nm* (for the normalized frequency $\omega_1=0.416(2\pi c/a)$), then the lattice constant *a* and the radius are 644.8*nm* and 193.4*nm*, respectively. The structure dimension becomes as $W_1=6.448\mu m \times L_1=7.737\mu m$. The focal point maximum shifting distance in the y-direction is equal to 1.289*μm*. The focal lengths for $\theta=30°$, $0°$, and $-30°$ are 1.2675*μm,* 0*μm* and 1.2675*μm*, respectively. Even though we outline the findings of square lattice dielectric crescent shapes in air background similar behavior can be obtained by utilizing the complementary structure (*i.e.*, air crescent shapes in dielectric background) patterned either by triangular or square lattice type.

## 2.2. Wavelength Demultiplexing by Star Shaped Photonic Crystal*

### 2.2.1. Introduction

Photonic crystals are periodic structures that are artificially designed in one-, two- or three-dimensions [2]. Regular symmetrical PCs display intrinsic dispersive characteristics in allowed frequency regimes, which can be listed as self-collimation [8], superprism and superlens [93] effects as well as absolute band gaps [4]. Conventional PCs are usually circularly shaped, whereas other types of unit cells are in various shapes such as rectangular, elliptical or annular [56-59]. In fact, introducing low-symmetry in the primitive cells present additional features in the dispersion contour engineering [94, 95] and enables enhancement of the photonic band gap properties [96]. Additionally, different optical applications have been investigated by employing low-symmetry in the PC structures, including: beam splitters [94], routers [97] and polarization insensitive waveguide designs [98]. In our present work, we propose a new type of PC structure named star-shaped photonic crystals (STAR-PC). To the best of our knowledge, this configuration has not previously been studied as a wavelength selective periodic dielectric structure. The designed STAR-PC enables us to manipulate the flow of light beams by exploiting the engineered dispersion diagrams at higher bands. By engineering iso-frequency contours, the proposed medium becomes sensitive to the incident wavelength changes and, in turn, can be thought as an alternative solution for the wavelength division applications. Wavelength Division Multiplexing (WDM) has been considered a promising concept for high capacity optical communication systems. This technology allocates each optical waveguide to different wavelengths by multiplexing them onto a single channel [99, 100]. An opposite system, which works in the reverse direction, is referred to as Wavelength Division Demultiplexing (WDDM).

Different techniques have been proposed especially for the design of WDDM systems, such as: arrayed waveguide gratings [101], Bragg gratings [102], multilayer thin film stacks [103], photonic crystal fibers [104], and graded index planar structures [105-107].

PCs can be modeled as a wavelength demultiplexer due to their intriguing dispersive properties, such as superprism [108] and negative refraction characteristics [109]. Utilization of planar PC waveguides [110], resonators [111], cavity, and defect-based PC filters [112] are among the other techniques that can be used as WDDM device. Moreover, one dimensional dielectric stack has been utilized to create a compact wavelength demultiplexer [113].

This study aims at investigating the wavelength selective capabilities of a newly proposed PC structure, STAR-PCs. We achieve a PC based wavelength-dependent spatial division without using specific materials, complex compounds, or introducing specifically optimized defects in the designed structure. Wavelength selectivity based on defects may expose difficulties on efficient input and output light couplings. Each defect should have different and finely tuned structural forms; consequently, there may be some stringent requirements for the fabrication process.

Wavelength division implementations using PC structures are mostly realized by employing dispersive property of PCs known as superprism. This phenomenon allows the separating of multiple wavelength channels with different separation angles [93]. Since the time that the superprism idea was proposed, researchers have gathered great interest in optimizing the superprism behavior to design feasible wavelength division devices [114-116]. However, wavelength division by means of superprism effect has limited wavelength resolution [117]. Spatially broadening the propagating beam brings the enlargement of structural dimensions in order to minimize inter-channel cross-talks [118].

To the best of our knowledge, this is the first time that the self-collimation concept is applied for the design of wavelength selective media. Self-collimated light beams propagate with almost no diffractions inside the PC structure, *i.e.* the spatial profile of the beam is almost preserved while it propagates inside the structure. Mixing of the adjacent channels can be avoided if the propagating beam has a small divergence angle. By means of diffraction-free beam propagation, enough spatial shifts between each channel can be introduced when the length of the device is increased.

**2.2.2. Geometrical Design of STAR-PC and Spectral Analyses of Extraordinary Self-collimation Effect**

The unit cell of low-symmetric STAR-PC in air background is shown in Fig. 2.2.1(a). The concerned structural parameters are demonstrated in the same figure. The proposed configuration has four vertices and eight edges. The width of each edge is denoted by *W* and the rotation angle of the unit cell in clock-wise (CW) direction is expressed by *α*, where in our case it equals to $0°$. Internal angle of each vertex is fixed at $\varphi=45°$. The schematic diagram of the square lattice STAR-PC with length *L* and width *H* are demonstrated in Fig. 2.2.1(b). The related lattice constant is defined by "*a*" and the refractive index of proposed PC is fixed to *n*=3.46 (Si-rods). Although the rotational symmetry is broken, the translational symmetry remains intact. Throughout this paper, the width of the edges is determined by *W*=0.30*a*. The dielectric filling factor is, then, equal to $F=3W^2=0.27$.

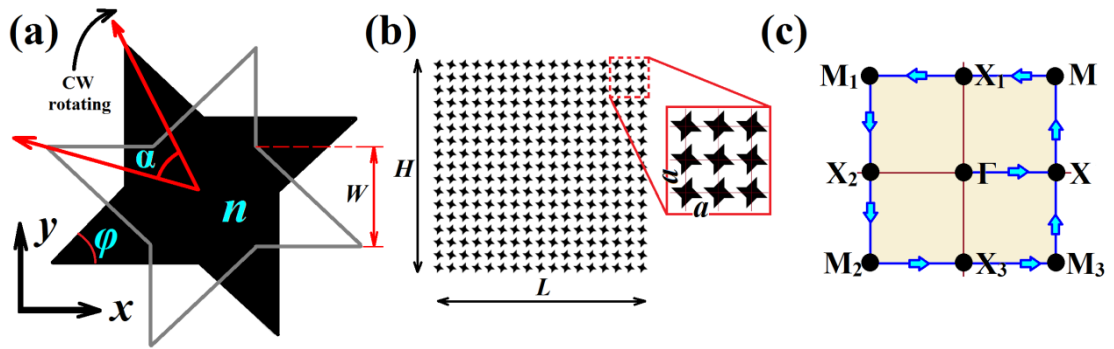

Figure 2.2.1. (a) Unit cell of proposed STAR-PC. It is composed of dielectric material (Si with the refractive index of 3.46) in air background. (b) Square lattice STAR-PCs covering a space of length *L* and width *H* and (c) corresponding Brillouin zone.

Breaking the structural symmetry in the unit cell causes striking effects on dispersion characteristics [8-10]. We have applied the plane-wave expansion method in order to obtain the band structure (dispersion relations) of Bloch modes and the IFCs regarding the designed STAR-PC structure. The dispersion analysis is performed by using MIT Photonic Bands (MPB) software [85]. Since the designed low symmetry STAR-PC has a complex geometrical shape, the total number of 16384 plane waves was used in the calculation to obtain convergent results. The PWM calculations are traced along all the edges of Brillouin zone due to low symmetry of the unit cell, as shown in Fig. 2.2.1(c). IFCs are created to quantify the allowed wavevectors in the designed structure and analyze the propagation behaviors of light beams through STAR-PCs.

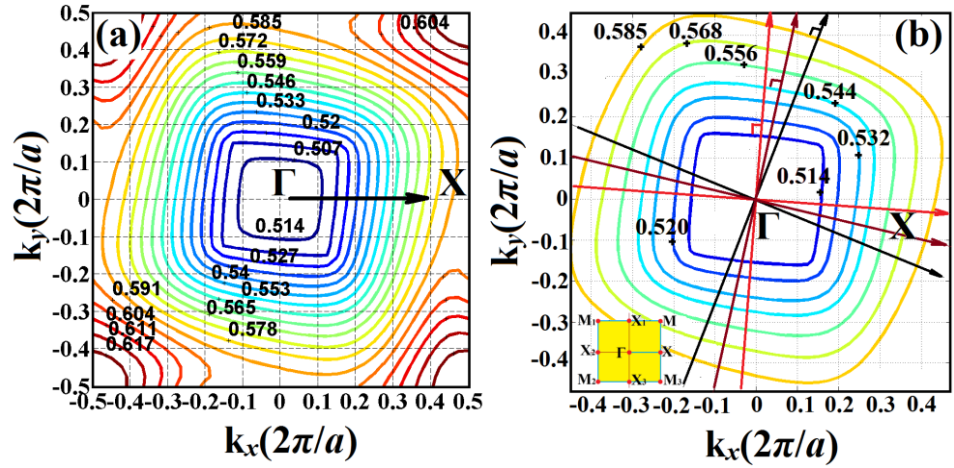

Figure 2.2.2. (a) Calculated IFCs of the fifth TM-band of square lattice STAR-PCs. and (b) detailed representation of the same band IFCs for the studied frequencies.

The propagated beam follows the direction which is determined by the IFCs according to the relation $\vec{v}_g(x,y) = \nabla_k \omega(k)$, where $k$ corresponds to wavevector and $\omega(k)$ denotes the related angular frequency. Here, the group velocity $\vec{v}_g$ of light inside the structure represents the velocity of energy transport in the direction perpendicular to the IFCs. Only transverse magnetic (TM) mode is considered. That means the electric-field has only $z$-component. The first band in the dispersion diagram behaves like an isotropic medium for lower frequencies. The magnitudes of the wavevector in the ΓM and ΓX directions are almost identical which results in circular shape IFCs. For the second TM band IFCs of STAR-PC, self-collimating phenomenon appears due to flattening of dispersion curves in the ΓX direction, which has already been studied in detail in Ref. 2. For higher bands, however, considerably different IFCs occur from that of conventional circular PC rods due to the lack of symmetry in the unit cell. The nontrivial structure of the cell, while affecting the first bands very little (as the first bands depend basically on the symmetry of lattice), can modify the higher bands more sensibly. The higher the bands, the stronger they are influenced by the symmetry of the cell. Hence, the corresponding fifth band IFCs, where wavelength division property emerges, are calculated and shown in Fig. 2.2.2(a). Interested IFCs for the study of wavelength selectivity behavior are represented in Fig. 2.2.2(b) and the corresponding Brillouin zone of STAR-PC is given as an inset. The designed structure is determined to be single-mode over the range of wavelengths by calculating and inspecting the dispersion diagram of the STAR-PC.

Interested wavelengths are carefully selected from the fifth band IFCs in Fig. 2.2.2(b). The tilt amount in the nearly flat IFCs increases while the normalized frequency gets larger. Therefore, light beam can follow different paths (directions) inside the periodic structure due to the self-collimating effect. This property enables spatially resolving the incident light beams with different wavelengths at the output of the structure.

Two dimensional (2D) FDTD simulation has been conducted via a freely available software MEEP developed at MIT to examine the wavelength division behavior of the STAR-PCs [27,119]. A grid of $\Delta x = \Delta y = a/30$ is implemented as a mesh size in 2D-FDTD calculations. Two types of input sources are used: either a continuous source which is needed in order to obtain spatial intensity distributions of the proposed structure, or a pulse with a Gaussian profile in time domain for computing the transmission spectra. The computational domain is surrounded by perfectly matched layers in order to eliminate back reflections at the boundaries. A square lattice structure having $H=180a$ and $L=400a$ is designed. A continuous source located in front of the structure is launched to excite the designed STAR-PCs.

The designed configuration is sequentially illuminated by a normal incident light beam with predefined wavelengths. The spatial size of the continuous source is taken to be 3*a*. FDTD results represented in Fig. 2.2.3(a) correspond to a collection of spatial intensity electric field distributions for selected three wavelengths within the interested region. The cascaded slices taken at the output of the field distributions are illustrated to demonstrate spatial shifts of the output signal. As can be seen in the same figure, the lateral beam shifting appears at the output interface due to the effect of tilted IFCs shape at the fifth TM band. The figurative representation of *n*-channel wavelength selective device is depicted as an inset in Fig. 2.2.3(a). As previously stated, the incident light with different wavelengths can be easily separated into the different channels without any structural corrugations or defects owing to dispersion characteristic of STAR-PC at the fifth band. The operating frequencies range from approximately $a/\lambda=0.520$ to $a/\lambda=0.568$ with a broad bandwidth of 8.82%. The concerning lattice constant is fixed to $a=843.2nm$ to allocate the desired wavelengths in optical communication range. In this case, the determined wavelengths {$\lambda_1, \lambda_2, \lambda_3$} are chosen in the range, where the wavelength selectivity property appears, and set to 1621.5*nm*, 1550*nm* and 1484.5*nm*, respectively. These wavelengths are selected in order to achieve high channel separation (low crosstalk). The calculated inter-channel crosstalk is around 19 dB, which implies low-level interference between the adjacent channels. We have calculated the input coupling efficiency that is above 97% and observed almost negligible insertion losses which are investigated in detail in the next section. These results demonstrate that our approach can be a prominent choice for the design of wavelength selective integrated devices.

Spatial separation of output signals in terms of their wavelengths may be considered another challenging issue. By means of diffraction-free beam propagation, enough spatial shifts between each channel can be introduced when the length of the device is appropriately arranged. The light beam is deflected from the optical axis (represented as a dashed line) by a propagation angle $\theta$ while travelling through STAR-PCs. The graph in Fig. 2.2.3(b) explains the relationship between the deviation angle $\theta$ and the wavelength $\lambda$ of the input signal. A total amount of $\Delta\theta=6°$ angle variation is observed in the wavelength range of 1484.5*nm*-1621.5*nm*. That result corresponds to a dispersion angle ratio of $\Delta\theta/\Delta\lambda$=0.044 (deg/*nm*). The almost linear relation between wavelength and inclination angle is apparent in the graph. This phenomenon strengthens the implementation of STAR-PCs as a wavelength selective medium.

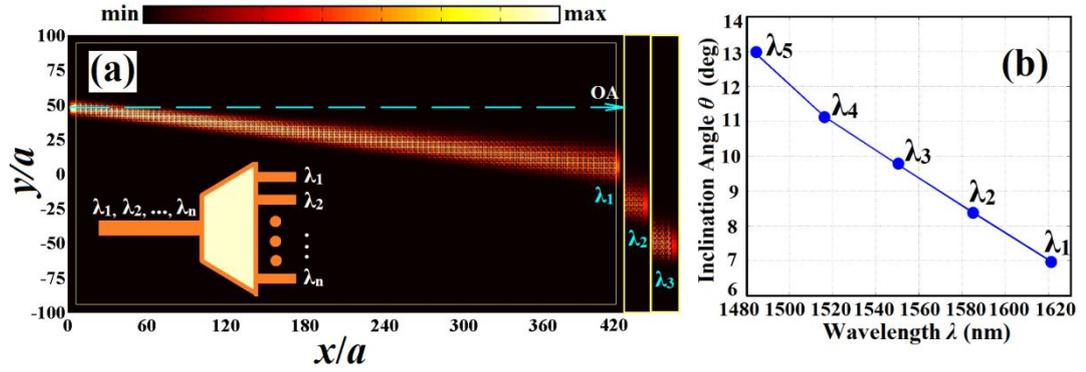

Figure 2.2.3. (a) The spatial intensity electric field distribution of square lattice STAR-PC with the structural parameters of (*L*, *H*) =(337.2*μm*,151.7*μm*). Inset shows illustration of n channel wavelength selective device. (b) Propagation angle *θ* within STAR-PC in terms of operating wavelength *λ* (in *nm*).

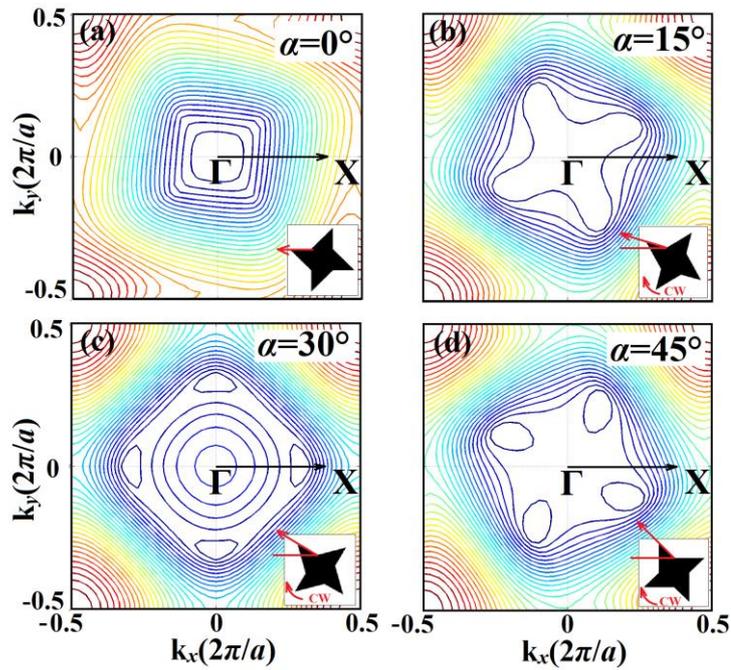

Figure 2.2.4. Fifth TM band IFCs of the rotated STAR-PCs. Corresponding angles are (a) 0°, (b) 15°, (c) 30°, and (d) 45°, respectively.

The newly proposed STAR-PC has four-folded ($C_4$) discrete rotational symmetry with intact translational symmetry. Dielectric distribution profile and rotational symmetry of the unit cell play a crucial role in the determination of the shape of IFCs [120]. As evident in Fig. 2.2.4, only the rotation of the four-folded symmetric primitive cell of STAR-PC in CW direction has a considerable influence on the contours' shape. An intrinsic wavelength sensitivity characteristic appears only for $\alpha=0°$ case, as shown in Fig. 2.2.4(a). However, in other cases, such as $\alpha=\{15°,30°,45°\}$, whose IFCs are shown in Figs. 2.2.4(b)- 2.2.4(d), this anomalous property either disappears (30° case) or loses its wavelength sensitivity strength (15° and 45° cases).

### 2.2.3. Discussion: Working Principle, Structural Imperfection Analyses, and Performance Comparisons

It is widely known that self-collimated light beams propagate with almost no diffractions inside the PC structure. Up until now, wavelength division implementations using PC structures are mostly realized by employing the superprism effect that produces high sensitivity to changes in both wavelength and incident angle. However, this propagating mode is subject to severe broadening due to diffraction. As far as we know, this is the first time that the self-collimation concept is applied for the design of wavelength selective media. Since the spatial beam profile is almost preserved as it propagates inside the structure, high crosstalk between adjacent channels can be avoided. By the help of the sub-diffractive beam propagation enough spatial shifts between each channel can be introduced provided that the length of the device is properly adjusted.

The designed structure operates in the fifth TM band and is a single mode over the operating frequency interval that is colored by a rectangle with dashed-line, as shown in Fig. 2.2.5(a). It should be noted that due to low symmetry in the primitive STAR-PC cell, the calculations of the photonic band structures should be traced along the [Γ-X-M-$X_1$-$M_1$-$X_2$-$M_2$-$X_3$-$M_3$-X-Γ] path of the Brillouin zone as demonstrated in Fig. 2.2.5(b). The shaded area in Fig. 2.2.5(a) indicates the linear band region with a group velocity of $v_g$=0.28$c$ in the frequency interval from $a/\lambda$=0.520 to $a/\lambda$=0.568, where $c$ is the speed of light in vacuum. As shown in Fig. 2.2.5(a), a linear slope at the fifth TM band is observed with respect to wavevector points. Investigating the dispersion bands with respect to all *k*-points in Fig. 2.2.5(a), one can see that the fifth TM band stays a single mode for all *k*-values. Thus, the modal dispersion due to multi-mode propagation within the same interval is inhibited. The linear structure at the fifth band supports the existence of the self-collimation characteristic for the designed structure [121]. For the analytical explanation of self-collimation effect in our PC design, the angular frequency of the studied TM band can be expressed as follows [122,123]:

$$\omega_n^{TM}(k_x, k_y) = d_0 + d_2 \mathbf{k}^2 + d_4 \mathbf{k}^4 + \cdots, \qquad (2.2.1)$$

where $\omega_n^{TM}$ denotes the angular frequency of $n^{th}$ TM band of the proposed configuration and $d_i$ (*i*=0, 2, 4, ...) parameter represents the diffraction coefficients. Only even powers of the wavevector $\mathbf{k}$=($k_x$, $k_y$) are considered since the designed structure possesses reciprocity. To calculate diffraction coefficients we can consider smaller values of $k_y$. Then $k_x$ can be expressed in terms of $k_y$ for a fixed angular frequency $\omega_n^{TM} = const$. In that case, the flat region of the interested spatial dispersion curve which denoted by $k_L$ in Fig. 2.2.6(a) in the propagation direction ($n$=5$^{th}$ band of the IFC) can be approximated by the corresponding equation:

$$k_x(k_y) = d_{0,n} + d_{2,n} k_y^2 + d_{4,n} k_y^4 + \cdots. \tag{2.2.2}$$

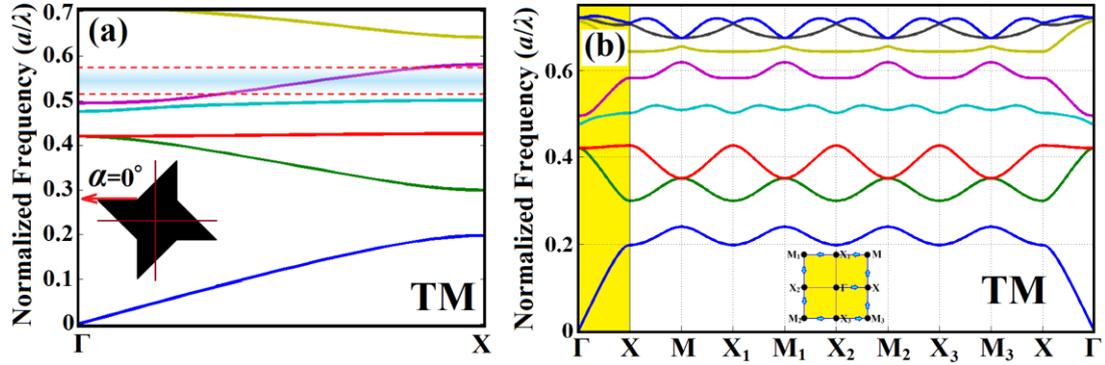

Figure 2.2.5. The representation of dispersion diagrams for STAR PC along the (a) Γ-X propagation direction and (b) all edges Γ-X-M-$X_1$-$M_1$-$X_2$-$M_2$-$X_3$-$M_3$-X-Γ of Brillouin Zone. Corresponding Brillouin Zone is given as an inset in the same plot.

Equation 2.2.2 is more precisely the approximation of the spatial dispersion for beam propagation along the ΓX direction for the fifth TM band.

The second-order diffraction parameter $d_2$ in Eq. 2.2.2 determines the type of diffractive behavior of propagating beam. In general, three types of diffraction exist, which directly depend on the values of $d_2$: positive (normal) diffraction, negative (anti-) diffraction and zero diffraction. In fact, the second-order diffraction coefficient $d_2$ is defined as the curvature of the transverse dispersion curve and it explains zero diffraction point phenomenon at a fixed frequency on a point in the $k$-domain. Thus, the non-diffractive propagation can be achieved only by an incident beam having infinitely broad beam widths. On the other hand, in the case when $d_2$ is set to zero (within the non-diffractive regime) the 4$^{th}$ order diffraction coefficient $d_4$ informs us of the finite transverse size of the light source that propagates without spatial broadening inside the PC structure.

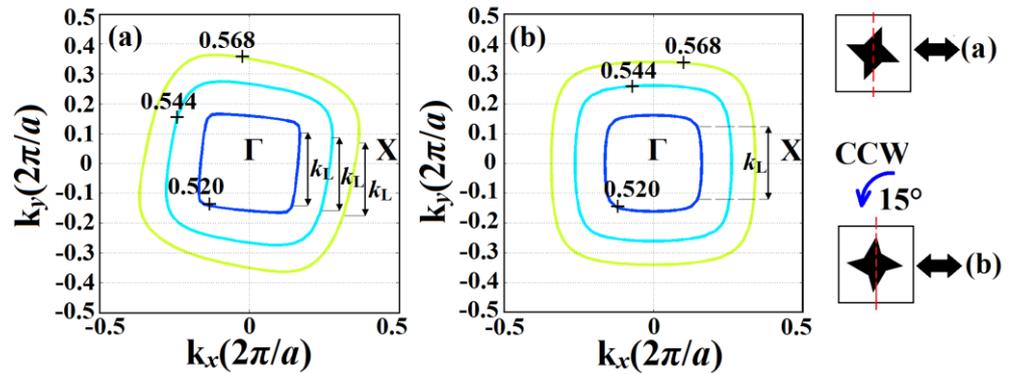

Figure 2.2.6. (a) and (b) representation of the IFCs of selected frequencies of square lattice STAR-PC and 15° degree counter clockwise rotated one for the fifth TM band, respectively. Corresponding STAR-PC unit cell schematic views are given as insets.

In order to calculate the diffraction coefficient parameters for the proposed low symmetric STAR-PCs, IFCs of the fifth TM band are employed. Then, $k_y$ dependence of angular frequency $\omega_n^{TM}$ is analyzed for fixed $k_x$ values. The determined second order diffraction coefficient $d_2$ ranges from 0.05737 to 0.07246 within the operating bandwidth, which corresponds to a nearly sub-diffraction regime [41]. To clarify the magnitude of the calculated second order diffraction coefficient inside the operating frequency interval we have also evaluated $d_2$ coefficient for the first band. The frequency interval from $a/\lambda=0.0919$ to $a/\lambda=0.195$ corresponds to $d_2$ values of 3.18 and 2.52, respectively. When we evaluate the same coefficient for the fifth band in the case of non-tilted IFCs the evaluated values become 0.05737 and 0.07246 for the normalized frequencies $a/\lambda=0.520$ and $a/\lambda=0.568$, respectively. The non-tilted IFCs appear after the unit cell is rotated 15° in the counter-clockwise direction as depicted in Fig. 2.2.6(b). The two operating regimes, *i.e.*, diffraction and sub-diffraction provide us an idea about the possible diffraction coefficient values [124]. In the case of zero-diffraction, the value of $d_2$ approaches to zero. The reason to have non-tilted IFCs in the calculation of $d_2$ coefficient is the concern about the diffraction/self-collimation feature of the STAR-PC without worrying about the frequency selectivity of the structure. It should be noted that, as shown in Figs. 2.2.6(a) and 2.2.6(b), the diffraction analyses are carried out for the selected operating frequency contours within the flat region, which is denoted by $k_L$ in the same plots.

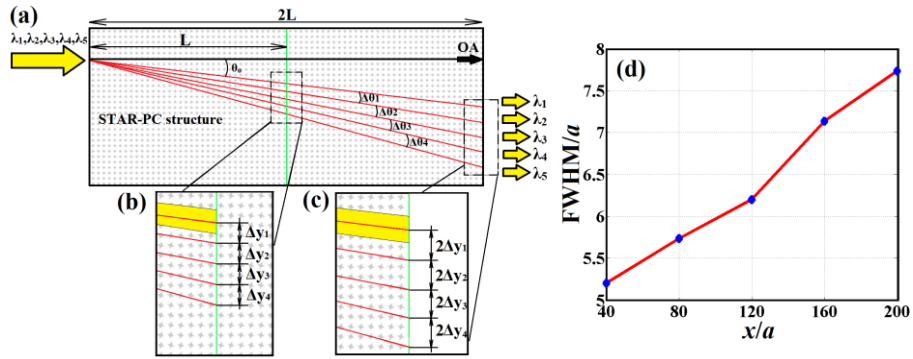

Figure 2.2.7. (a) Schematic representation of designed wavelength selective medium with ray trajectories of individual channels. (b) The magnified view of the ray paths at propagation distances of $L$ and $2L$ in (b) and (c), respectively. (d) Dependence of FWHM of the propagating beam on the propagation distance.

By means of diffraction-free beam propagation enough spatial shifts between each channel can be introduced when the length of the device is purposely increased. The tilted self-collimation property observed in the fifth TM band provides light propagation while experiencing limited diffraction inside the PC structure. Therefore, we can formulate the dependence of the vertical beam shifting on the propagation angle and propagation distance as follows: $\theta_n=\theta_{n-1}+\Delta\theta_n$, $n=1,2,3...$ and $\Delta y_n=L(tan(\theta_n)-tan(\theta_{n-1}))$. Figure 2.2.7(a) is the pictorial demonstration of the lateral separation at different wavelengths after propagating certain distances. The vertical shifting of the separated wavelengths depends on the propagation distance as well as the propagation angle, which can be inferred from last two expressions. In both expressions, spatial broadening of the beam is neglected. Figures 2.2.7(b) and 2.2.7(c) demonstrate the separations of incident signals at two propagation distances, *L* and 2*L*, respectively. Due to the self-collimation phenomenon of the STAR-PC structure, the incident beam propagates with limited diffraction behavior. The values of full width at half maximums (FWHMs) of beam inside the structure at different locations are observed in Fig. 2.2.7(d). FWHM increases by a factor of 1.29 when beam propagates from 100*a* distance to 200*a*. On the other hand, doubling the length of the structure induces a vertical shifting of 2Δ*y*. This implies that the spatial separation of different wavelengths can be accomplished by extending the lateral dimension of the proposed STAR-PC structure.

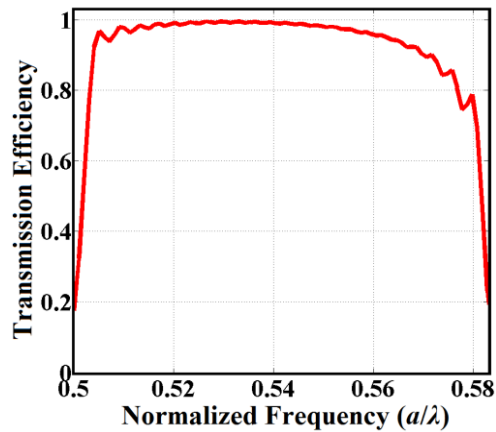

Figure 2.2.8. Representation of the transmission efficiency plot including the input and output coupling losses.

Another important factor is the efficient excitation of the relevant mode and ensuring that the collimated mode stays single mode. To underpin that feature of the STAR-PC structure we conducted additional numerical simulations with the Gaussian source located at the outside of the PC structure. The group index of the band at the flat section is ~3.60. Therefore, it is expected that insertion loss is small and the absence of internal reflection at the output face would produce relatively high power transmission. Figure 2.2.8 demonstrates the input coupling efficiency of STAR-PC when the source is located at the outside of the structure. That transmission graph explains that the special form of the unit cell provides higher coupling efficiency in the frequency range between $a/\lambda$=0.515 and $a/\lambda$=0.570. The measured input coupling efficiency is over 97% within the operating range and almost all the incident power can be transmitted inside STAR-PC. The zigzag-shaped unit cell of the STAR-PC structure here may lead to efficient direct coupling of incident beam from the air [125]. Therefore, we can conclude that the insertion loss is almost negligible in the case of STAR-PCs, as can be seen in Fig. 2.2.8.

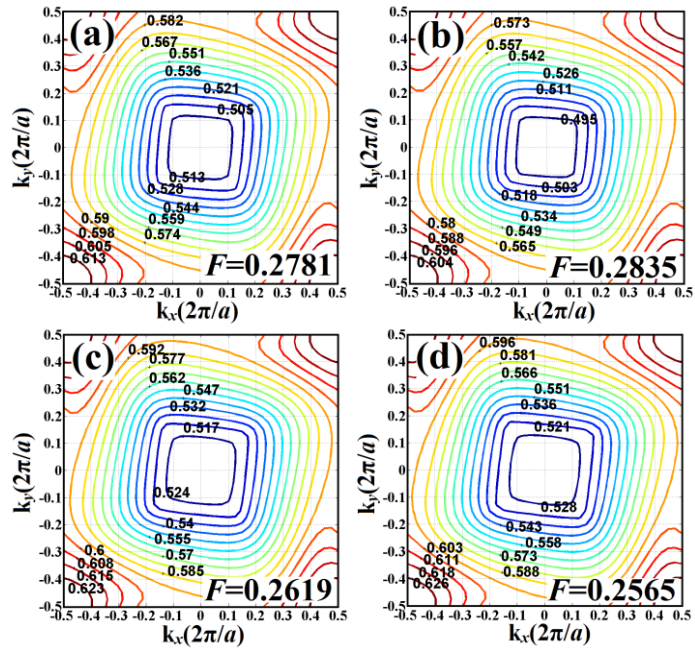

Figure 2.2.9. (a) IFCs of fifth TM band of the STAR-PC by varying the filling factor that is increased by (a) 3% and (b) 5% from the original filling factor $F$=0.27. Similarly (c) and (d) present the cases of 3% and 5% decreasing the filling factor. Regarding filling factor values depicted as an inset in the same plots.

It is worthy to check the sensitivity of structure against structural imperfections possibly caused by the fabrication process. For this reason, we have altered the filling factor of the STAR-PC unit cell by changing the width of the edges $W$. Figure 2.2.9 demonstrates the response of the fifth band IFCs to the changes of filling factor which deviates from the original one, *i.e.* $F=0.27$. Firstly, in order to increase corresponding dielectric filling factor $F$ of the STAR-PC we increase the width of the edges $W$ in Fig. 2.2.1(a). Figures 2.2.9(a) and 2.2.9(b) represent the cases where dielectric filling factors of STAR-PC unit cell are increased by 3% and 5%, respectively. The increment in the dielectric amount of the PC unit cell gives rise to the shifting of the operating frequency toward lower frequency regions. In the opposite case, Figs. 2.2.9(c) and 2.2.9(d) represent the IFCs of the STAR-PC having diminished filling factors of $F=0.2619$ (-3%) and $F=0.2565$ (-5%), respectively.

Table 2.2.1. Operating frequency changes with respect to the filling factor $F$ deviation.

| | Change amount in $F$(%) | Filling factor ($F$) | $W(a)$ | Operating frequencies ($a/\lambda$) | Operating wavelengths (nm) | Band width (%) |
|---|---|---|---|---|---|---|
| No change | 0 | 0.27 | 0.3000 | 0.520 0.544 0.568 | 1621.5 1550.0 1484.5 | 8.82 |
| Increasing | 3 | 0.2781 | 0.3044 | 0.516 0.540 0.564 | 1634.1 1561.4 1495.0 | 8.88 |
| Increasing | 5 | 0.2835 | 0.3074 | 0.512 0.536 0.560 | 1646.8 1573.1 1505.7 | 8.95 |
| Decreasing | 3 | 0.2619 | 0.2954 | 0.528 0.552 0.576 | 1596.9 1527.5 1463.8 | 8.69 |
| Decreasing | 5 | 0.2565 | 0.2924 | 0.532 0.556 0.580 | 1584.9 1516.5 1458.8 | 8.28 |

As evident in Figs. 2.2.9(a) - 2.2.9(d), the fifth band IFCs preserve its original rotated/tilted form and we can only observe small deviations on the range of operating frequencies. We have realized that the tilted self collimation property is still applicable with an acceptable tolerance limits of ΔF=±5%. The calculated results are collectively represented in detail in Table 2.2.1. While *F* is varied between 0.2781 (+3%) and 0.2835 (+5%), the operating frequencies change from 0.516 to 0.564 and from 0.512 to 0.560, respectively. Furthermore, the increment of filling factor leads to 1.5% variations in the WDM operating bandwidth. In the case of filling factor decrease by 3% (*F*=0.2619) and 5% (*F*=0.2565), the operating frequencies deviate between 0.528 and 0.580. The result in the table implies that even though there may be undesired deviation of the STAR-PC unit cell size during the fabrications process, the rotated form of the fifth band IFCs can be still preserved. Moreover, the tilted self-collimation property (flatness is not affected) and light beam deflecting angle *θ* also remained unchanged. Hence, even though there is a deviation in filling factor in the limit of ±5% by appropriate tuning of the operating frequencies, wavelength separation can still be achieved.

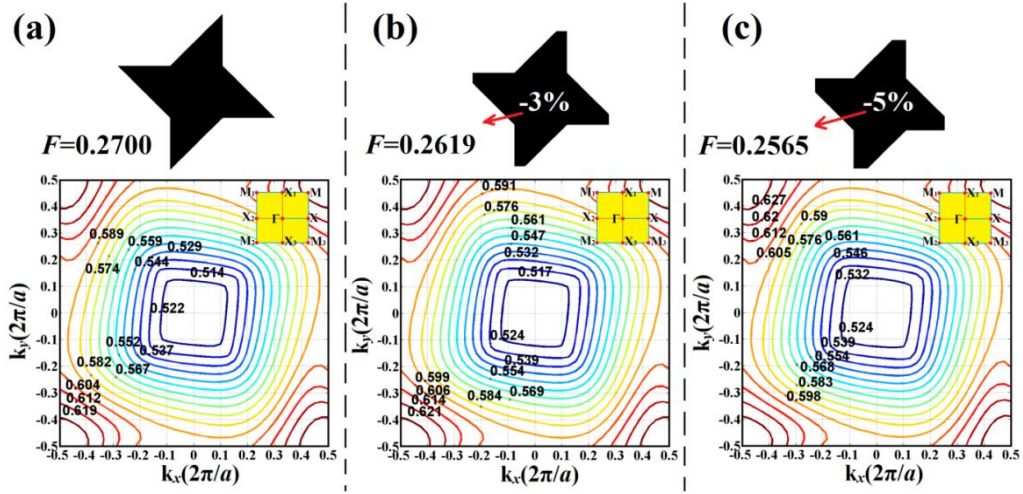

Figure 2.2.10. The analyses of the sensitivity of designed STAR PC to any kind of rounding of the tips. (a) The schematic view and the corresponding fifth band IFCs of STAR PC unit cell without corrugations. (b) and (c) are of corrugated cases, in which the regarding filling factors are reduced by 3% and 5%, respectively. Regarding Brillouin zone boundaries are given as insets in the same plots.

In order to investigate the imperfection condition, i.e. sensitivity to any types of rounding at the STAR-PC vertices, the following scenario is considered. Figure 2.2.10(a) shows the schematic of the proposed STAR-PC in its original form with corresponding fifth band IFCs. In the cases of Fig. 2.2.10(b) and 2.2.10(c), the vertices of the STAR-PC are intentionally flattened so that regarding filling factors are reduced by 3% and 5%, respectively. While comparing with the original one as in Fig. 2.2.10(a), the intrinsic forms of dispersion contours do not sense the structural deformations at the tips of STAR-PC. Hence, the rotated shape of the fifth band IFC remains unchanged. Since flattening of the edges of the unit cell means reducing the dielectric filling factor, the operating frequency range moves to higher frequencies than expected. The unit cell symmetry plays an important role in constructing the shape of the dispersion contours in higher bands [120]. In our case, even though the vertices of STAR-PC are intentionally deformed so that filling factor is varied in the limit of 5%, the overall $C_4$ rotational symmetry in proposed PC structure is maintained. These results imply that our proposed design is durable to the possible undesired imperfections that may occur during manufacturing process.

It is necessary to include a comparison of the present work with other related demultiplexing approaches in terms of bandwidth, crosstalk, and implementation issues such as source alignment and oblique incident case. Wavelength division using PC structures are mostly realized by employing dispersive property of PCs called the superprism effect [93]. This phenomenon allows the separating of multiple wavelengths with different separation angles according to the anomalous dispersion curves. However, wavelength division by means of superprism effect has a limitation due to the diffraction of the light upon propagation [117]. On the other hand, the presented demultiplexing approach maintains spatial mode distribution of the separating beam intact within the STAR-PC, whereas, in the superprism based demultiplexing devices, the incident source's spatial mode profile becomes deformed, which may adversely affect efficient coupling to output channels. In the case of the superprism based WDM configurations, additional focusing apparatus such as superlenses are required to overcome the diffraction of output signals and make them collimated [108]. Reduced crosstalks between adjacent output channels are another concern to be carefully considered. Superprism based devices are highly exposed to diffractions so that the regarding crosstalks inevitably take higher values [109, 115]. The related crosstalk in our design, however, gets a value around 19 dB due to anomalous sub-diffractive property of STAR-PCs. In terms of bandwidth values, the present work possesses large bandwidths reaching maximum value of 8.82 %. On the other hand, Gerken *et al*. studied a concept to obtain wavelength multiplexing and demultiplexing devices by exploiting group velocity effects in thin-film filters [126]. Although high spatial shift at the exit interface is achieved, the structure operates within the narrower bandwidth wavelength region that equals to 1.67 %. Finally, it is desirable to have easy implementation of design in practice. There should be no strict restriction in terms of the source location and incident angle value. In this regard, there is no alignment issue of the input source position in the current work and oblique incidence angle satisfying the condition $|\alpha| \leq 12.62°$ is essential for proper operation of the WDM structure [127]. On the other hand, in case of GRIN PC based WDM device, the incident angle deviation can be considered as an important issue [106,107]. Thus, the regarding ray path in the GRIN medium depends highly on the light incident angle and the source location.

## 3. THE GRADED INDEX PHOTONIC CRYSTALS

The terms gradient index and graded index in literature are often used to describe an inhomogeneous medium in which the refractive index varies from point to point [128, 129]. The GRIN media occur commonly in nature. Examples are the crystalline lens and the retinal receptors of the human eye, and the atmosphere of the earth. The atmosphere of the earth has a refractive index that decreases with height because the density decreases at higher altitudes. Many unusual atmospheric phenomena, a mirage being the best-known example, result from the bending of the rays of light by this gradient.

There are three basic gradient index types [130]. The first is an axial gradient where the refractive index varies in a continuous way along the optical axis of the medium. The second is a radial gradient where index profile varies continuously from the optical axis to the periphery along the transverse direction in such a way that iso-indicial surfaces are concentric cylinders about the optical axis. The last type is the spherical gradient where index changes symmetrically around a point so that iso-indicial surfaces are concentric spheres. Historically, Maxwell [131] was among the first who considered inhomogeneous media in optics, when, in 1854, he described a lens with gradually varying of refractive index, known as Maxwell's fish eye, of spherical symmetry with the property that points on the surface, and within the lens are sharply focused at conjugate points. In 1905, Wood [132] designed a cylindrical lens by a dipping technique whereby a cylinder of gelatin is produced with refractive index axial symmetry.

Next 40 years later, Luneburg [128] investigated ray propagation through inhomogeneous media and he also analyzed light propagation through a GRIN medium with hyperbolic secant refractive index profile. In 1951 Mikaelian [133] analytically described using of hyperbolic secant refractive index profile to provide focusing effect of cylindrical rod type lens. It is proved that in a GRIN medium, the optical rays have curved trajectories. By an appropriate choice of the refractive index distribution, a GRIN medium can have the same effect on light rays as a conventional optical component, such as a prism or a lens. The possibility of using GRIN media in optical systems has been considered for many years, but the manufacture of materials has been the limiting factor in implementing GRIN optical elements until the 1970s [129]. In the last years, however, many different gradient index materials have been manufactured. The revival of GRIN optics has not been casual; it is connected to a considerable degree with the enormous development of optical communications systems, integrated optics, and micro-optics. However, these recently evolved fabricating processes are limited by the small variation of the index, the small depth of the gradient region, and the minimal control over the shape of the resultant index profile. Merging of two different disciplines such as Graded Index Optics and Photonic Crystals can be considered as a plausible solution for such types of fabricating and designing difficulties. In this chapter of the thesis we have proposed PCs based GRIN configurations to analyze focusing and mode order transformation phenomena.

## 3.1. Design of Flat Lens-Like Graded Index Medium by Photonic Crystals: Exploring Both Low and High Frequency Regimes*

### 3.1.1. Introduction

The interaction of photons with the dielectric structures creates unique properties in the electromagnetic spectrum if the structure has wavelength-scale geometrical features along with a high-contrast refractive index variation instead of uniform/homogenous medium.

When the structural appearance of dielectric medium possesses certain type of spatial periodicity, then, these artificial structures are termed as photonic crystals [2]. The manipulation of photons becomes relatively easy due to the crystalline nature of the PCs. Since the pioneering works in Refs. [2] and [3] the PCs become increasingly popular, especially, in the applications of photonic integrated circuits. In the dispersion relation, certain number of forbidden frequency intervals can be created so that no matter what is the incidence angle, the electromagnetic wave is blocked to penetrate inside the structure. Within the forbidden frequency interval, cavity and waveguide modes can occur if artificially created defects are introduced inside the periodic structure. Therefore, utilizing photonic band gap peculiarity gives birth to the broad range of PC applications such as optical mirrors, switches, and filters [4]. Moreover, due to the high tunability of the dispersion behaviors in allowed frequency regions, some of the anomalous properties of PCs which are highly dispersive prism, super-prism, self-collimation, self-guiding, routing, negative refraction phenomena can be revealed [8, 55, 93,134-136].

In the last fifty years, the ability of controlling the flow of light further improved by introducing the idea of a graded index optics (GRIN) [128,129,137]. The GRIN medium can be characterized as an inhomogeneous medium in which refractive index varies in gradual manner along radial, axial or spherical directions [130]. In general, we witness GRIN media in nature via different means: some examples are crystalline lens of the eye, the atmosphere of the earth and the mirage effect. As mentioned before, the possibility of employing GRIN media in optic and photonic systems has been attracted a great attention of scientists in that area. The widely employed version of a GRIN medium can be found in the fiber optic technology that is known as GRIN fibers [138-140]. The index distribution imitates a quadratic form (parabolic refractive index profile) that has maximum value at the core of the fiber and decays to a lower refractive index value along the radial direction. The unique feature of GRIN fibers is the nature of rich modal dispersion relation. The non-uniform index distribution enables different order of modes to travel different distances at equal times. As a result, modal dispersions can be compensated in a GRIN fiber [141-144].

It is known that light-rays follow curved trajectories in a GRIN medium. Consequently, curving the light path gives birth to the optical effects same in conventional optical elements with curved interfaces such as focusing, diverging or collimation. As a matter of fact we should mention the great work of J. C. Maxwell who was the first that considered GRIN media as an optical lens known as Maxwell's fish eye in 1854 [131]. After decades, the ray propagation through the GRIN media is mathematically formulated and deeply analyzed by Luneburg [128]. These pioneering works paved the way for future developments in the field of GRIN optics and its optical applications [137].

Several methods of producing and manufacturing GRIN materials such as ion diffusion, chemical vapor deposition, copolymerization and monomer diffusion, silver ion exchange, *etc.*, have been developed. However, several challenging issues can arise during fabrication processes such as limitation of the index gradient variation, the small depth of the gradient region and high precision control over the shape of the resultant index profile. In order to overcome such types of difficulties, PCs based GRIN configurations can be considered as an alternative solution.

It is feasible to design a GRIN if the parameters of the two dimensional (2D) PCs is rearranged appropriately. These structures are known as graded index photonic crystals and can be designed by engineering of PC parameters, such as gradual changing of filling factor, lattice period, and/or material index [67, 70, 145]. In recent years, the GRIN PC structures have been contributed to numerous nano-photonic and optical applications that prolong from optical mode couplers [66, 146] to the design of effectively focusing lenses [67-69,147]. Moreover, revealing the mirage effect, efficiently guiding and manipulating the flow of light can be achieved by the help of GRIN PC concept [65, 71, 148-149].

In this study, a novel type of GRIN PC lens structure whose refractive index distribution is adapted to the hyperbolic secant (HS) function is introduced. In order to design such a GRIN PC medium the location of dielectric rods with constant radii are modified according to the desired index distribution. Dispersion engineering method has been applied to attain the appropriate effective refractive index form. Focusing, defocusing and collimation behaviors of the proposed HS GRIN structure is deeply investigated by analytical approach based on the geometrical optics, i.e., ray theory where the corresponding analytical conditions for focusing, collimation and diverging cases are mathematically formulated. Moreover, proposed GRIN PC structure is investigated under both low and high frequency regions which yield novel outcomes in terms of light manipulation such as strong and selective focusing of light properties. Frequency and structural length dependence of focusing properties of GRIN PC are extensively analyzed and affirmative conclusions are conducted.

In our previous publications, we have indicated the presence of a high transmission window and its potential for different applications such as waveguide coupler, bending and mode transformer designs [150-152]. On the other hand, when the literature is carefully searched, one can encounter to the limited amount of works where GRIN concept is also investigated under non-homogenization region [106,153-155]. In particular, the super-bending and mirage effect in GRIN PCs are explained by using photonic band anisotropy phenomena, which occurs at higher bands, i.e., engineering of regarding IFCs by modification PC unit cell parameters (dielectric filling ratio) [153]. Furthermore, a special attention is paid for light propagation within the GRIN PC at short-wavelengths in Ref. 154. In the same work, contrary to long-wavelength regime of GRIN PC, the dispersive phenomenon at high frequency bands is considered, where the concept of light path curving with changing wavelength is exploited. Also this approach is experimentally demonstrated for optical wavelengths [106]. Recently, flat lensing concepts occurring at higher bands are deeply investigated [156-158]. The focusing feature is explained by means of convex-curved frequency contours of the PC structures that yield near field focusing. To the best of our knowledge, the lensing and waveguide property of proposed HS GRIN PC at non-homogenization region are deeply studied for the first time in the present work considering light focusing behavior and aberration dependency aspects.

The findings of the current work are expected to have important implications in many fields especially in the photonics area. For instance, the discrete lenses can be replaced by more compact flat devices. The bulky lenses have curved surfaces, small focusing power and are susceptible to alignment problems. The designs can be easily transformed into other frequency regimes. Another critical point is that only the nonlinear optics and fabrication procedures which are complex and expensive may create graded index materials. However, the proposed approach in the present work has a potential to take over the design tasks of GRIN media. In that way, previous remarks will directly influence the aforementioned research fields positively. Lastly, findings of the present work can be applied to acoustic waves with graded sonic crystals.

## 3.1.2. Geometrical Explanation of Light Propagation through the GRIN Medium Using Ray Theory

In general, Ray theory works as a powerful tool when the scientists in optics and photonics area deal with propagating light waves in inhomogeneous media such as in graded index medium. In order to geometrically explain the progression of light waves through the object whose dimensions are much greater than the wavelength, the Ray theory can be considered as a plausible solution. Unlike wave optics, ray theory does not assume any wave characteristic of light and treats the propagation of light as a straight-line except for changes of the propagation direction induced by reflection or refraction. In particular, Ray theory does not describe phenomena such as interference, diffraction, and polarization effects which require wave theory description. The theory dictates that light ray follows a line or curve that is perpendicular to the light's wave fronts. As long as the wavelength is very small compared with the size of structures with which the light interacts then Ray theory gives an excellent approximation.

It is known that when light propagates in GRIN at large wavelength scale, it experiences refraction and reflection gradually [129, 159-160]. Therefore, its behavior can be adequately described by rays obeying a set of geometrical rules. In this section, we want to give a rapid insight on the modeling of light propagation through an optical system, *i.e.,* GRIN medium, by the help of Ray theory/Geometrical Optics.

The ray behavior inside a GRIN medium can be explained by Ray equation. It determines the electromagnetic wave propagation by the use of geometrical optics approximation and it can be mathematically expressed by the following equation:

$$\frac{d}{ds}\left[n\frac{dr}{ds}\right] = \nabla n, \qquad (3.1.1)$$

where $n$ is the refractive index of GRIN medium, $r$ is a vector representation of position $(x, y)$ and $ds = \sqrt{dx^2 + dy^2}$ is the differential arc length along the ray path. Moreover, the light rays can be considered as the orthogonal trajectories to the geometrical wave fronts. So that relation between surface of equal phase and ray (wave and ray optics) can be expressed by Eikonal equation as follows:

$$\left|\vec{\nabla}S\right|^2 = n^2, \qquad (3.1.2)$$

where $S$ is an Eikonal function and $\vec{\nabla}$ is a gradient operator. The surfaces where Eikonal function $S$ is constant represent the surfaces of equal phases, *viz.* geometrical wavefronts, which in turn dictate the shape of the propagating electromagnetic field.

In this study, two types of graded index configuration are considered: a GRIN structure having continuously varying refractive index distribution and PC based GRIN medium whose index profile is approximated to that of continuous medium at a long wavelength region. The detailed design approach and regarding numerical results of the latter configuration are reported in the next sections. The Ray theory approach is applied for the case of continuous GRIN medium in order to better understand the nature of the light propagation. The refractive index distribution of the continuous GRIN medium is adjusted to have a Hyperbolic Secant (HS) profile and it is formulated by the given equation:

$$n(y) = n_0 \, sech(\alpha y), \qquad (3.1.3)$$

where $n_0$ is the refractive index at the optical axis and $\alpha$ is a gradient parameter that represents the depth of the index distribution. Throughout the study we abbreviated the GRIN media having hyperbolic secant index profile by HS GRIN. The reason why HS refractive index profile is preferred is that compared to more common form of GRIN profiles such as quadratic (parabolic) one, HS profile demonstrates the following advantages: First, it is a more general form of quadratic profiles which can be approximated by a parabolic function. Second, the Eikonal equation featuring light propagation can be solved free of any types of approximation and third the designed structure is free of aberration for meridional rays [130]. Besides the mentioned advantages, the HS profile parameters $n_0$ and $\alpha$ are also easily tunable. In the light of the above statements, one can deduce that optimum structural parameters could be determined without challenging restrictions for HS GRIN medium.

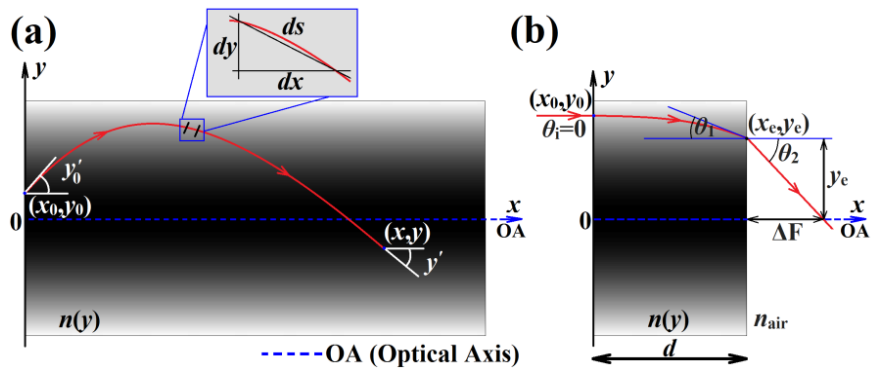

Figure 3.1.1. (a) Ray propagation in a GRIN medium having an HS index profile $n(y)$ and (b) the schematic view of back focal length $\Delta F$ calculation at the output.

In order to calculate the ray trajectories in a continuous HS GRIN medium the optical ray path calculation is schematically represented in Fig. 3.1.1(a). As can be seen in the figure, the ray is described by its position $y$ and the slope $\dot{y}$ (derivative with respect to *x*). The initial ray position $y_0$ and its slope $\dot{y}_0$ at the input plane $x_0 = 0$ are connected with the ray position *y* and slope $\dot{y}$ at the plane *x* by solving the Eikonal equation (Eq. 3.1.1) [130] for the continuous HS GRIN medium (details of the derivation can be seen in Appendix A):

$$y(x) = \frac{1}{\alpha}\sinh^{-1}\left(\dot{u}_0 \frac{\sin(\alpha x)}{\alpha} + u_0 \alpha \cos(\alpha x)\right), \tag{3.1.4}$$

$$\dot{y}(x) = \frac{-\alpha u_0 \sin(\alpha x) + \dot{u}_0 \cos(\alpha x)}{\alpha \cosh\left\{\sinh^{-1}\left[u_0 \alpha \cos(\alpha x) + \dot{u}_0 \frac{\sin(\alpha x)}{\alpha}\right]\right\}}, \tag{3.1.5}$$

where $y(x)$ is a trajectory that is obtained by the incident position $u_0$ and incident angle $\dot{u}_0$ and $\dot{y}(x)$ provides the slope information of the trajectory. Note that the propagation of rays in an HS index medium obtained in hyperbolic coordinate (*u*) by using following transformation $u = \sinh(\alpha y)$ (see Appendix A). The ray trajectories are obtained without any approximation and the full analytical solution is also given in Appendix A. The optical rays follow a curved trajectory while exiting the structure as illustrated in Fig. 3.1.1(b). The same plot also explain the plane wave excitation mechanism of GRIN medium by showing incidence angle ($\theta_i$=0) and size of the beam. When the light ray enters the free space after exiting the HS GRIN structure it refracts obeying Snell's law and travels in a straight line. As illustrated in Fig. 3.1.1(b), the output ray intersects with the optical axis (OA) and the distance between that point and end face of the structure is defined as the back focal length ΔF, which is formulated by the following (the detailed derivations are given in Appendix):

$$\Delta F = \frac{y_e \sqrt{\dot{y}(d)^2[1 - n_0^2 \sec h^2(\alpha y_e)] + 1}}{n_0 \sec h(\alpha y_e)\dot{y}(d)}, \tag{3.1.6}$$

where ($x_e$, $y_e$) is the position of the ray at the end face of the HS GRIN medium and according to Fig. 3.1.1(b) the length parameter *d* is set to $x_e$. Incident angle $\theta_1$ of the light ray at the ($x_e$, $y_e$) position and the refracted angle $\theta_2$ are given as an inset in the same figure.

The ray trajectories are calculated by the derived Eqs. 3.1.4 and 3.1.5 and the regarding ray paths are plotted inside the continuous HS GRIN medium as shown in Fig. 3.1.2. As can be seen in Fig. 3.1.2(a) the light rays oscillate in a sinusoidal manner along the optical axis [132, 161] and thus, the critical parameters such as the oscillation period and the amplitude of the oscillation can be calculated. In Fig. 3.1.2(a) the period parameter is defined as Pitch (*P*) and given as insets in the same figure. Light rays oscillate along the optical axis with a pitch of $P = (2\pi/\alpha)$ where $\alpha$ is a gradient parameter of the HS refractive index profile. Hence, knowing the pitch enables to obtain focusing, collimation and diverging effects by appropriately arranging the length of the HS GRIN medium. The ray trajectories for focusing and collimation cases are illustrated in Figs. 3.1.2(b) and 3.1.2(c), respectively. When the length of the HS GRIN structure is terminated at the positions with $L_x<0.25P$, the incident rays converge and focus at the output of structure, in which case the back focal length ΔF is calculated by Eq. 3.1.6. If the lens length is exactly equal to $L_x=0.25P$ then the focal point appears at the back end face of the structure and in this case ΔF is equal zero. In this regards, one can say that the exact length size $L_x=0.25P$ is like a transition parameter between focusing and diverging property of HS GRIN structure. On the other hand, if the length parameter $L_x$ is fixed at $L_x=0.50P$, then, the output rays follow trajectories parallel to optical axis as can be observed in Fig. 2(c). In the case of collimation, the slope $\dot{y}(L_x)$ at the distance $L_x=0.50P$ equals to zero and hence, ΔF goes to infinity according to Eq. 3.1.6, which describes collimating behaviour of the configuration. The detailed conditions for focusing, de-focusing and collimation behaviours of the HS GRIN medium is explained in the next sections. The overall ray theory calculation implies that a GRIN medium can be utilized for the light converging, diverging and collimation purposes by adjusting the thickness of the GRIN structure.

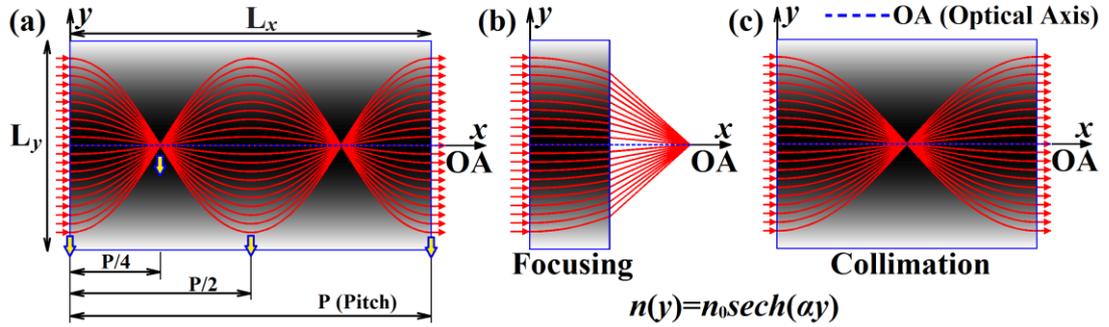

Figure 3.1.2. (a) Calculated ray trajectories in a HS profile GRIN medium. Special cases for the designed configuration such as (b) focusing and (c) collimation when the structure is appropriately terminated at distances $L_x<0.25P$ and $L_x=0.50P$, respectively.

Geometrical optics can provide enough information about light-ray behavior through the GRIN structure to provide the design of a GRIN element with adequate performance. However, it is also necessary to analyze the performance of the GRIN structure physically by utilizing wave theory concept.

It is known that producing the continuous GRIN media can be considered as a challenging issue due to fabrication difficulties during the diffusion process, the need of planar faces and restriction or limitation of the index gradient (usually smaller gradient occurs). To overcome these difficulties, the use of PCs for the approximation of continuous GRIN media is considered. Hence, design approach of the approximation continuous GRIN medium by PCs and time-domain analyses as well as focusing performances are investigated in the next section.

### 3.1.3. Design Approach of Hyperbolic Secant Index Profile GRIN PC Medium

Photonic crystals can be considered as a powerful compound in order to imitate continuous GRIN media that have any index profiles. There are several methods that can be followed to design GRIN PC medium. One method is based on appropriate arrangement of the radii of the dielectric rods (air holes) in air (dielectric) background so that filling factor of the elementary PC unit cell can be varied [64]. However, this process needs precise and small increments on the rod radii. Moreover, it can also limit the range of index deviation that can be needed. The second method lies on the infiltration of air holes with different substances or requirement of different types of material to construct the GRIN PC. Finally, the last method which seems to be more practical than the mentioned approaches is appropriate adjustment of size of PC unit cells while keeping the material type and rod radii the same [70]. Therefore, in the presented study the modulation of the lattice spacing is utilized in order to approximate continuous GRIN media by PCs. For that purpose, as a first stage, the dispersion relations of PC unit cells with different lateral sizes are calculated by exploiting plane wave expansion (PWE) method [85]. Calculated dispersion diagrams of the first band in the ΓX direction are depicted in Fig. 3.1.3(a). As can be seen in the figure when the dimensions of the unit cells increases in the transverse *y*-direction the related bands move to higher frequencies. Note that the vertical dimension of the unit cells ranges from $0.40a$ to $2.0a$ with a $0.20a$ step size. The lattice constant is represented by *a*. Moreover, the distance between rods along the propagation *x*-direction is fixed to *a*, *i.e.* gradient of effective index profile changes along only the transverse *y*-direction, and the radii of the rods are equal to $r=0.20a$. Corresponding variations of cell sizes are depicted in Fig. 3.1.3(a) as an inset. As stated before it is practically difficult to implement rods with different materials, hence we keep refractive indices of them at $n=3.13$ (considered as Alumina rods).

In the next stage, as presented in Fig. 3.1.3(b) the corresponding group index ($n_g$) curves for each band are extracted by using the slope information of the regarding first bands. In Fig. 3.1.3(b) one can observe that for lower frequencies, $n_g$ curves are linear and closely spaced, which provides a slight variation in group indices. As a final stage, the proposed structure having a specified effective refractive index profile is designed at a fixed frequency lying in the region where small alterations occur in the group indices. It is worth noting that in order approximate the continuous GRIN medium with HS index profile, the proposed GRIN PC structure is designed at the normalized frequency of $a/\lambda=0.10$. In order to demonstrate the extensions of the obtained group indices the $n_g$ curves are zoomed out in the design frequency region around $a/\lambda$=0.10 and it is given as an inset in Fig. 3.1.3(b). As can be seen, the calculated group indices at frequency of $a/\lambda$=0.10 cover values between 1.39 and 2.28. To obtain GRIN PC structure with any type of stair-step index distribution within those index values one needs intermediate index values to generate smooth variation. Hence, interpolation method is applied to determine intermediate points by fitting the calculated $n_g$ profile of rectangular cells that have lateral sizes which deviates from 0.40$a$ to 2.0$a$. Then, the concerning structure is formed by sequentially placing the rectangular cells having those intermediate $n_g$ values in such a way that the desired step-stair index distribution is revealed. Note that the effective medium theory coincides with the exploited dispersion engineering method at the long wavelength regimes, *i.e.* below the normalized frequency of $a/\lambda$=0.10 [71]. Determination of the long wavelength region boundary is crucial, in other words, the lattice constant *"a"* should be much smaller than the wavelength to ensure staying in the effective medium region. Therefore, within those long wavelengths ($\lambda$ >10$a$), the proposed HS GRIN PC scheme can be considered as an effective homogeneous medium.

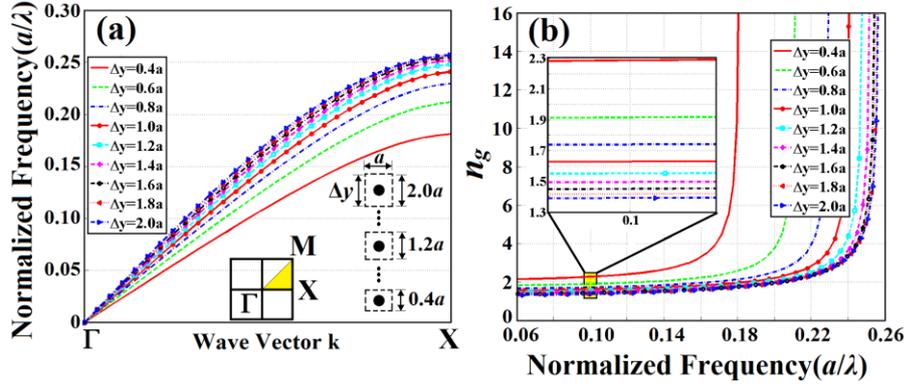

Figure 3.1.3. (a) The dispersion diagram of the first band along the Γ-X direction is shown. (b) Group index curves corresponding to the each dispersion bands shown in (a). Also the direction of ΓX is depicted by giving irreducible Brillouin zone as an inset in (a).

In this study, it is aimed to design GRIN PC medium by mimicking of continuous GRIN medium which has HS index profile. For this reason, firstly the schematic view of continuous HS GRIN medium with the important parameters such as length $L_x$ and width $L_y$ are presented in Fig. 3.1.4(a). Corresponding refractive index profile plot is given in Fig. 3.1.4(b) where the refractive index varies in the range of 1.30 and 2.20. We should note that the mathematical formulation of relevant refractive index distribution $n(y)$ of the continuous HS GRIN medium is given in previous section in Eq. 3 and HS GRIN PC designed by keeping HS index profile parameters equal to $n_0$=2.2 and $\alpha$=0.112$a^{-1}$. Exploiting the previously described method to engineer GRIN PC structure, one can easily approximate continuous GRIN medium by PC rods. The designed GRIN PC medium and its imitated stair-step (discrete) version of HS effective index profile are presented in Figs. 3.1.4(c) and 3.1.4(d), respectively. The important parameters such as lateral $L_y$ and longitudinal $L_x$ dimensions of the proposed HS GRIN PC medium are given in Fig. 3.1.4(c). As can be inferred from the stair-step effective index profile in Fig. 3.1.4(d), effective index changes from 1.44 to 2.20. It increases toward the optical axis due to the decrement of the lattice spacing between adjacent rods. The dimension of the considered HS GRIN PC structure have total width of $L_y$=20$a$. The longitudinal dimension $L_x$ of the design may vary depending on intended purposes (focusing, diverging or collimation lens designs) which will be explained in detail in the next sections. While comparing both index profiles in Figs. 3.1.4(b) and 3.1.4(d), the shape and gradient of the both profiles match quite well which approves that the imitation method properly works. We should note that the main purpose of giving of continuous HS GRIN medium schematic in this section in Fig. 3.1.4(a) and 3.1.4(b) is only to show how properly index distribution is approximated by the help of PCs. Hence, in overall numerical calculations in the next sections the only latter one (designed GRIN PC) is conducted.

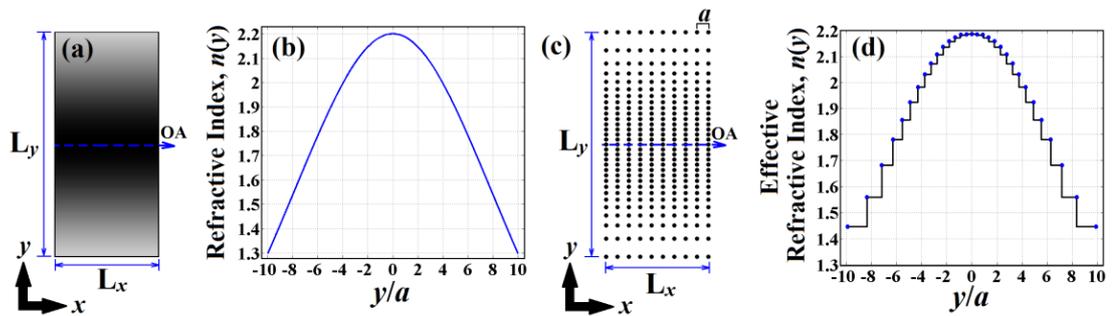

Figure 3.1.4. (a) Schematic representation of the continuous HS GRIN medium and its (b) refractive index distribution. (c) Schematic view of designed GRIN PC structure and its (d) stair-step (discrete) version of the hyperbolic secant index profile at a fixed frequency of $a/\lambda$=0.10.

The designed HS GRIN PC structure is numerically modeled in 2D and its time-domain analyses are conducted by the help of FDTD method [119]. The computational domain is concerned only in 2D spatial domain and the third dimension is taken to be uniform. The boundaries of the computational domain are surrounded by perfectly-matched layers (PMLs) [87] in order to eliminate undesired back-reflections. Furthermore, in all numerical simulations only transverse magnetic (TM) polarization employed where the concerned non-zero electric and magnetic field components are $E_z$, $H_x$, and $H_y$, respectively. Through all the FDTD calculations, a grid size arranged to be $\Delta x = \Delta y = a/32$. Also, two types of input sources: either a continuous source or a pulse with a Gaussian profile in time is used. The former one is needed to obtain spatial intensity distributions of the lensing features of the HS GRIN PC. On the other hand, the latter type of input source is preferable to compute the power transmission spectrum of the proposed structure. Hence, a pulse with a Gaussian profile is launched to the front side of the HS GRIN PC. A detector is located at the end of the structure in order to measure the transmission spectra. Note that, the structural length of the investigated HS GRIN PC structure is fixed to 100$a$. Since, the length of 100$a$ should be enough to present spatial domain characteristics at chosen operating frequencies. The transmission efficiencies are calculated and normalized by taking the ratio of detected and incident power. The carefully inspection of transmission spectrum in Fig. 3.1.5 shows that there are two high transmission regions as well as forbidden gaps throughout the whole frequency bands. The first frequency window is placed between $a/\lambda$=0.10 and $a/\lambda$=0.20 where transmission efficiency oscillates between %95 and %60, respectively. This oscillation in the graph originated from the back-reflections at the front and back faces of the HS GRIN PC structure which is called as Fabry-Perot oscillations. On the other hand, the second window shows more stable transmission efficiency between the frequencies of $a/\lambda$=0.43 and $a/\lambda$=0.55 where corresponding efficiency deviates between %80 and %99. While comparing two high transmission windows one can deduce that the relatively low transmission in the spectrum for small normalized frequencies (below $a/\lambda$=0.20) is due to the leakage of light along the transverse direction (some of the light cannot reach the end face of the GRIN medium, weak guiding causes some amount of power to be lost). The almost perfect transmission occurring of HS GRIN PC at higher frequencies is worth noticing. In order to demonstrate operating frequency bands where numerical analyses will be implemented the related transmission windows are denoted as low and high frequency regions in Fig. 3.1.5.

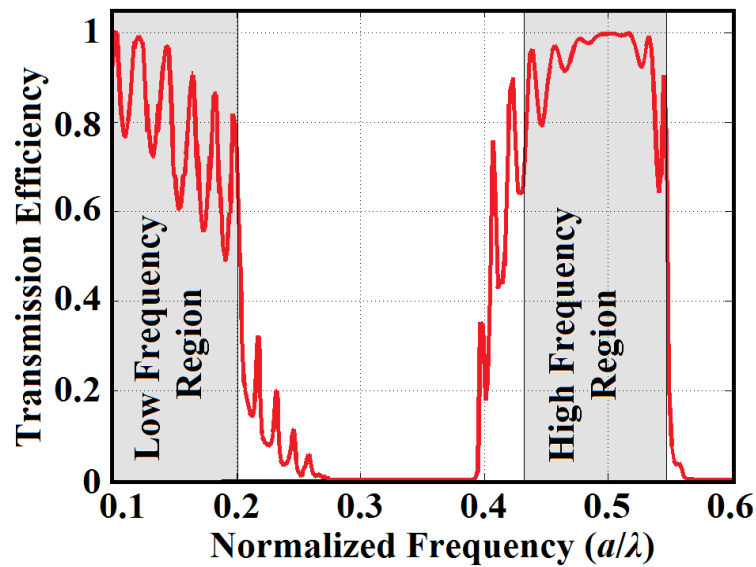

Figure 3.1.5. Calculated transmission efficiency of the HS profile GRIN PC structure is given. High transmission regions are defined by shading and denoted by low and high frequency regions in the figure.

As mentioned above a homogenization procedure is implemented at the low frequency region. For that purpose, the design frequency is arranged to $a/\lambda=0.10$ as can be declared in the inset of Fig. 3.1.3(b). Afterwards, the FDTD analyses are utilized in order to calculate the regarding transmission over a broad bandwidth and to explore possible additional transmission windows for the designed GRIN structure. We expect transmission windows as well as band gaps due to the fact that the lattice periodicity along the propagation direction is kept intact. As can be seen from Fig. 3.1.5, two high transmission windows are observed within the frequency interval $a/\lambda=0.10$-$0.20$ and $a/\lambda=0.43$-$0.55$ where transmission efficiencies oscillate between %60 and %99, respectively. It is worth to note that the design of the GRIN medium is performed only considering the low frequency region. On the other hand, the designed medium is tested both for low and high frequency regions. Therefore, we do not extract and report effective refractive index values for the high frequency window in this work.

Next, FDTD method is carried out to inspect the electromagnetic field propagation inside the HS GRIN PC structure that is tested under both low and high frequency intervals. In order to observe the oscillation dynamics of the propagating beam, the length of the structure is taken to be relatively long and it is equal to $100a$. There are two reasons to investigate the electromagnetic field interaction with the GRIN PC structure: the calculation of pitch lengths for the GRIN PCs and their dependence on the frequency variations. For that purpose, the structure is illuminated by continuous source having a Gaussian profile. Then, the regarding steady state fields are extracted and shown in Fig. 3.1.6. The incident beam propagates in the direction of arrows which are given as insets in Figs. 3.1.6(a) - 3.1.6(h). We investigated the variation of the oscillation periods which can be defined as pitch lengths and previously denoted by $P$ for different frequency values picked up from lower and higher frequency regimes as indicated in the transmission spectrum in Fig. 3.1.5. In order to observe the movement of the first and the second focal points (that points are critical for pitch-length calculations) inside the structure, the HS profile GRIN PC is excited by an incident source at the normalized frequencies selected in the low and high frequency intervals. Figures 3.1.6(a) - 3.1.6(d) represent the steady-state field distributions at lower frequencies, i.e. at $a/\lambda=\{0.10,0.12,0.14,0.16\}$, respectively. At lower bands, while the frequency is increased slowly from $a/\lambda=0.10$ up to $a/\lambda=0.16$, the focal point shift is almost negligible for the selected frequency values. The recorded half-pitch ($P/2$) values are around $30a$. In addition, Figs. 3.1.6(e)-3.1.6(h) are prepared in order to observe how frequency variations at higher bands affect the first focusing characteristic of GRIN PC medium. The corresponding frequency values of propagating beams in Figs. 3.1.6(e) - 3.1.6(h) are $a/\lambda=\{0.46,0.48,0.50,0.52\}$, respectively. In higher frequency levels, the response of the GRIN PC focusing characteristics to different incident wavelengths slightly varies staying within the values of $28a$ and $30a$. While the incident frequency increases, the focal point shifts to right-side. On the other hand, stronger focusing takes place when the incident frequency is increased, which means the regarding spot size decreases and propagating beam becomes tighter. The focusing property is more complex at the cut-off frequencies (band edges).

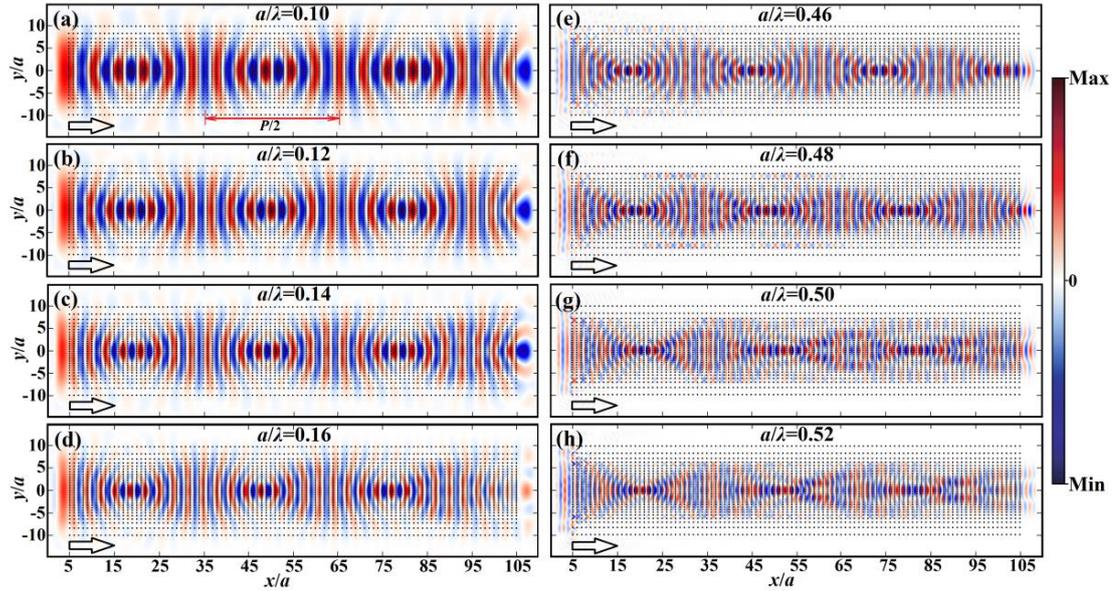

Figure 3.1.6. List of steady state electric field distributions for lower and higher frequency regions. In lower frequency bands, the operating frequencies are set to (a) $a/\lambda$=0.10, (b) $a/\lambda$=0.12, (c) $a/\lambda$=0.14 and (d) $a/\lambda$=0.16. On the other hand, the incident beam operates at (e) $a/\lambda$=0.46, (f) $a/\lambda$=0.48, (g) $a/\lambda$=0.50 and (h) $a/\lambda$=0.52 within the higher frequency interval. Half-pitch is represented in (a) for convenience. Arrows denote the light propagation direction.

In order to better understand half-pitch size dependence on wavelength variations within the long and short wavelength regimes, the graphs in Figs. 3.1.7(a) and 3.1.7(b) are prepared. As can be inferred from Fig. 3.1.7(a), slight oscillations in the half-pitch size from $P/2$=28$a$ to $P/2$=30$a$ occur within the frequency range of $a/\lambda$=0.10-0.17. That region is indicated by a dashed rectangle in Fig. 3.1.7(a). On the other hand, since the transmission window is surrounded by cut-off frequencies (lower and upper bounds), operating near the frequencies of $a/\lambda$=0.18-0.20 makes the half-pitch lengths highly wavelength dependent ($P/2$ reduces down to 23$a$).

The higher frequency region shows rather complicated behavior. One feature is that $P/2$ value increases while operating wavelength decreases, which can be seen in Fig. 3.1.7(b). At the same time, the spatial field distribution covers smaller area as shown in Figs. 3.1.6(e) - 3.1.6(h), which implies the compression of beam spot size inside the HS GRIN PC. At some frequencies, there also appears a region as in lower frequencies where the P/2 value stays nearly constant and that interval is enclosed by a dashed rectangle in Fig. 3.1.7(b). The *P*/2 value varies from *P*/2=29*a* to *P*/2=30*a* within the frequencies of *a*/*λ*=0.46-0.53. In the light of above results, several important remarks can be inferred from the graphs in Figs. 3.1.7(a) and 3.1.7(b): at some wavelengths, the pitch value remains nearly constant. In other words, the oscillation period remains unchanged for frequency variation at these intervals. The *P*/2 value becomes nearly constant (*P*/2=30*a*) at around *a*/*λ*=0.51, where the least aberration effect is observed. The increase in the *P*/2 value signifies the reduction in the strength of structural focusing characteristics. When the incident light has wavelength that is close to the edges, in the cases of *a*/*λ*=0.40-0.45 and *a*/*λ*=0.54-0.56, the light propagation provides complex periodical oscillation at the proximity of cut-off regions. Therefore, the calculated half-pitch value is highly dependent on operating wavelengths as can be seen in Fig. 3.1.7(b). Although field localizations at both low and high frequency regimes display different characteristics, designed GRIN PC medium may provide similar *P* values, which can be observed in Figs. 3.1.7(a) and 3.1.7(b). Hence, GRIN PC medium provides rich light manipulation characteristics depending on the operating wavelength.

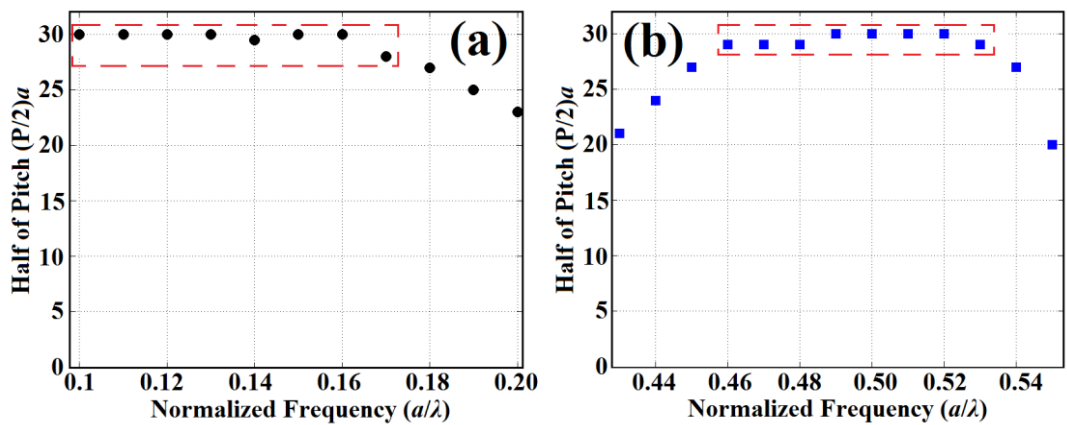

Figure 3.1.7. The variations of half-pitch value (P/2) depending on the operating frequencies are shown for low and high frequency regions in (a) and (b), respectively.

### 3.1.4. Discussion and Evaluation: Collimation, Focusing and De-focusing of Designed HS GRIN PC

In previous section, we observed the oscillation pattern of the propagating beam by looking at the time-domain snap shots of the electric-fields in Fig. 3.1.6. The incident light first converges and reaches the smallest beam width. After that de-focusing and focusing follows each other. At the transition regions while moving from de-focusing to focusing or vice versa the beam width reaches fullest spatial width. This fact allows us to study special cases such as collimation, focusing and diverging behaviours in considered HS GRIN PC configuration. Therefore, we specially terminated the lengths of the GRIN media based on the previously extracted $P/2$ values. Then the planar HS GRIN PC structure can focus, collimate or de-focus the incident light while exiting the medium.

In this section, we presented the collimation, focusing and de-focusing behaviours of designed structure by exploiting the oscillatory nature of the light travelling within the GRIN medium. The HS GRIN PC medium has been tested under two different frequencies, which are selected as representative values at lower and higher frequency regimes such as $a/\lambda=0.10$ and $a/\lambda=0.48$, respectively. It is worth noting that these operating frequencies are chosen with respect to the graphs in Figs. 3.1.7(a) and 3.1.7(b). As previously defined, the propagating beam oscillates along the optical axis of the GRIN medium with a period/pitch "$P$". Knowing the pitch, it is possible to obtain conditions for collimation, focusing and de-focusing behaviours when the structure is illuminated by a line source right in front of the structure. For the special cases the following conditions should be satisfied:

$$L_x = \begin{cases} L_{\text{coll}} & \text{if} \quad L_x = \left(\dfrac{m}{2}\right)P, \\ L_{\text{focus}} & \text{if} \quad \left(\dfrac{m}{2}\right)P < L_x < \left(\dfrac{2m+1}{4}\right)P, \\ L_{\text{de-focus}} & \text{if} \quad \left(\dfrac{2m+1}{4}\right)P < L_x < \left(\dfrac{m+1}{2}\right)P, \end{cases} \qquad (3.1.7)$$

where $L_{coll}$, $L_{focus}$ and $L_{de\text{-}focus}$ are length of the HS GRIN PC structure for the collimation, focusing and de-focusing cases, respectively. In addition, $m$ is an integer and equals to $m=0,1,2,3...$

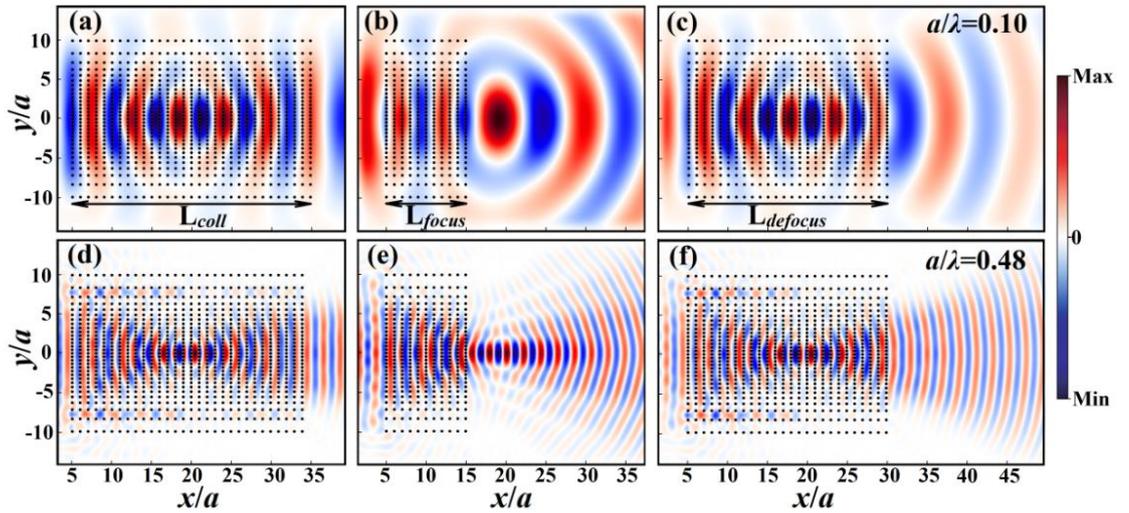

Figure 3.1.8. Special cases of the proposed HS GRIN PCs for low and high frequency bands are represented. In (a), (b) and (c), collimation, focusing and de-focusing behaviours operating at a fixed frequency of $a/\lambda=0.10$ are shown, respectively. (d), (e) and (f) correspond to the same characteristics for the normalized frequency of $a/\lambda=0.48$. The lengths of the structures are defined as $L_{coll}$, $L_{focus}$ and $L_{de\text{-}focus}$ and depicted as an inset in the figures.

In order to observe collimation, focusing and de-focusing effects, the lengths of HS GRIN PCs $L_x$ are intentionally terminated according to Eq. 3.1.7. Figure 3.1.8 represents the regarding steady-state electric field distributions of HS GRIN PC structure for collimation, focusing and de-focusing cases operating at the normalized frequencies of $a/\lambda=0.10$ and $a/\lambda=0.48$. To be more specific, the length of the design is set to a half of period $L_{coll}=30a$ and $L_{focus}=10a$ for the collimation and focusing cases at the lower frequency of $a/\lambda=0.10$, whose snapshots of field profiles are represented in Figs. 3.1.8(a) and 3.1.8(b), respectively. Moreover, the de-focusing condition appears when the structure is terminated at the length of $L_{de\text{-}focus}=25a$ and the corresponding steady state field profile is shown in Fig. 3.1.8(c). While analyzing the same task at higher frequency of $a/\lambda=0.48$, the lengths of the proposed PC structure are adjusted to $L_{coll}=29a$ for collimation and $L_{focus}=10a$ for focusing conditions. The corresponding field distributions are presented in Figs. 3.1.8(d) and 3.1.8(e). Finally, Fig. 3.1.8(f) is prepared to demonstrate de-focusing behaviour of the proposed configuration where its length is set to $L_{de\text{-}focus}=25a$. Elaborating on the above figures, one can observe that in the case of collimation effect in Figs. 3.1.8(a) and 3.1.8(d) the wave fronts of incident beam converges and diverges. When beam reaches the end of the structure it has flat wave fronts so that collimated beam appears at the output. For the focusing case, converging wave fronts is obtained at the exit of the structure and hence, a real focal point exists at the output as presented in Figs. 3.1.8(b) and 3.1.8(e). On the contrary, in the de-focusing case a virtual focal point occurs inside the structure and the propagating beam starts to diverge at the exit as shown in Figs. 3.1.8(c) and 3.1.8(f). Considering Fig. 3.1.8, another important remark is that increasing the normalized frequency from $a/\lambda=0.10$ to $a/\lambda=0.48$ creates similar oscillations. Operating at higher frequencies, the half of pitch stays almost constant at around $P/2=30a$ and field distribution around the transverse direction diminishes for smaller wavelengths. In other words, the same type of source with a different frequency gets strongly confined as compared to the lower frequency cases. The simple way to explain such phenomenon is that the propagating beam's wavelength reduces at higher frequencies and thus, guided mode gets intensively modulated by the underlying periodic lattice. The light oscillates at a smaller transverse space and hence, the transverse field confinement becomes stronger at $a/\lambda=0.48$ than $a/\lambda=0.10$.

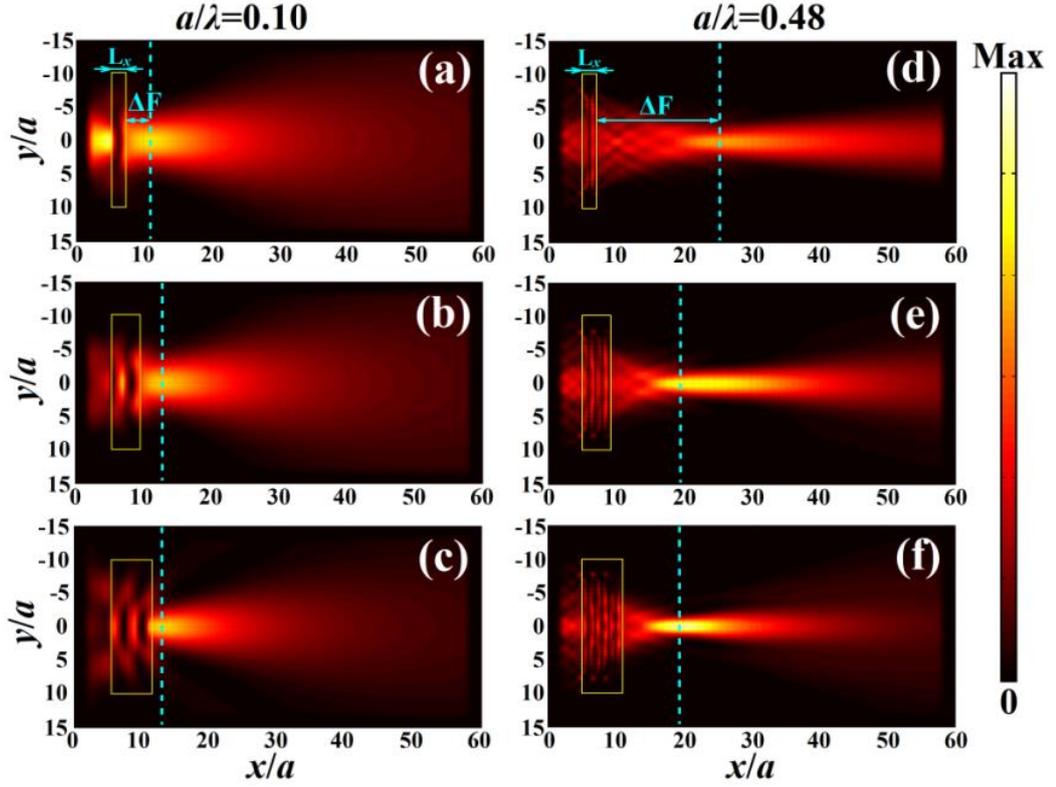

Figure 3.1.9. The spatial e-field distributions of GRIN PCs having different lengths such as (a) $L_x=2a$, (b) $L_x=4a$ and (c) $L_x=6a$ are investigated at operating frequency of $a/\lambda=0.10$. Similarly, field distributions are presented for the operating frequency of $a/\lambda=0.48$ where the lengths of the structures are taken the same.

One can replace conventional focusing optical elements such as focusing apparatus (lenses) with a GRIN medium. The additional benefits of artificially designed HS GRIN structure are flat front and back surfaces, frequency selectivity of the structure and strong focusing capability. Thanks to the above results, we have shown the designed HS GRIN PC performs as an optical element that can be implemented for focusing purposes. In the next steps, detailed investigation of the focusing characteristics is performed. The focal point dynamics are investigated depending on different structural lengths and the output field patterns are compared for the lower and higher operating frequency cases. As mentioned before the structure length should satisfy the condition given in Eq. 3.1.7 to provide focusing effect. Previously the oscillation periods, *i.e.* pitches are obtained for propagating beams at the low and high frequencies of $a/\lambda=0.10$ and $a/\lambda=0.48$, respectively. Then the corresponding $P/2$ values are $30a$ and $29a$, respectively. Therefore, considering the focusing condition in Eq. 3.1.7 the lengths of the HS GRIN PC structures are adjusted. Figure 3.1.9 provides a collection of spatial intensity profiles for different lengths of GRIN PC structure equal to $L_x$={2$a$, 4$a$, 6$a$} at selected frequencies of $a/\lambda$=0.10 and $a/\lambda$=0.48. The regarding field patterns at the frequency of $a/\lambda$=0.10 are presented in Figs. 3.1.9(a) - 3.1.9(c), respectively. Under the same structural parameters, the HS GRIN PC is excited with a source having a higher normalized frequency of $a/\lambda$=0.48 and the corresponding field distributions are illustrated in Figs. 3.1.9(d) - 3.1.9(f). The distance from the output surface of the HS GRIN PC to focal point is defined as the back focal length $\Delta F$ and it is defined in Fig. 3.1.9. According to the field distributions, one can deduce that while increasing the structural length focal point moves closer to the output surface and thus, the corresponding $\Delta F$ decreases at both low and high frequencies. It can also be inferred from Figs. 3.1.9(d) - 3.1.9(f) that the smaller wavelength senses the structural modification even more and the effective index gradient of the GRIN PC medium may increase. For the higher frequency region, there are no side lobes and the field focusing creates more like a pencil-beam as can be seen in Figs. 3.1.9(d) - 3.1.9(f).

To provide quantitative analyses of the focusing behaviour at the two different operating frequencies of $a/\lambda$=0.10 and $a/\lambda$=0.48, cross-sectional intensity profiles are taken at the focal points of field distributions in Fig. 3.1.9 along the transverse $y$-direction. The corresponding cross-sectional intensity profiles at the fixed frequency of $a/\lambda$=0.10 for different structural lengths are obtained and superimposed in Fig. 3.1.10(a). Based on the cross-sectional intensity plots, the determined full width of half maximum values at the focal points equal to FWHM={5.78$a$, 5.56$a$, 3.98$a$}. That alteration in FWHM values implies that while the focal point moves towards the end facet of the HS GRIN PC structure and side lobes start to appear, the focusing capability of the configuration strengthens, *i.e.* corresponding spot sizes becomes smaller. In the limit when $L_x$ approaches to $P$/4 then beam width becomes smallest and $\Delta F$ should take zero value. The analytical derivations based on Ray theory confirm these observations.

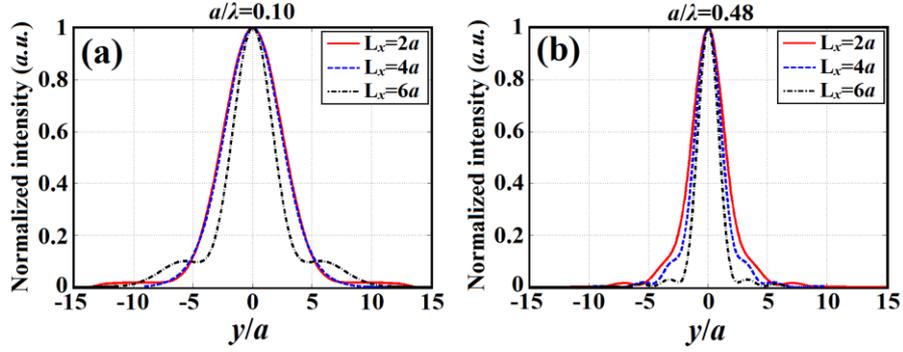

Figure 3.1.10. The cross-sectional intensity profiles of the focal point in the cases of (a) low frequency of $a/\lambda=0.10$ and (b) high frequency of $a/\lambda=0.48$ for different structural lengths of $L_x=\{2a, 4a, 6a\}$.

In the high frequency case of $a/\lambda=0.48$, the structural lengths are kept the same as in Fig. 3.1.10(a) and the calculated cross-sectional intensity plots are depicted in Fig. 3.1.10(b). As expected from Figs. 3.1.9(d) - 3.1.9(f), the spot size of the focused light is smaller at higher frequency of $a/\lambda=0.48$ than lower frequency due to higher confinement and strong modulation mechanism. The corresponding FWHM values are shown in Fig. 3.1.10(b) and equal to FWHM=$\{3.14a, 2.56a, 2.02a\}$. While comparing the calculated cross-sectional intensities at lower and higher frequencies, even though the HS GRIN PCs has the same length of $L_x=6a$, the FWHM at the higher frequency of $a/\lambda=0.48$ is 1.97 times narrower than that at $a/\lambda=0.10$. We should note that these comparisons are made in the spatial domain. In the next paragraph we have tabulated the values in terms of wavelength.

In order to finely explore the focusing properties of the HS GRIN PC medium, we extracted two features: ΔF and FWHM values variations with respect to $L_x$ at two different frequencies $a/\lambda=0.10$ and 0.48, respectively. The structural length $L_x$ dependences of the back focal lengths ΔF and corresponding FWHM values are analyzed and the result are demonstrated in Table 3.1.1. As stated before to provide focusing condition Eq. 3.1.7 can be considered. For this reason, the lengths of the HS GRIN PC structures are changed between $1a$ and $15a$ with step size of $2a$.

Table 3.1.1 presents variations of FWHM and ΔF in terms of both unit distance ($a$) and wavelength ($\lambda$). There are general trends applicable for both wavelength regions:

ΔF values decrease as we increase the length L$_x$. Consequently, the corresponding FWHM value reduces. As depicted in Table 3.1.1, for a longer wavelength when L$_x$ changes from 1*a* to 15*a* corresponding ΔF values reduce from 6.0*a* to 0.05*a*. Meanwhile, when L$_x$ is between 11*a* and 15*a*, ΔF becomes nearly zero. We should point out that the initial values of ΔF at two selected frequencies are very different, 6.0*a* vs. 22.95*a*. Similarly in the case of *a*/*λ*=0.48, the length L$_x$ varies from 1*a* to 15*a* and regarding ΔF value decreases from 22.95a to 0.50*a*. It can also be observed from Table 3.1.1 that ΔF for low frequency approaches to zero when L$_x$ is greater than 11*a* whereas in the case of high frequency at *a*/*λ*=0.48 the HS GRIN PC still succeeds the focusing task with ΔF greater than zero. Similar trends are valid for ΔF if we present the results in terms of wavelength units. When we present FWHM values in terms of both unit distance and wavelength, different trends appear as L$_x$ changes.

For instance, in the case of distance units "*a*" low frequency region has a larger FWHM compared to that of in high frequency case (7.26*a* vs. 3.99*a*). On the other hand, if we calculate FWHM in terms of *λ* units, opposite trend emerges, so that FWHM values turn out to be 0.72*λ* and 1.92*λ* for low and high frequencies, respectively. The further increasing length of the structure reveals strong suppressing of beam spot size. The minimal FWHM values are gathered for the structural length of L$_x$=15*a*, which are determined as 3.18*a* and 1.41*a* (corresponds to 0.31*λ* and 0.67*λ* in terms of *λ* units) for the selected two frequencies of *a*/*λ*=0.10 and *a*/*λ*=0.48, respectively. It is worth noting that low frequency region provides sub-wavelength focusing when L$_x$ lies between values of 3.0*a* and 15*a*. On the other hand, there is no indication of sub-wavelength focusing for high frequency region. As a result it can be deduced that the lower frequencies provide larger spot-size conversion ratio than higher frequency case. This is an important distinction of HS GRIN PC medium operating at longer wavelengths. To maintain the same spot-size conversion ratios at lower and higher frequencies, the required lengths of structure are smaller/larger for the low/high frequency parts. The ability to manipulate focal point dynamics with respect to GRIN length becomes fruitful when we investigate the structure at both lower and higher frequency region.

Table 3.1.1. FWHM and ΔF values at two different normalized frequencies, *a/λ*=0.10 and 0.48 are presented when L$_x$ varies from 1*a* to 15*a*.

| L$_x$ | *a/λ*=0.10 | | | | *a/λ*=0.48 | | | |
|---|---|---|---|---|---|---|---|---|
| | FWHM (*a*) | ΔF (*a*) | FWHM (λ) | ΔF (λ) | FWHM (*a*) | ΔF (a) | FWHM (λ) | ΔF (λ) |
| 1*a* | 7.26*a* | 6.00*a* | 0.72λ | 0.60λ | 3.99*a* | 22.95*a* | 1.92λ | 11.01λ |
| 3*a* | 5.30*a* | 4.25*a* | 0.53λ | 0.42λ | 2.52*a* | 12.40*a* | 1.21λ | 5.95λ |
| 5*a* | 4.43*a* | 1.90*a* | 0.44λ | 0.19λ | 2.36*a* | 11.30*a* | 1.13λ | 5.42λ |
| 7*a* | 4.13*a* | 1.40*a* | 0.41λ | 0.14λ | 1.85*a* | 6.95*a* | 0.89λ | 3.33λ |
| 9*a* | 3.03*a* | 0.20*a* | 0.30λ | 0.02λ | 1.64*a* | 4.65*a* | 0.79λ | 2.23λ |
| 11*a* | 3.12*a* | 0.15*a* | 0.31λ | 0.015λ | 1.43*a* | 1.55*a* | 0.69λ | 0.74λ |
| 13*a* | 3.11*a* | 0.15*a* | 0.31λ | 0.015λ | 1.42*a* | 0.45*a* | 0.68λ | 0.21λ |
| 15*a* | 3.18*a* | 0.05a | 0.31λ | 0.005λ | 1.41*a* | 0.10*a* | 0.67λ | 0.04λ |

## 3.2. Mode Transformation Using Graded Photonic Crystals With Axial Asymmetry*

### 3.2.1. Introduction

Photonic crystals are periodic dielectric structures which have an extraordinary ability to control the flow of light at optical wavelength scales [4]. The uniqueness of these structures is that they exhibit photonic band gaps (PBGs), *i.e.* frequency bands in which propagation of light are suppressed [4]. PC structures with fairly large PBGs are of paramount importance in designing polarization independent photonic designs [57, 95-96]. Besides, large PBGs are also required in a variety of optical applications such as waveguides [50,162] and defect-mode microcavities [163]. Such types of designed PC devices can be assembled into a single integrated device, which in turn provides additional improvements in terms of system size, power consumption, controllability, reliability, and cost.

Those composite devices are named as photonic integrated circuits (PICs). In PIC applications, it is important to manipulate and transform the fundamental state of propagating mode to the higher order modes. This transforming property can be used in mode multiplexing-demultiplexing, mode filtering, and multiport optical switching applications [164]. There have been reported some mode conversion studies by phase matching, *i.e.* introducing structural restrictions to the waveguide dimensions in the transverse direction [165-167]. Another mode conversion method is multichannel branching using adiabatic transition regions where fundamental mode is transformed adiabatically and output signals are observed in the separated branches [168]. Non-adiabatic technique for mode conversion by using the irregular metallic waveguide structure is also demonstrated in Ref. 169.

While comparing both methods, the adiabatic transformation can be efficiently achieved over a relatively long distance while the other approach (non-adiabatic) is highly sensitive to variations in the input signal. Nevertheless, in both cases fabrication disorders might be a serious problem. The mode conversion process may also be accomplished by using asymmetric waveguide gratings [170, 171] or Y-junctions [172]. In a recent study, an add-drop filter has been implemented involving a linear mode conversion cavity and control of excitation symmetry for an odd TE-like mode [173, 174]. Another study experimentally reported that the second harmonic generation could be useful to control and generate spatial mode of the beams [175]. In addition, the design of a mode selector device that supports propagation of the first anti-symmetric mode of conventional silicon waveguide is reported in Ref. 176. The role of the mode selecting device is blocking all modes except the desired specific one (predefined modal pattern). Finally, a further mode conversion approach was suggested by Leuthold and his colleagues where modes of different orders were transformed one into another after propagation in a specially designed multimode waveguide [177].

Interestingly, the above mentioned schemes for mode order conversion are designed by homogenous medium and more often mode conversion is applied only for guided modes. In this study, we propose a novel method for mode conversion operation by using two-dimensional asymmetric graded index (A-GRIN) structures. GRIN PC structures are the compositions of PCs whose refractive index, the lattice period of the unit cells or the filling factor gradually alter along the propagation or transverse to propagation direction [70, 64]. The use of GRIN PCs can be an effective approach for achieving mode-order conversion task. Considering GRIN PCs' gradually varying effective refractive indices and controlling the property of group and phase velocities due to anomalous dispersion characteristic make such structures powerful tool for the mode conversion structure design. In recent years, the GRIN PC structures have been contributed to numerous nano-photonic and optical applications that prolong from optical mode couplers [66, 146] to the design of effectively focusing lenses [67, 69, 147]. Moreover, revealing the mirage effect, efficiently guiding and manipulating the flow of light can be achieved by the help of GRIN PC concept [65, 71, 148-149].

In this paper, a novel type of 2D A-GRIN PC structure is proposed to achieve mode conversion by manipulating discrete PC structure. The designed mode conversion structure is a dielectric medium whose effective refractive index gradually varies along the transverse *y*-direction. We analyzed two different 2D A-GRIN PC media as candidate environments for mode converters. The locations of PC rods have been adjusted in order to obtain an Exponential and Luneburg lens refractive index profiles. Only transverse-magnetic guided mode is considered and generated even/odd modes are denoted as $TM_0$/$TM_1$. Phase profiles of outputs for the specified GRIN PC based mode converters have been compared with that of the externally excited sources. Moreover, mode conversion concept is deeply investigated by analytical approach based on the relation between ray theory and wave optics. The conversion performances of the mode converters have been analyzed and affirmative conclusions have been conducted. Note that the effective medium theory (EMT) that is valid at the long wavelength regimes has been applied to attain the appropriate index distribution [71]. It is important to determine the boundary of the long wavelength region. The lattice constant *a* should be much smaller than the wavelength to ensure staying in the effective medium region. As a result, in the presented study the design frequency satisfies the long wavelength regime.The proposed mode converter approach is compact and has flexibility due to its two-dimensional periodical adjustment of the structure. The designed medium can be positioned in front of the incident light at a certain location to adjust the desired phase retardation between the upper and lower sections of the even mode. The multimode nature of the GRIN medium allows broadband operation. It is expected that input and output coupling of light is not a formidable task due to easy beam width alteration in GRIN medium. Finally, planar front and back surfaces of the medium provide some misalignment tolerances [150].

### 3.2.2. Problem Definition: The Need of Mode Conversion Device

In modern telecommunication systems based on optical devices, optoelectronic components and optical fibers, it is crucial to deliver information over a long range distance with high speed and protection ratio. To achieve those requirements, scientists are forced to design cheap, compact, efficient, reliable and multifunctional integrated circuits known as PICs. It comprises multiple optical elements which have either different or common operating tasks. The role of PICs is tremendous in designing interconnection devices in optical communication systems. Moreover, controlling the spatial field distribution of light wave and providing efficient optical coupling between two distinct layers are unavoidable challenges in PIC designs. In addition, operating within the fundamental mode regime is a general requirement for the optoelectronic devices and semiconductor lasers. To guarantee stable fundamental mode operation, special care is taken to suppress higher order modes which are considered unfavorable for the systems to work properly. However, in some cases, operating in higher-order mode regime can be more useful than the fundamental mode regime. To exemplify, in Ref. 178 coupling strength of lateral gratings operate more effectively in the multi-mode regime because the lateral profile of the higher-order modes shows an adequate overlap with the grating area. Another benefit of the higher-order modes in the case of high power laser generations is demonstrated in Refs. 179 and 180. Moreover, dispersion properties of higher-order spatial modes in optical fibers can be used to compensate the positive dispersion in conventional single-mode fiber spans [181]. In this regard, the design of optical mode converters as an interface becomes a crucial task in order to allow efficient conversion of fundamental mode signal into a predefined higher-order mode or vice versa. Such an interface could be employed as a junction between two different types of waveguides or photonic crystal fibers.

In this study, a novel method exploiting A-GRIN concept is proposed to solve the mode conversion problem. The schematic of the proposed A-GRIN based mode converter and the corresponding index profiles are shown in Fig. 3.2.1. A-GRIN medium is designed such that the value of refractive index beginning from the upper edge of the structure increases until reaching to the optical axis (OA) according to a predefined index profile, which is depicted in Fig. 3.2.1. The index variation of the lower part of the structure is a duplication of the upper region. The introduced asymmetry exhibits high index-contrast between the edges of the upper and bottom part in proximity of the optical axis. Consequently, the incident beam centered at the optical axis exposes to different modulations by the upper and lower parts of the structure and thus incident beam divides into the two branches. Since upper and lower halves of beams propagate with different phase velocities, the proposed structure demonstrates phase shifting, *i.e.* phase retardation, phenomenon at the output.

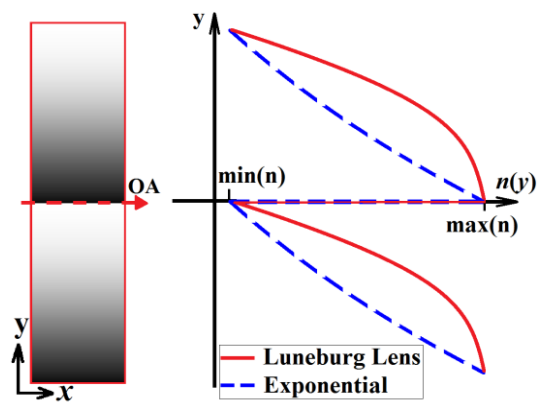

Figure 3.2.1. Schematic representation of mode converting structure with corresponding refractive index variation plots. Inhomogeneous and asymmetric GRIN media are shown on the left and the transverse profiles of the Exponential and Luneburg Lens refractive index profiles are presented on the right.

We consider two different index profiles which are shown in Fig. 3.2.1. The first one is an Exponential index variation while the second one is inspired from the Luneburg lens index profile. As known, implementing the mode conversion process by A-GRIN continuous media can be considered as a challenging issue due to fabrication difficulties during the diffusion process, the need of planar faces and restriction or limitation of the index gradient (usually smaller gradient occurs). To overcome these difficulties, the use of PC for the approximation of continuous GRIN media is considered. Details of these profiles and calculation steps to determine the effective index distribution will be explained in the next section.

### 3.2.3. Mode Converter Structure: Design Approach by Asymmetric GRIN PC

In this study, the design of A-GRIN structure is achieved by relocating PC dielectric rods. Continuously graded refractive index media can be approximated by changing the locations of the PC dielectric rods having fixed refractive indices. To achieve this firstly, the dispersion diagrams of PC unit cells with different lateral sizes are calculated by exploiting plane wave expansion method in the frequency domain [85]. Obtained dispersion relations of the first band are shown in Fig. 3.2.2(a). The related bands move to higher frequencies while the unit cell size increases (lateral dimension increases and longitudinal dimension is kept constant). Note that the lateral dimension of the unit cells ranges from $\Delta y=0.40a$ to $\Delta y=2.0a$ with a $0.20a$ step size. The lattice constant is represented by $a$. Moreover, the distance between rods along the propagation $x$-direction is denoted by $\Delta x$ and fixed to $a$, *i.e.* gradient of index profile changes along only the transverse $y$-direction, and the radii of the rods are equal to $r=0.20a$. Corresponding variations of cell sizes are depicted in Fig. 3.2.2(a) as an inset. Due to the difficulty of implementing rods with different materials in practice, we keep refractive indices of them at $n=3.13$. Each rectangular cell consists of cylindrical alumina rods residing in air background. The second stage in the design is calculation of the group indices of each band ($n_g$) by using the slope information of the relevant curves presented in Fig. 3.2.2(a). The obtained group index variation is illustrated in Fig. 3.2.2(b). For longer wavelengths (lower frequencies), curves are closely spaced and thus provide a slight variation in group indices which can be understood from the Fig. 3.2.2(b). However, when we move towards the edges (cut-off region) the dispersive effect provides a nonlinear behavior in group index plots. Each curve enters the cut-off region at different frequencies and strong dispersion occurs at around these regimes. As a final stage, the proposed structure having a specified refractive index profile is designed at a fixed frequency lying in the region where small alterations occur in the group indices. We should note that A-GRIN structure is designed at the normalized frequency of $a/\lambda=0.10$. The $n_g$ curves are zoomed out in the design frequency region around $a/\lambda=0.10$. The zoomed plot given as an inset in Fig. 3.2.2(b) demonstrates the extension of the calculated indices. The calculated group indices at determined normalized frequency $a/\lambda=0.10$ covers values between 1.39 and 2.28. To obtain GRIN PC structure with any type of stair-step index distribution within those index values (ranging from 1.39 to 2.28) one needs intermediate index values to generate smooth variation in the desired index profile. Hence, interpolation method is applied to determine intermediate points by fitting the calculated $n_g$ profile of rectangular cells that have lateral sizes from $\Delta y=0.40a$ to $\Delta y=2.0a$. Then, the concerning structure is formed by sequentially placing the rectangular cells having those intermediate $n_g$ values in such a way that the desired step stair index distribution is revealed. It should be noted that there is a restriction on the displacement of dielectric rods. Since the diameter of each rod is $0.40a$, the minimum separation between adjacent rods should be greater than $0.40a$ to inhibit touching each other. In this regard, the minimum distance starts from $\Delta y=0.44a$.

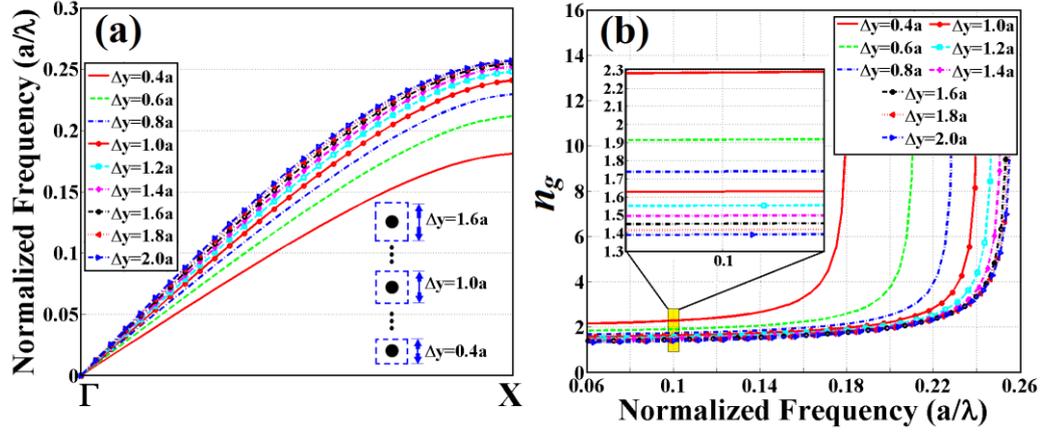

Figure 3.2.2. (a) The dispersion curves corresponding to the first band are shown. (b) Group index ($n_g$) dependency of each dispersion bands is shown. The index variation at $a/\lambda=0.10$ is also given in same plot as an inset.

As mentioned before, two different index profiles for mode converter design are considered. Specifically, they are Exponential and Luneburg lens index profiles. The Luneburg lens profile is mathematically expressed as follows:

$$n_{L\text{-}lens}(y) = \begin{cases} \sqrt{b - \delta(y/y_{max})^2} & y \geq 0 \\ \sqrt{b - \delta((y + y_{max})/y_{max})^2} & y < 0, \end{cases} \quad (3.2.1)$$

where the regarding parameters are $b=4.84$, $\delta=3.15$ and half of maximum length of the structure in the lateral $y$-direction $y_{max}=10.36a$. An Exponential profile formula is given in Eq. 3.2.2 as follows:

$$n_{\exp}(y) = \begin{cases} n_0 \exp(-\alpha y) & y \geq 0 \\ n_0 \exp(-\alpha(y + y_{max})) & y < 0, \end{cases} \quad (3.2.2)$$

where the expressed parameters are denoted by $n_0=2.20$ (refractive index at the optical axis), gradient parameter $\alpha=0.0526a^{-1}$, and half of maximum length of the structure in the lateral $y$-direction $y_{max}=10.36a$. Note that the parameters ($b$, $\alpha$, $\delta$) are determined after performing sequential optimization process for fixed $y_{max}$ and $n_0$ values. In Fig. 3.2.3(a), the schematic view of the proposed A-GRIN PC structure having Luneburg lens index profile is presented and related parameters are depicted also in the same plot. The height and width of the structures are equal to $h=20.72a$ and $w=5.20a$, respectively. Also, the input and generated output mode profiles are figuratively superimposed in the same figure. To reduce the Fresnel reflections at the front and back interfaces, additional layers are introduced with smaller radii $r=0.10a$. Figure 3.2.3(b) shows continuous and approximated stair-step index distribution for both continuous GRIN and GRIN PC based configurations, respectively. In a similar fashion, the schematic of A-GRIN PC having Exponential index profile and regarding effective refractive index profile is presented in Figs. 3.2.3(c) and 3.2.3(d), respectively.

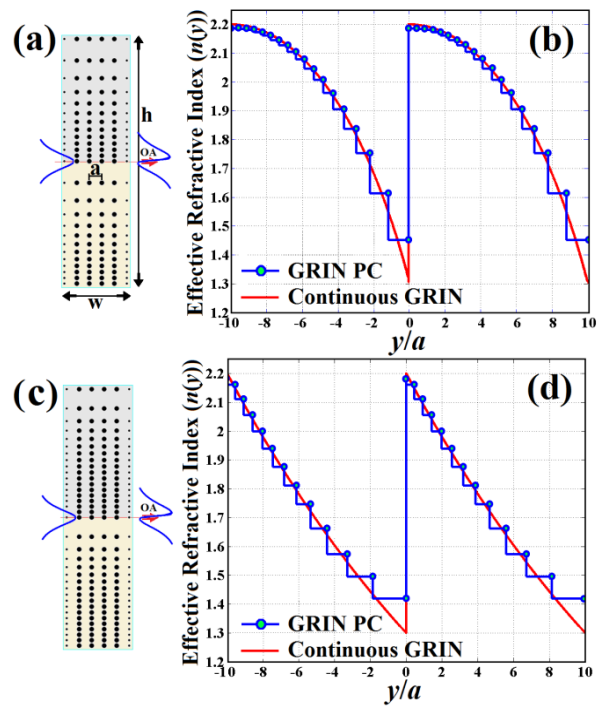

Figure 3.2.3. Schematic view of proposed GRIN PC structures where (a) and (b) demonstrate GRIN structure with the Luneburg lens index profile and its index distribution plots (continuous and discrete versions (stair-step plots)), respectively. Similarly, the Exponential index profile GRIN PC structure is represented in (c) and the related index variation plot is shown in (d).

All the effective index plots in Fig. 3.2.3 demonstrate that the effective refractive index of the A-GRIN PC decreases towards the upper part of the structure due to the increment of lattice spacing between the adjacent rods. The opposite case occurs for the lower section of the structure. In the case of exponential index distribution that varies from 1.42 to 2.20, the first rectangular cell placed at the optical axis (at the location $y=0a$) has a lateral size of $\Delta y=0.44a$ and the last placed rectangular unit cell possesses a lateral size $\Delta y=1.88a$ (at the location $y=10a$). On the other hand, the A-GRIN structure with Luneburg lens index profile comprises of rectangular unit cells with lateral sizes ranging between $\Delta y=0.44a$ and $\Delta y=1.93a$. Based on the above described method, the minimum effective refractive index values for the generated A-GRIN media with the Exponential and Luneburg lens profiles are calculated as 1.42 and 1.45, respectively. However, the highest refractive index value is the same for both profiles and equals to $n=2.20$. The main difference between the two index profiles is the gradient variation. The shape of the Exponential profile dictates that profile's gradient increases by moving from $y/a=-10$ to $y/a=0$ which is represented in Fig. 3.2.3(d). On the contrary, the gradient of the Luneburg lens profile in the same direction decreases.

Time-domain data is obtained by the help of FDTD method to calculate transmission efficiency and analyze mode conversion performances [27,119]. Perfectly matched layers construct the boundaries of the structure in order to serve as an absorbing boundary condition [119]. Moreover, only TM polarization has been assumed thus non-zero electric and magnetic fields are $E_z$, $H_x$, and $H_y$, respectively. A grid size of $\Delta x = \Delta y = a/32$ is implemented as a mesh size in FDTD calculations. We should note that two types of input sources: either a continuous source or a pulse with a Gaussian profile in time is used. The former one is needed to obtain spatial intensity distributions of the mode conversion features of the A-GRIN PC. On the other hand, the latter type of input source is preferable to compute the power transmission spectrum of the proposed structure. A detector is located at the end of the structure in order to measure the transmitted light. The corresponding transmission efficiencies for the two different A-GRIN PC cases are detected and normalized by calculating the ratio of the detected and incident power. The result is presented in Fig. 3.2.4. We should note that the computational domain is concerned only in 2D spatial domain and the third dimension is taken to be uniform. The inspection of the transmission plots in Fig. 3.2.4 shows that the high transmission region lies between $a/\lambda=0.193$ and $a/\lambda=0.204$. In case of Exponential profile transmission efficiency is around 80% while for the Luneburg lens case the calculated transmittance is around value of 70%. Slight fluctuations in the transmission plots are originated due to the reflections between the front and back faces of the A-GRIN PC structures. Operating within the selected low frequency region introduces additional leakage of light along the transverse *y*-direction (some amount of light cannot reach the end face of the GRIN PC medium, weak guiding causes some amount of power to be lost). Hence, for both cases the transmission efficiencies are not reaching the maximum transmittance level.

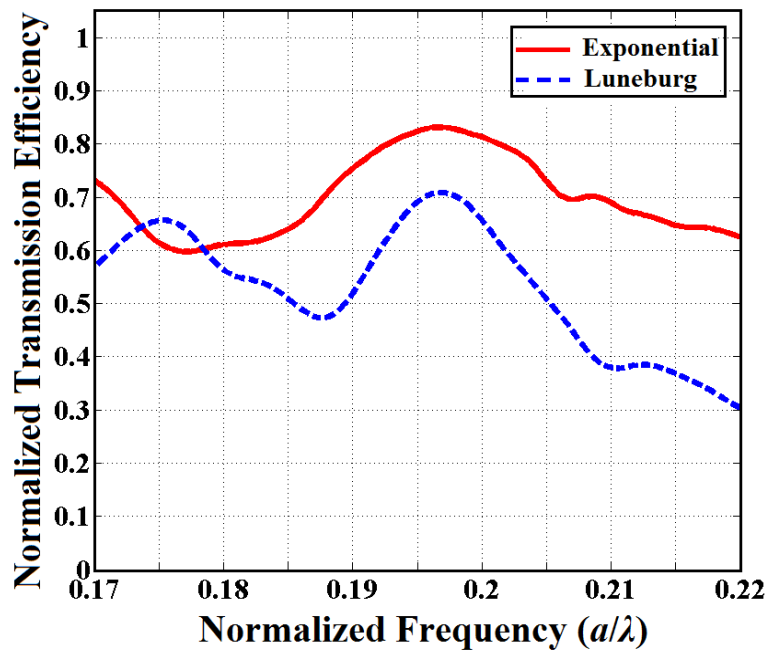

Figure 3.2.4. Calculated transmission efficiencies of A-GRIN PC structures with Exponential and Luneburg lens index distributions.

### 3.2.4. Discussion and Evaluation of the Numerical Simulation Results

The difference between the lateral and longitudinal dimensions of the individual PC cell provides the construction of the anomalous spectral features. Desired manipulation of light propagation can be achieved by properly arranging the individual cells' dimensions. Moreover, to qualitatively analyze light propagation behavior inside the PC structure the information about the phase and the group velocity is deployed. In this regard, the frequency domain analyses should be performed at the individual cell level where the importance of the iso-frequency contours (IFCs) is invaluable. These contours are the intersections of band surfaces and planes at the particular frequencies. The calculation of the IFCs will give us information about group velocity and/or phase velocity characteristics of propagating light within the PC structure. In order to calculate dispersion relations of Bloch modes within the designed GRIN PC structure we have utilized PWE method. The propagating beam follows the direction normal to the IFCs and has a group velocity, which is determined according to the relation $v_g = \nabla_k \omega(k)$, where $k$ corresponds to wave-vector [4]. On the other hand, the phase velocity $v_p = \omega(k)/|k|$ characterizes the speed of propagating wavefronts. Note that the phase velocity and group velocity are scalar and vector quantities, respectively.

We start with the computation of the dispersion contours for the individual PC cells with different lateral sizes $\Delta y$ to compute dispersion relation of the whole GRIN PC structure. Figure 3.2.5(a) is a collection of intentionally selected dispersion curves for PC unit cells having lateral sizes of $\Delta y = \{0.44a, 0.57a, 0.71a, 0.98a, 1.26a\}$ at a fixed normalized frequency $a/\lambda = 0.20$. The reason for selection of that frequency will be explained later in the text. The selected frequency contours correspond to the first band of TM polarization mode. Related individual cells are figuratively depicted in the same plot as an inset. We know that in a square $(a \times a)$ unit cell PC the phase and group velocities are independent of the orientation since the IFCs are in a circular shape. However, in case of GRIN PCs, due to the gradually changing size of the rectangular cell along $y$-direction, both phase and group velocities are inconstant and depend upon incident waves' direction [65, 182]. Therefore, changing the aspect ratio induces prominent transformation of IFCs from circular to the elliptical shape dispersion surfaces, which can be observed in Fig. 3.2.5(a). These shape transformations can be explicated as a variation of the effective index amount in rectangular cell. Note that the aspect ratio $\rho$ is defined as the ratio of the longitudinal size $\Delta x$ to the lateral size $\Delta y$ of the a rectangular PC unit cell. The aspect ratios of the selected unit cells described in Fig. 3.2.5 are determined as $\rho=\Delta x/\Delta y=\{2.27, 1.75, 1.40, 1.02, 0.79\}$. In Fig. 3.2.5(a), the straight solid line is defined as a construction line, which indicates the conservation of tangential component of the wavevector inside PC medium. The normal vectors that are illustrated by arrows at the intersection between the construction line and dispersion curves determine the flow directions of light inside the structure for different rectangular cells. It can be deduced that the gradual change in aspect ratio produces expected but yet dramatically change in IFC shapes (minor and major axes of contours swap) and thus leads to transformation of the light direction. The sequence of the lateral size $\Delta y$ alternation of rectangular PC unit cells is in opposite direction for the lower region of the A-GRIN PC mode converter structure. Similarly for upper region, the corresponding collection of selected dispersion curves at fixed frequency $a/\lambda=0.20$ are presented in Fig. 3.2.5(b). While moving from the optical axis toward the lower edge of the structure the corresponding aspect ratio decreases; thus, the elliptical contours become narrower according to the longitudinal axis.

It is worthy of note that the IFC analysis is implemented to contribute the physical explanation of light trajectories inside the inhomogeneous PC medium. The contours represent the corresponding curves of each rectangular cell picked up at both upper and lower part of the structure. The layer is designated as an interior part of the configuration. Even though the input source is incident only along the Γ-X direction, right after light interacts with the first layer of the A-GRIN structure, non-zero $k_y$ component arises. As a result, we draw a representative construction line for a fixed $k_y$ value (positive and negative values for upper and lower sections, respectively) and inspect the change occurring in the gradient of the contours. That information gives us an insight about the light paths.

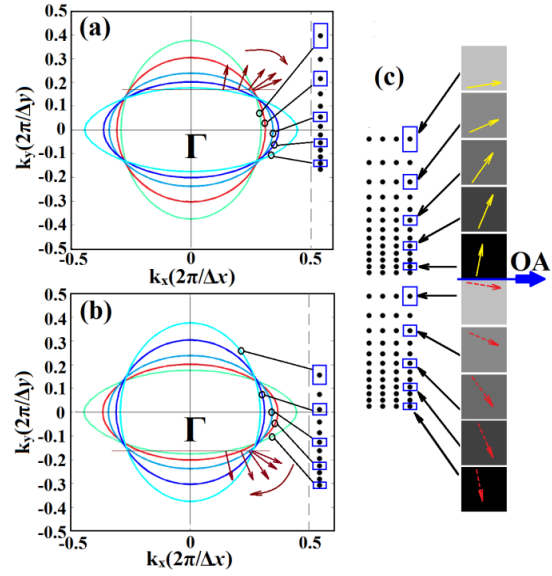

Figure 3.2.5. Plots (a) and (b) demonstrate the collections of IFCs of different lateral sizes (rectangular cells are depicted as an inset) rectangular PC cells. The corresponding $\Delta y$ ranges from 0.44a to 1.26a and vice versa regarding to the optical axis (upper and lower part of the structure), respectively. The arrows which are normal to corresponding IFCs in same plots represent light flow direction. (c) The corresponding light propagation directions within each cell are superimposed by the help of two different types of arrows (solid and dashed). The medium has an Exponential index profile.

By inspecting the collected IFCs (for the upper and lower regions) we can demonstrate the direction of flow of light inside the GRIN PC structure by means of group velocity vector directions. Designed A-GRIN PC mode converter structure and figurative interpretation of the interior part of the configuration is schematically presented in Fig. 3.2.5(c). The interior part is illustrated in such a way that the rectangular PC unit cells with different lateral $\Delta y$ sizes correspond to the rectangular cells with different gray color level. The representation in this form is intended to schematically demonstrate effective index variation in A-GRIN PC structure. The qualitative analyses of IFCs testifies that the propagating beam inside the PC structure separates into two branches and follows different optical paths (solid and dashed arrows) as shown in Fig. 3.2.5(c). The propagation of light within the PCs can be represented by Bloch waves. Even though strong scattering is imposed to propagating Bloch waves by the periodic structure, they have definite propagation directions. The propagation direction after any interface is determined by a boundary condition, hence it is required the tangential component of the wave-vector $k$ to remain constant. In this regard, we can conclude that the phase velocity directly determines the behavior of light at the boundary because of its direct relation with the *k*-vector through $v_p = \omega(k)/|k|$ expression.

Light propagation in inhomogeneous medium can also be characterized by geometrical optics or ray theory. As already known the ray behavior within the inhomogeneous medium can be explained by the eikonal equation. The eikonal equation determines the electromagnetic wave propagation by the use of geometrical optics approximation. The eikonal equation can be mathematically expressed as follows:

$$\frac{d}{ds}\left[n\frac{dr}{ds}\right] = \nabla n, \tag{3.2.3}$$

where *n* represents refractive index, *r* is a vector representation of position (*x, y*) and $ds = \sqrt{dx^2 + dy^2}$ is the differential arc length along the path of the ray. Furthermore, the geometrical light rays can also be defined as the orthogonal trajectories to the geometrical wavefronts. So that relation between surface of equal phase and ray (wave and ray optics) can be expressed as follows:

$$\left|\vec{\nabla}S\right|^2 = n^2, \tag{4}$$

where *n* is the refractive index of the inhomogeneous medium, *S* is an eikonal function and $\vec{\nabla}$ is a gradient operator. The surface where eikonal function is *S=constant* represents the surfaces of equal phases or geometrical wavefronts, which in turn dictate the shape of the propagating electromagnetic field.

In order to calculate the ray trajectories in a continuous A-GRIN medium we figuratively represent in Fig. 3.2.6(a) the illustration of ray path calculation. As can be seen in the figure, the ray is described by its position $y$ and the slope $\dot{y}$ (derivative with respect to *x*). The ray position $y_0$ and the slope $\dot{y}_0$ at the input plane $x_0=0$ are connected with the ray position $y$ and slope $\dot{y}$ at the plane *x* by solving the Eikonal Eq. 3.2.3 [130] for the half of space ($y > 0$) of continuous GRIN medium having Exponential refractive index profile (details of the derivation can be seen in Appendix):

$$y(x) = \frac{1}{\alpha} \ln\left( \frac{u_0 \alpha \cos(\alpha x) + \dot{u}_0 \sin(\alpha x)}{A\alpha} \right), \tag{3.2.5}$$

$$\dot{y}(x) = \left( \frac{-u_0 \alpha \sin(\alpha x) + \dot{u}_0 \cos(\alpha x)}{u_0 \alpha \cos(\alpha x) + \dot{u}_0 \sin(\alpha x)} \right), \tag{3.2.6}$$

where $y(x)$ is a trajectory that defines the incident position $u_0$ and incident angle $\dot{u}_0$ and $\dot{y}(x)$ provides the slope information of the trajectory. Note that the propagation of rays in an Exponential index medium obtained in exponential space ($u$) by using following transformation $u = Ae^{\alpha y}$ (see Appendix B). The ray trajectories are obtained without any approximation and the full analytical solution is given in Appendix B. The calculated ray paths within the exponential asymmetric continuous GRIN medium are depicted in Fig. 3.2.6(b). As stated before, geometrical wavefronts orthogonal to ray trajectories are also depicted as an inset in Fig. 3.2.6(b) by dashed curved lines which explain the relation between ray and wave optics. Because of lateral asymmetry of the proposed GRIN medium as a mode converter, according to the optical axis in Fig. 3.2.6(b), ray trajectories show that the propagating wave through the proposed GRIN mode converter can be divided into the two different branches of rays with different wavefronts. This behavior of the beam coincides with the prediction stated before in this section by the use of PWE based iso-frequency contours. The ray paths in Fig. 3.2.6(b) are obtained according to the Eqs. 3.2.5 and 3.2.6. The detailed derivation of the equations is described in Appendix.

To explain how the wavelength of the propagated wave gets affected by the change of the refractive index at different positions, we use the "Optical Path Length" as a physical quantity. According to the optical path length, an expression is given below [42]

$$OPL = \int_{x_0}^{x} L \, dx = \frac{1}{I_0} \int_{x_0}^{x} n(s)^2 \, ds, \tag{3.2.7}$$

where $n(s)$ is the local refractive index as a function of distance $s$ along the ray. The difference in refractive indices induced by lateral shift $h/2$ (upper part $n(y)$ and lower part $n(y+h/2)$ shown in Fig. 3.2.6(b) as an inset) imposes on travelling of light with different optical path lengths. The discrepancy on the optical path lengths in turn results in wavefront deformation and wave phase retardation of the propagated beam. Moreover, based on expressions (3.2.3), (3.2.4) and (3.2.7), we can conclude that the eikonal and, therefore, the phase of the geometrical field undergo changes dictated by the optical path length. In addition, we can give a mathematical expression of the phase difference by

$$\Delta\varphi = kn(y)ds, \qquad (3.2.8)$$

where $k=2\pi/\lambda$ is the wave-vector and $\lambda$ is the vacuum wavelength of the incident beam. Consequently, asymmetric configuration of the proposed A-GRIN structure as a mode converter splits the propagating beam into two branches as shown in Fig. 3.2.6(b). Propagating light rays within the upper and lower parts of the structure experience different optical path lengths.

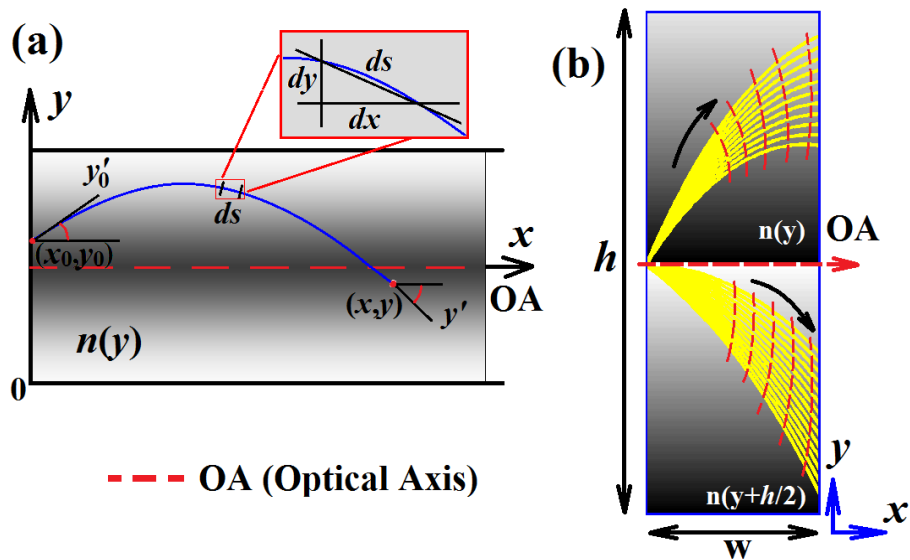

Figure 3.2.6. (a) Ray propagation in a GRIN medium with $n(y)$ index profile (b) ray paths and regarding geometrical wavefronts in an asymmetric GRIN structure with an Exponential index profile.

This peculiarity provides the phase difference between them. Note that to introduce and clarify the physical concept of the A-GRIN mode conversion process, detailed analytical investigation of the only exponential index profile structure is performed. Next, we will discuss this phenomenon and demonstrate phase shifting in detail by utilizing time-domain analyses.

In this study, the mode conversion performances of the Exponential and Luneburg lens index profile A-GRIN PC structures are investigated. To obtain A-GRIN PC structures homogenization procedure is implemented at the low frequency region ($a/\lambda=0.10$) where a slight group index variation occurs depending on the variations of PC unit cell lateral $\Delta y$ sizes. Afterwards, the FDTD analyses are utilized in order to calculate the regarding transmission over a broad bandwidth as demonstrated in Fig. 3.2.4. By choosing the lattice spacing in *x*-direction equal to $a$=310 nm we can obtain operating wavelength as $\lambda_0$=1550*nm* which is frequently used for optical communication. In this case, the length of the converters can be determined as *w*=1860*nm*. Besides, the heights of the Exponential and Luneburg lens profiles become $h$=6.423*μm*. As stated in Fig. 3.2.4, a high transmission is observed within the frequency interval $a/\lambda$=0.193-0.204. Hence, the operating frequency is selected as $a/\lambda$=0.20 that lies inside that interval. According to the transmission efficiency results, we launch an even mode continuous source with a Gaussian profile at a normalized frequency of $a/\lambda$=0.20 where high transmittance is observed and acquire the corresponding instantaneous spatial field profiles in Figs. 3.2.7(a) - 3.2.7(c). Figure 3.2.7(a) shows the field pattern of incident even mode source while Figs. 3.2.7(b) and 3.2.7(c) represent the transformed even to odd mode field profiles by exploiting A-GRIN media with Exponential and Luneburg lens index profiles, respectively. The dashed and solid arcs in Figs. 3.2.7(a) - 3.2.7(c) represent the radial sections. The reason for the selection of circular shaped cross sections is to represent the corresponding phase differences among the radiating lobes. As can be deduced from the radial phase profile in Fig. 3.2.7(d), for the incident even mode source, no phase differences occur among upper and lower power lobes. On the other hand, while exciting A-GRIN structure by an even source, $\pi$ phase shift is produced for the cases of both Exponential and Luneburg lens refractive index distributions so that efficient even to odd mode conversion occurs, depicted in Figs. 3.2.7(e) and 3.2.7(f), respectively. The calculated phase profiles strengthen the use of A-GRIN configurations for the design of compact and efficient mode order converter devices. It should be remarked that in even to odd mode conversion process only even mode source, not odd mode source is incident to the A-GRIN PC structure.

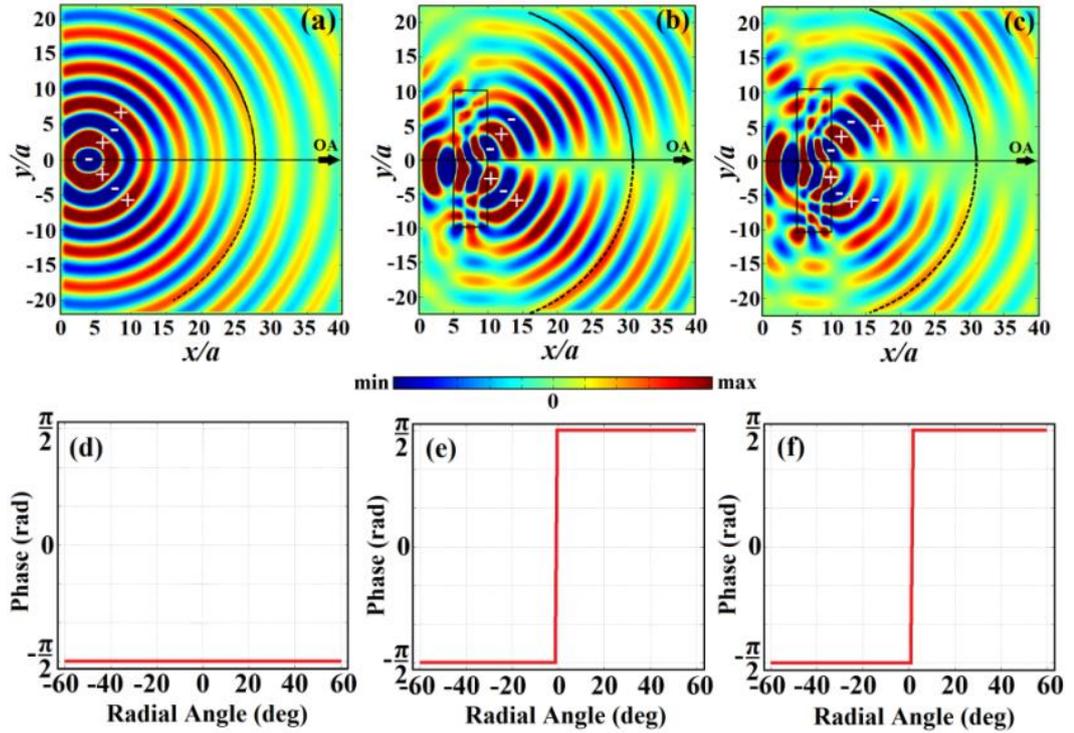

Figure 3.2.7. Instantaneous electric fields of an (a) incident even mode source without structure and converted even to odd mode source utilizing asymmetric GRIN PC structures with (b) an exponential and (c) Luneburg lens index profiles. In plots the OA indicates the optical axis and the signs "+" and "−" represent the mismatch in the phase fronts of the decomposed propagating beam. The radial phase profiles of (d) an ideal incident even mode source and configurations with (e) exponential and (f) Luneburg lens profiles. Phase profiles are extracted over predefined radial sections.

One may question the reason for the selection of two different frequencies, $a/\lambda=0.10$ and $a/\lambda=0.20$. Since effective medium theory is applied in the design of A-GRIN structure, we have conducted that stage of the work by selecting an appropriate frequency such as $a/\lambda=0.10$. The important criterion here is to stay inside the effective medium region. After that, the created asymmetric photonic structure is tested under a broadband interval thanks to the FDTD method. It is seen that higher frequencies such as $a/\lambda=0.20$ provides superior performance in terms of mode conversion. Hence, the design and operating frequencies are differently selected. As seen in Fig. 3.2.2, PC is highly dispersive at around $a/\lambda=0.20$ so that it is difficult to implement a gradual variation for the refractive index profile. Small lateral size change of the unit cell induces large refractive index variation.

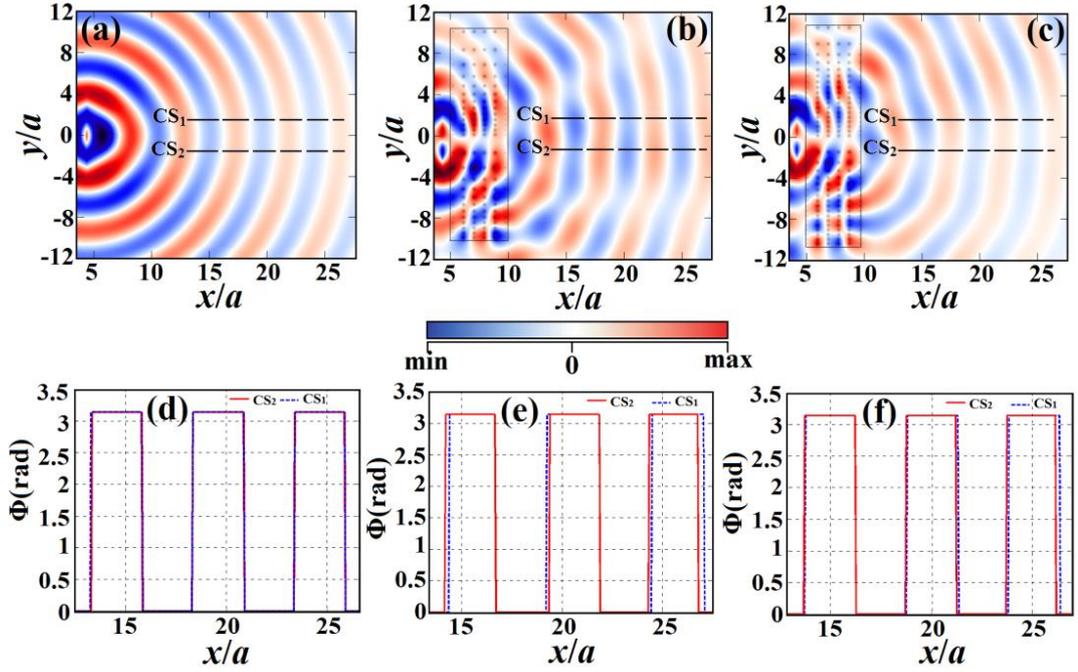

Figure 3.2.8. Instantaneous electric fields of an (a) ideal even mode source without any structure, (b) an Exponential index A-GRIN PC and (c) Luneburg lens index A-GRIN PC. The corresponding phase profiles of the three cases are presented in (d), (e) and (f), respectively.

The A-GRIN PC structure also performs well for transforming odd mode to even mode. To obtain backward mode conversion process, A-GRIN structures are illuminated by an incident odd mode source at the backplane of the configurations. The spatial intensity profile snapshots (instantaneous e-fields) at the predefined time steps are collected in Fig. 3.2.8. In this case, to show odd to even mode conversion process, different type of representation is preferred. The regarding phase differences along the longitudinal directions are also extracted in each case. The instantaneous field pattern of the ideal even source is shown in Fig. 3.2.8(a). An ideal even source spatial and phase profiles are given in order to compare with generated ones by odd to even mode conversion process. The instantaneous spatial distribution of the propagating ideal source with odd mode for Exponential and Luneburg lens profile A-GRIN PC structures are presented in Figs. 3.2.8(b) and 3.2.8(c), respectively. The emanating wavefronts of the higher order mode source experience transformation induced by A-GRIN PC configuration and the odd mode is converted back to even mode with slight wavefront deformations. To quantitatively evaluate the odd to even mode conversion efficiency, *i.e.* inverse conversion efficiency, we extract the phase information along the dashed lines in Figs. 3.2.8(d) - 3.2.8(f). In Figs. 3.2.8(a) - 3.2.8(c) the dashed lines denoted as $CS_1$ and $CS_2$ define cross-section locations. As can be seen in Fig. 3.2.8(f), phase profiles at the cross-section locations of an ideal even source are completely overlapping to each other. Besides, the phase profiles of the converted beam in the back planes of A-GRIN structures having the Exponential and Luneburg lens index profiles are taken at the same location and demonstrate high overlap performance. Hence, the regarding FDTD results show that an odd to even mode conversion can also be performed with a high efficiency. It is also important to note that in inverse odd to even mode conversion process, only odd mode source is incident to the A-GRIN PC structure.

It is worthy of note that in present work efficiently converted odd mode is coupled into air. Nevertheless, one can manage to guide the converted mode into either a regular dielectric waveguide or GRIN PC waveguide by concatenating an appropriate block structure behind the A-GRIN design. By means of such a lens component, the unguided exiting beam having odd mode becomes guided while maintaining its odd mode phase pattern.

# 4. TWO DIMENSIONAL LIMITED DIFFRACTION BEAM GENERATION BY ANNULAR PHOTONIC CRYSTALS*

## 4.1. Introduction

The wavelike behavior of light can be elucidated with the diffraction of electromagnetic waves that occurs whenever a light beam encounters an aperture or an obstacle. The degree of spatial spreading of light is inversely proportional to the dimensions of the aperture. It is well known that the plane wave is a diffraction-free mode solution of the wave equation derived from Maxwell's equation. Diffraction is responsible for spreading of Gaussian intensity profile of light beam generated on a laser or a similar light source as it propagates through free space. Hence, the on-axis intensity decays quickly upon traveling a certain propagation distance. The Rayleigh range $Z_r$ is used as a criterion for determining the spreading of a Gaussian beam along its propagation direction. This parameter is a measure of the distance over which a Gaussian beam increases its cross-sectional area by a factor of two: $Z_r = \pi \omega_0^2 / \lambda$, where $\omega_0$ is the beam waist size at the focal point and $\lambda$ is the wavelength of light.

Durnin *et. al.*, was the first who pointed out that one could obtain a set of solutions for the free-space scalar wave equations that were "non-diffracting" [183]. It was shown in the same study that zero-order Bessel function could be used to obtain an exact solution to the free-space wave equation expressed as:

$$\left(\nabla^2 - \frac{1}{c^2}\frac{\partial^2}{\partial t^2}\right)E(r,z,t) = 0, \qquad (4.1)$$

where $c$ is the speed of light, $\nabla^2$ is the Laplace operator.

---

*This chapter is based on: H. Kurt and M. Turduev, "Generation of a two-dimensional limited-diffraction beam with self-healing ability by annular-type photonic crystals," J. Opt. Soc. Am. B, vol. 29, pp. 1245-1256 (2012)91

Searching for a possible solution to the aforementioned second order differential equation other than the plane wave solution gives us the following electric field (e-field) expression,

$$E(r,t) = E_0 \exp(i(-\omega t + k_z z)) J_0(k_r r). \tag{4.2}$$

The phase term oscillates at an angular frequency $\omega$ and e-field propagates along the $z$-direction with a wave vector component $k_z$. The amplitude of the solution along the transverse to propagation direction varies according to $J_0$ zero-order Bessel function of the first kind. The wave vector components obey the relation $k = (k_x^2 + k_y^2 + k_z^2)^{1/2} = 2\pi/\lambda$, where the transverse wave vector is defined as $k_r = (k_x^2 + k_y^2)^{1/2}$ and $r = (x^2 + y^2)^{1/2}$. If we look at the power intensity variation of the solution upon propagation, we realize that it is proportional to $J_0^2(k_r r)$ which is independent of the $z$ propagation direction. Hence, the diffraction-free nature of such a solution becomes apparent. However, this solution exists in 3D space and to generate an idealized Bessel beams necessitates an infinite amount of energy which is physically impossible. In practice, however, one may approximately generate quasi-Bessel beams using finite apparatus [184-195]. The integral form of $J_0$ can be written as follows [196]:

$$J_0(k_r r) = \frac{1}{\pi} \int_0^{\pi} \exp(i k_r r \cos(\theta)) d\theta. \tag{4.3}$$

A series of side lobes accompany the narrow central lobe which conveys some part of total power. We should note that plane wave is a special case of the above solution, i.e., $k_r = 0$, $J_0 = 1.0$. The radial intensity of the Bessel beam decreases as the argument of the Bessel function, $(k_r r)$ increases. For the interval $0 < k_r < \omega/c$, the solution is a propagating wave. The evanescent wave corresponds to the interval $k_r > \omega/c$. Finite-apertures contribute diffraction due to limitation on the radial extent of the Bessel beam. In addition to being limited-diffraction over certain extend, Bessel-like beams exhibit self-healing property as well. If they encounter an obstacle in their propagation path, they can re-construct themselves [185, 197-200].

Integrated photonic circuitry populates the use of planar light-wave technology. The functionalities of optical elements usually occur in the plane and the out-of-plane direction has usually uniform variation with a limited height's of certain types of slabs. The 2D counterpart of less diffracted beams generated in 2D space may contribute to some important applications in semiconductor lasers and optical on-chip data transfer. As a result, there is also need to create limited diffraction beams in 2D. In other words, discrete optical elements occupying huge volumes in 3D can be used to create pseudo-nondiffracting beams which can be useful in many applications including microscopy, optical tweezers, imaging, nonlinear optics and laser micromachining [185, 201-205]. In the 2D counterpart, by compromising from the third dimension, diffraction can be tailored in-plane using configurations that occupy compact areas at will. This is one of the motivations of the current work. Suppressing light spreading in 2D can also be achieved by utilizing different approaches such as surface modes and tapered waveguide exits [206-213]. All these methods more or less aim a similar goal, *i.e.*, carrying optical source to longer distances without serious diffraction. In 2D configuration where the out-of-plane direction $x$ is assumed to be uniform, the following conversions occur: $k_r=k_r$, $k_x=0$, and $k = \sqrt{k_y^2 + k_z^2} = 2\pi/\lambda$.

In this study, computational domain is concerned only in 2D spatial domain and third dimension taken to be uniform. The generation of limited diffraction beam restrictively governs the diffraction dependence only along propagation direction. Moreover, the generated limited diffraction beam's transversal intensity profile bears resemblance to the zero-order Bessel function profile. In order to avoid any misunderstanding, we should make clear distinction between generated limited diffraction beam and Bessel-like beam. As it is known, Bessel-like beams are solutions to the scalar wave equations with two transverse coordinates. However, generated limited diffraction beam is characterized only in one transverse direction.

## 4.2. 2D Axicon Shape Photonic Crystals and Numerical Results

Limitations of creating non-diffracting beams which have infinitely large diffraction-resistant propagation distance encouraged researchers to propose novel solutions to create pseudo-nondiffracting beams. Aperture lenses, conical lenses, holographic technique and spatial light modulators are among some of these techniques [185, 189-192, 214, 215]. In a recent study, 2D limited diffraction beams with Bessel intensity profiles were generated via axicons made of square-lattice photonic crystal structure [195]. Advantages of generated beam over Gaussian profile were highlighted in that study. In Ref. 216, the authors investigated the round axicon tip instead of sharp one, both theoretically and experimentally. The influence of the wave refracted by the round tip of the axicon and the axial oscillations of the central peak's power variation in the propagation direction occurred by the off-axis part of the axicon were explored. It is also possible to generate high-order Bessel beams which have a central minimum and a non-diffracting bright ring. To generate such beams Laguerre-Gaussian light beam launched on axicon in order to transform that beam into an approximation of a high-order Bessel beam is investigated in Ref. 217.

In the present work, we construct an axicon shape PC using an annular PC (APC) [57]. There are additional flexibilities in the design stage thanks to APC configuration. It has more features in terms of structural parameters within the unit cell such as the two radii and refractive indices of each region (inner and outer rings). Besides, the circular shape of the unit cell can be transferred to elliptical one which increases the design parameters. Hence, it is expected that light interaction in APC may create rich spectral and temporal wave features which may be exploited for less diffracted beams purposes. In-plane configuration of proposed study is also advantageous due to compactness of the structure. In Ref. 218, APC were proposed to engineer the dispersion property of the structure which is made of low-refractive index contrast materials. The modulation of dispersion curves was achieved by means of liquid infiltration process.

The designed axicon-shape APC and band diagram of the unit cell for TM polarized light (e-field is perpendicular to the crystal plane) are shown in Figs. 4.1(a) and 4.1(b), respectively. The structure consists of circular dielectric rods perforated at the center. The external and internal radii of the circular rods are denoted by $R = 0.30a$ and $r = 0.15a$, respectively. The lattice constant is represented by $a$. The group indices of the background environment and the dielectric material are $n_1 = 1.0$ and $n_2 = 3.13$, respectively. The axicon-shape APC structure is divided into two sections as shown in Fig. 4.1(a). The first part has a length of $6a$, where "$a$" is the lattice constant. Length of the second part that lies between the tip of the axicon and the right bound of the first part is initially $20a$. That parameter will be altered later in the paper in order to see the impact of the structure lateral extent to the generated two-dimensional beam. The square-lattice APC is terminated by an arrow form with the apex angle of $\tau$=900.

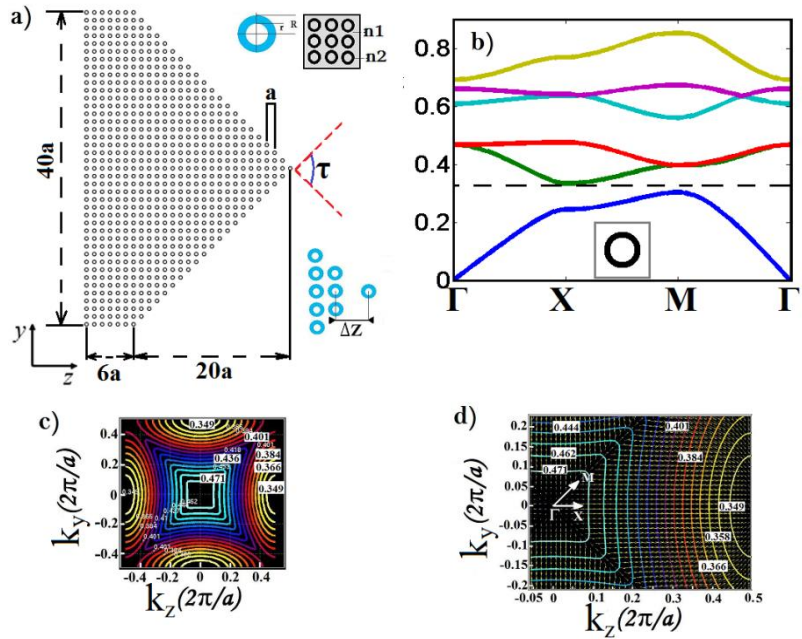

Figure 4.1. (a) The schematic presentation of 2D axicon-shape square lattice annular PC and (b) its dispersion diagram calculated by plane wave expansion method. In (c) and (d) iso-frequency contours of the second band and its gradient distribution are shown by small arrows.

The dispersion diagram is calculated by plane-wave expansion method [85]. The center frequency of the input source is set at $\omega a/2\pi c$=0.35 which is placed at the lower part of the second photonic band of the APC. The iso-frequency contour corresponding to that frequency resembles a convex shape as shown in Figs. 4.1(c) and 4.1(d). In addition, we present the gradient distribution of the iso-frequency contours in Fig. 4.1(d), where one can observe the focusing effect on the operation frequency of $a/\lambda=0.35$ along ΓX direction at second TM band. Contrary to Ref. 13 which uses different photonic crystal parameters and operating frequency in the first band, we focus second band in the present work. That ensures additional focusing on top of the geometrical arrangement. Hence, longer diffraction-free beam propagation is expected to be achieved. It is important to point out that if the operating frequency corresponds to the first band in the dispersion diagram, the PC closely acts as an isotropic medium.

We continued on numerical experiments with the FDTD simulations [27, 119]. In order to eliminate reflections originated from the ends of the finite computational window, the boundaries are surrounded by perfectly matched layers [119]. Input source with a Gaussian distribution in the time domain is launched. The normalized bandwidth of the input Gaussian distributed beam is $\Delta\omega/\omega_c=0.007$, where $\omega_c$ is the center frequency and $\Delta\omega=0.0025$ is the full-width at half-maximum of the beam.

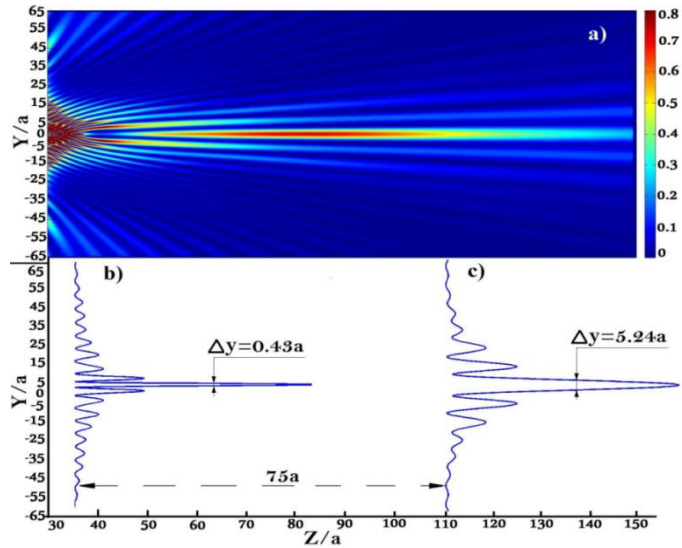

Figure 4.2. (a) The spatial intensity profile of the generated limited diffraction beam. The cross-sectional intensity profiles are shown at the focus point and $75a$ propagation distance in (b) and (c), respectively.

The two-dimensional limited diffraction beam is obtained by the transformation property of the axicon-shape APC. The spatial intensity profile is depicted in Fig. 4.2(a). By the help of side lobes, the main lobe intensity of the two-dimensional limited diffraction beam stays fairly constant along the propagation direction. The center power value of the limited diffraction beam decays to 65% of the maximum on-axis intensity after beam propagates a distance of 75*a*. As mentioned before, cross-sectional view of the generated limited diffraction beam's intensity profile resembles zero-order Bessel function as shown in Figs. 4.2(b) and 4.2(c). The transverse beam profiles at the two selected points are presented in Figs. 4.2(b) and 4.2(c), respectively. The central lobe of the limited diffraction beam stays well collimated and diffraction-resistant light propagation is apparent. The results show that beam's central lobe size at the FWHM enlarges gradually from $\Delta y=0.43a$ to $\Delta y=5.24a$. Therefore, the waist of the central lobe is increased by 12.8 times after 75*a* propagation distance. When we compare this result with the diffraction of the conventional Gaussian beam, the FWHM value varies from 0.43*a* to 159.1*a* when a measurement is taken after a propagation direction of 75*a*. As we can see from this comparison the obtained beam is much more resistive to diffraction. It is expected that the presented results will be enhanced further if some optimization procedures, such as changing the inner and outer rings radii, location and size of the apex rod are executed. Therefore, we proceed to the next section with a goal of achieving longer diffraction-limited light propagation.

We should note that we report limited diffraction beam generation at a normalized frequency value of $\omega a/2\pi c=0.35$. What makes this frequency value special is based on the following information. It is the region that occupies the bottom of the second band in the dispersion diagram (X-point) as shown in Fig. 4.1(b). When we inspect the iso-frequency contours at around these frequencies in Figs. 4.2(b) and 4.2(c) we see that contours have convex shapes. The gradient of these curves dictates the flow of light propagation according to the relation: $v_g=\Delta_k\omega(k)$, where $v_g$ represents the group velocity of beam. Due to convex shapes of the curves, beam is expected to be focused at around these frequencies. When beam reaches the inclined surfaces of the axicon-shape structure, it is strongly focused so that emerging waves at the lower and upper sections of the structure interfere and limited diffraction beam emerges. As we move to higher frequencies above 0.35 the interference is no longer achieved at the end of axicon-shape structure. Consequently, limited diffraction beam cannot be realized. Similar reasoning can be proposed for the lowest band (first band). A beam with a frequency corresponding to first band is susceptible to diffraction upon propagation. Again, the absence of interference is responsible for absence of the limited diffraction beam creation in the first band.

## 4.3. Improving Limited-Diffraction Propagation of Generated Diffraction Limited Beam

In this part of the study, we investigate alternation of the central lobe propagation distance by optimizing APC element at the apex point. As shown in Fig. 4.3(a), the central lobe intensity of the limited diffraction beam is preserved along propagation direction within different $Z_{max}$ distances for various $\Delta Z$ values that denotes apex shifting amount. The beam could propagate for a $Z_{max}$ value of $92a$ when $\Delta Z=1.4a$. Therefore, applying an appropriate optimization prompts to enlargement of the beam propagation distance for approximately 23%. While determining the $Z_{max}$ value we take the decay of the central lobe intensity to 65% of the maximum intensity of the power. Various cases for several $\Delta Z$ values are depicted in Fig. 4.3(b).

When the APC axicon apex location is optimized, two different behaviors of the limited diffraction beam's central lobe power intensity distributions are observed. In the first case ($\Delta Z < 0$), the maximum power intensity of the central lobe is shifted towards APC axicon tip as shown by the dashed lines in Fig. 4.3(b). On the other hand, in the second case ($\Delta Z > 0$), the central lobes' maximum power intensity dislocates far away from the APC axicon tip. The longest propagation distance in our case occurs when $\Delta Z=1.4a$. We investigated this case further in Fig. 4.4. The amount of side lobes of the limited diffraction beam determines the degree of the diffraction–resistive feature of the generated beam. Therefore, because of supporting side lobes, longitudinal intensity profile of the main lobe decays very slowly. The limited dimension of the computational domain in the *y*-direction artificially damps some amount of the side lobes. As a result, the axial intensity diminishes quickly. In the case of Fig. 4.4, by increasing the computational domain in the *y*-direction, we allowed the occurrence of more side lobes. As a result, the propagation distance is enlarged by an additional 4%.

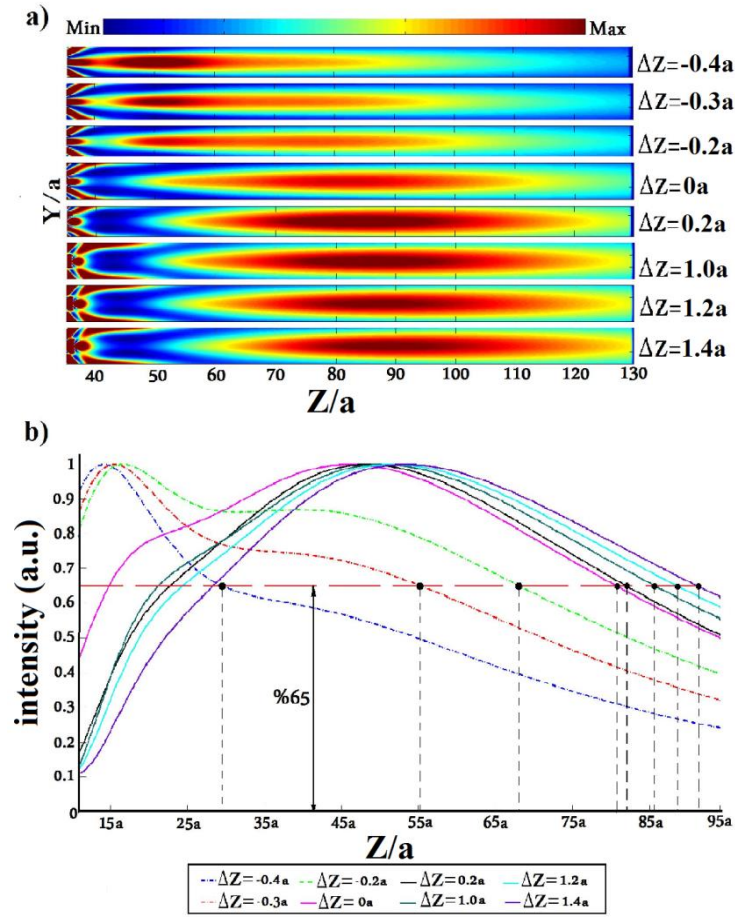

Figure 4.3. Optimization of the maximum propagation distance $Z_{max}=92a$ of the limited diffraction beam's central lobe by shifting the APC element at the apex location. (a) Spatial intensity distributions of generated diffraction limited beam. (b) On-axis intensity distributions for different longitudinal shifts of apex rod.

If we want to convert the normalized values into measurable physical quantities, the following numbers should arise. The operating wavelength can be tuned to telecom wavelength 1550*nm*. In this case, the lattice constant, inner and outer radii of rods become 542.50*nm*, 81.37*nm*, 162.75*nm*, respectively. The limited diffraction occurs over the propagation distance of ~50*μm* after the beam waist has 0.23*μm* width. The self-healing or self-reconstruction property of the generated limited diffraction beam is investigated in the next section.

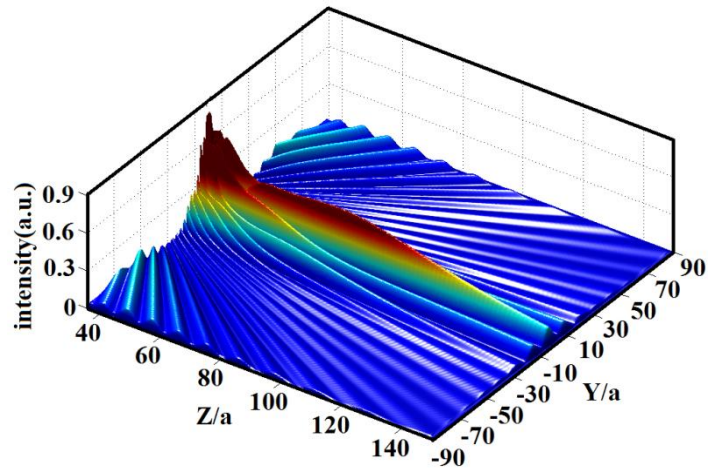

Figure 4.4. The representation of the spatial intensity profile that corresponds to apex shifting value of ΔZ=1.4*a*.

## 4.4. Self-healing Ability of Two-dimensional Limited Diffraction Beam

One of the striking features of the non-diffraction beam is its self-healing property [185, 197-200]. In this section of the article, we present a similar behavior found for our limited diffraction beam. Obstructions with different sizes and shapes (rectangular and circular) are placed along the optical axis at a certain distance. The interactions between the beam and the obstacles are investigated. The refractive index of the obstacle is *n*=1.5. The presence of the obstacle causes back reflections, which interfere with the forward propagating beam. We compared the transverse intensity profiles of the beam at two different locations, *viz.* just behind and far away from the obstacle. The two profiles exhibit disparate features. Even though the first one does not resemble Bessel function profile (there is no central lobe which is basically blocked by the obstruction), the second profile, which is measured at a certain distance away from the obstacle, fairly mimics zero-order Bessel function profile. The figures supporting these observations are presented below.

Figs. 4.5(a) and 4.5(b) show the spatial intensity profile of generated limited diffraction beams and with the effect of the presence of obstacles on the radiation pattern. The sizes of the obstacles are set to be ($dz$=0.4$a$)×($dy$=4$a$) and ($dz$=1$a$)×($dy$=8$a$). Both rectangular dielectric objects are placed at the same location at $z$=80$a$. The interference patterns between the tip of the axicon and the obstacle are seen in the above-stated figures. The amount of reflected and transmitted light depends on the size of the obstruction. As can be seen in Fig. 4.5(c), the existence of the obstacle causes disturbance of the main lobe of the pattern and an increase in the strength of side lobes. Furthermore, whenever the size of the obstacle is enlarged, the beam profile will be more influenced due to the increasing effect of back-reflection. The transverse intensity profiles for different cases in the event of the obstacle at $z$=85$a$ (40$a$+45$a$) are presented in Fig. 4.5(c). Another measurement is taken at $z$=101$a$ (40$a$+61$a$) within the same initial conditions and the resultant intensity distribution is represented in Fig. 4.5(d). When the size of the object gets larger, less light passes through and for both cases, the central lobes appear as double-peaked. The obstacle blocks the main lobe of the limited diffraction beam. After propagating certain distance such as 16$a$, the beam reconstruct itself as can be seen from Fig. 4.5(d).

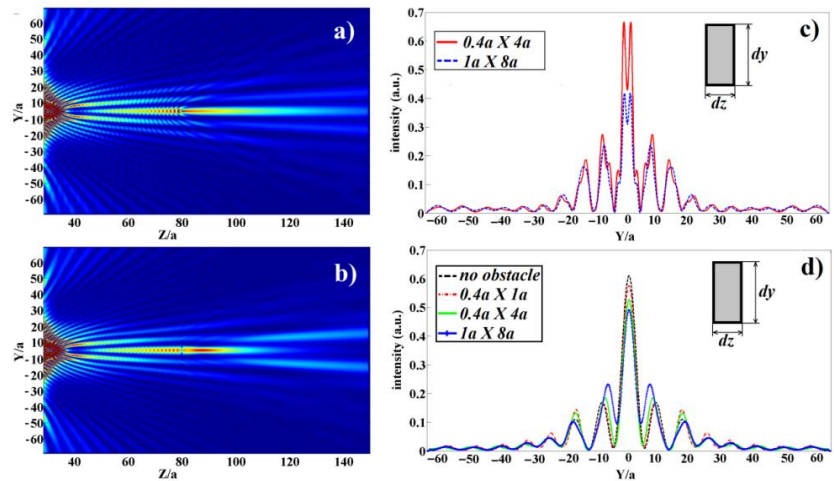

Figure 4.5: The self-healing properties of the generated diffraction limited beam. (a) and (b) show spatial intensity distributions of the generated diffraction limited beam facing obstacles with dimensions ($dz$=0.4$a$)×($dy$=4$a$) and ($dz$=1$a$)×($dy$=8$a$), respectively. (c) and (d) show comparison of the cross-sectional intensity profiles.

In order to prove the superiority of the generated two-dimensional limited diffraction beam in terms of reconstruction feature as compared with Gaussian one, FDTD simulation results are demonstrated in Figs. 4.6(a) and 4.6(b). The spatial intensity profile of free-space propagation of Gaussian beam is demonstrated in Fig. 4.6(a). A similar plot for the same beam is prepared under the presence of an obstacle with dimensions ($dz$=0.4$a$)×($dy$=1$a$). As can be seen in Fig. 4.6(b), the beams' transverse-mode intensity profile is significantly distorted. The cross-sectional profiles of the beam for different obstacles' sizes show the inability of the wave to reconstruct after propagating certain distance. Figs. 4.6(c) and 4.6(d) indicate field's transverse profiles at two different distances, respectively (i.e., 85$a$ and 101$a$).

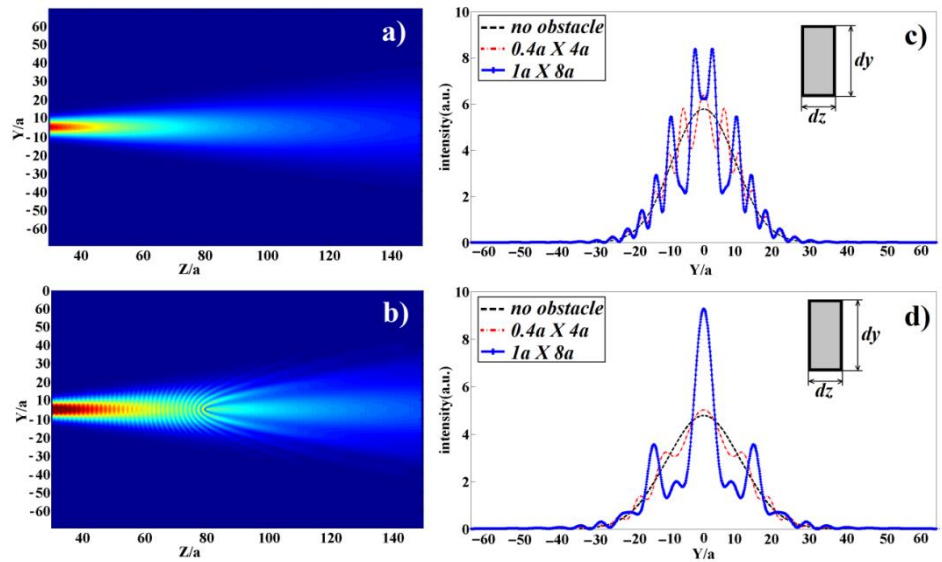

Figure 4.6. (a) Spatial intensity distribution of the tapered Gaussian beam in free-space with 0.8$\mu m$ FWHM. (b) Spatial intensity distribution of the propagation of the tapered Gaussian beam with obstacle of ($dz$=0.4$a$)×($dy$=1$a$) dimensions which settled at 80$a$ distance from source. (c) A comparison of the cross-sectional intensity profiles of the tapered Gaussian beams just after the obstacle at 85$a$ distance from the source with different obstacle dimensions. (d) A comparison of the cross-sectional intensity profiles of the tapered Gaussian beams at 101$a$ distance from the source. The different types of lines correspond to different obstacle dimensions.

Intentionally we subtracted spatial intensity profiles for both cases with and without obstacle to analyze the responses of the obstacles to the limited diffraction and

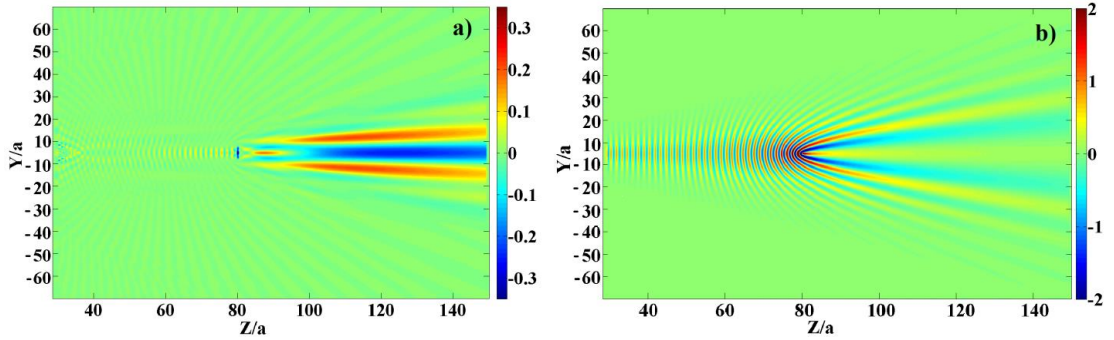

Figure 4.7. (a) The subtracted spatial intensity distribution of the generated diffraction limited beam propagation with and without an obstacle whose dimensions are ($dz=1a$)×($dy=8a$) (b) The subtracted spatial intensity distribution of the Gaussian beam propagation with and without an obstacle whose dimensions are ($dz=0.4a$)×($dy=1a$).

Gaussian beams. The resultant field distribution for Gaussian and limited diffraction beams depicted in Figs. 4.7. Mainly, central lobe gets affected for limited diffraction beam and side lobes do not sense the obstacle as shown in Fig. 4.7(a). On the other hand, in comparison with limited diffraction beam the absence of any supporting side lobes for Gaussian beam greatly modifies the forward propagating beam after the obstacle as shown in Fig. 4.7(b). The interference effect occurs for both cases due to the back-reflected light. On the contrary of our limited diffraction beam with limited-diffraction and self-healing ability, the Gaussian beam with a waist larger than that of generated limited diffraction beam is dramatically suffers from the diffraction and the appearance of an obstacle. In conclusion, we have shown re-generation capability of generated limited diffraction beam if certain size-obstacles are inserted along the wave path.

## 4.5. Discussions: Spot Size, Lateral Extent and Self-healing Phenomena

The current work reports limited diffraction light propagation distances over 75$a$ and 92$a$ by implementing a 2D axicon-shape APC with a lateral dimension of 40$a$. The FWHM of the central spot increases from 0.43$a$ to 5.24$a$ and the intensity on the central line stays fairly constant. The spot size of 0.43$a$ corresponds to a focusing of light down to $\lambda/6.6$ and it expands up to 5.24$a$ (1.8$\lambda$). The intensity level drops and the decay rate is proportional to spot size and inversely proportional to number of side lobes. It is feasible to perform that result with alternative nano-photonic devices.

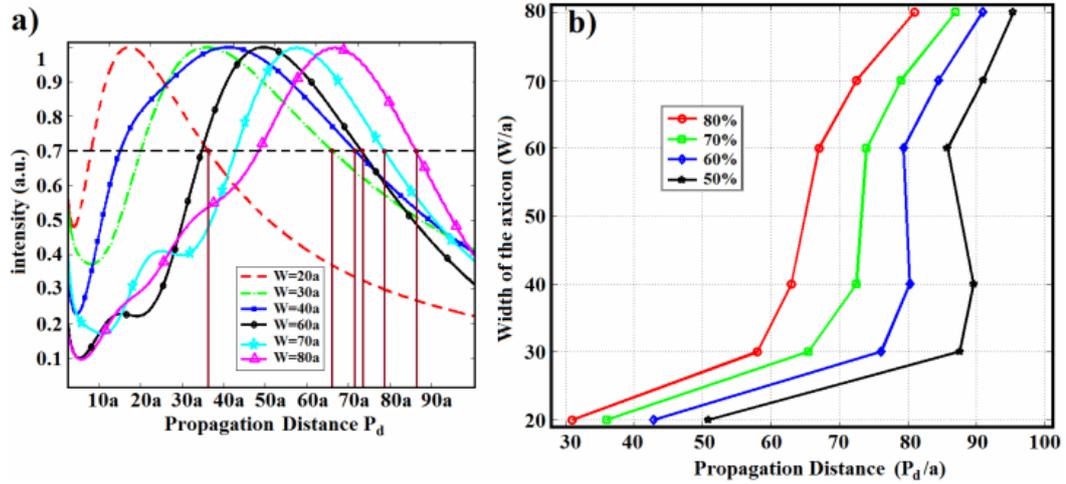

Figure 4.8. (a) The on-axis intensity distributions for different widths of axicon APC. (b) The width versus diffraction limited propagation distance.

However, most of them include metallic components. That in turn yields problems associated with the bandwidth and optical loss (absorption). On the other hand, the current photonic structure is made of all-dielectric materials. Hence, it can be operated over different frequencies without suffering from metallic losses. It is known that regular conical shape axicons with higher lateral extent are able to produce longer non-diffracting beams in 3D space. In other words, doubling the transverse dimension of axicons produces a non-diffracting beam that reaches two-times longer distance by maintaining its transverse profile constant.

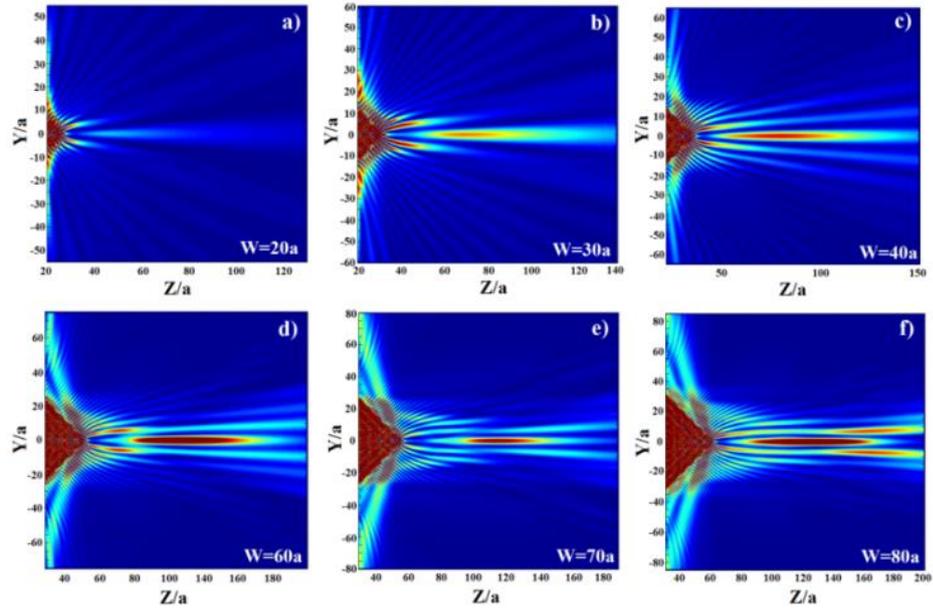

Figure 4.9. The spatial intensity profiles of the generated diffraction limited beams for different width of the axicon PC. (a) to (f) correspond to width values of W=20$a$, 30$a$, 40$a$, 60$a$, 70$a$, and 80$a$, respectively.

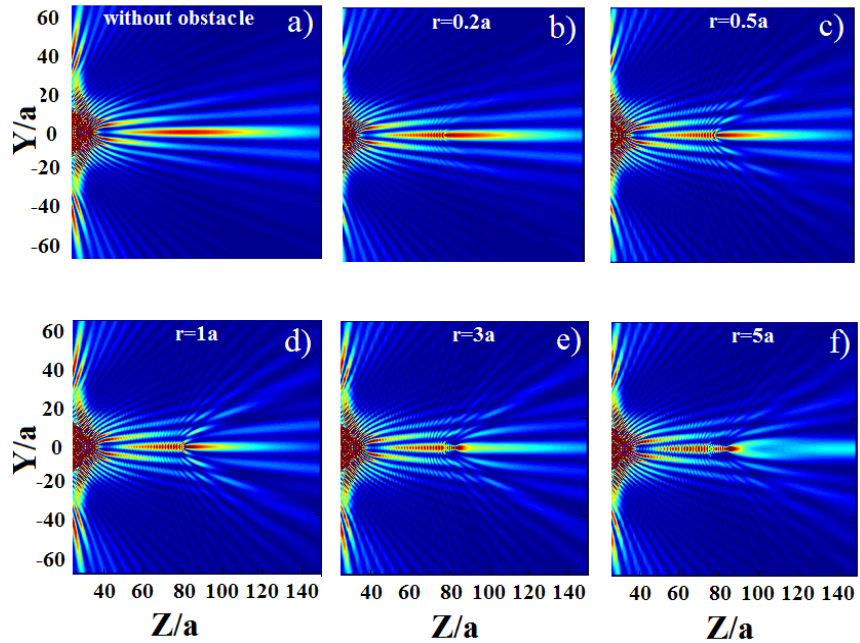

Figure 4.10. The spatial intensity profiles of the generated diffraction limited beams for different radii of obstacle placed behind the axicon PC. Frame (a) corresponds to absence of obstacle and frames (b)-(f) correspond to radii values of $r$=0.20$a$, 0.50$a$, 1.0$a$, 3.0$a$, and 5.0$a$, respectively.

However, in 2D case the dependency of axicon dimensions to the transformed beam is quite different. In this study, we explored the dependency of axicon-shape APC's lateral extent to beam profile. Therefore, the lateral dimensions of the axicon-shape APC is varied from 20*a* to 80*a* and Fig. 4.8(a) demonstrates axial intensity variations for each width value. When the normalized intensity drops to 80% of the maximum value, the corresponding propagation distance denoted as $P_d$ is read from Fig. 4.8(a).

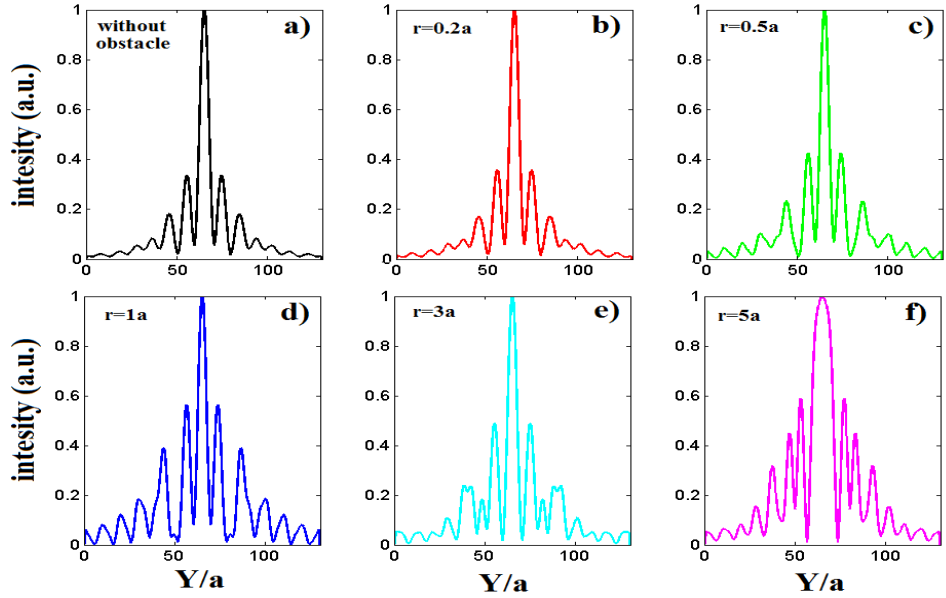

Figure 4.11. The cross-sectional intensity profiles of beams in Fig. 4.10 are shown at a distance of 85*a* from the tip of the axicon PC.

A similar data extraction is performed for 70%, 60%, and 50% cases. When we increase the width, the propagation distance shows a non-linear dependency as shown in Fig. 4.8(b). This result starkly contrasts with the result of a regular conical shape axicon in 3D space that possesses linear dependency between width and propagation distance. To underpin the nonlinear dependency we prepared spatial intensity distributions in Figs. 4.9(a) - 4.9(f). When we increase the lateral width of the axicon-shape APC, the length of the structure also increases due to the constraint that the apex angle is kept at 90°. As a result, the input light propagates a longer distance inside APC and reaches the sides of the structure with a different phase terms. The interference of lower and upper parts of the field yields different beam profile along the transverse direction. Hence, even though the number of side lobes increases when width increases the diffraction limited propagation distance shows enhancement with unequal increments.

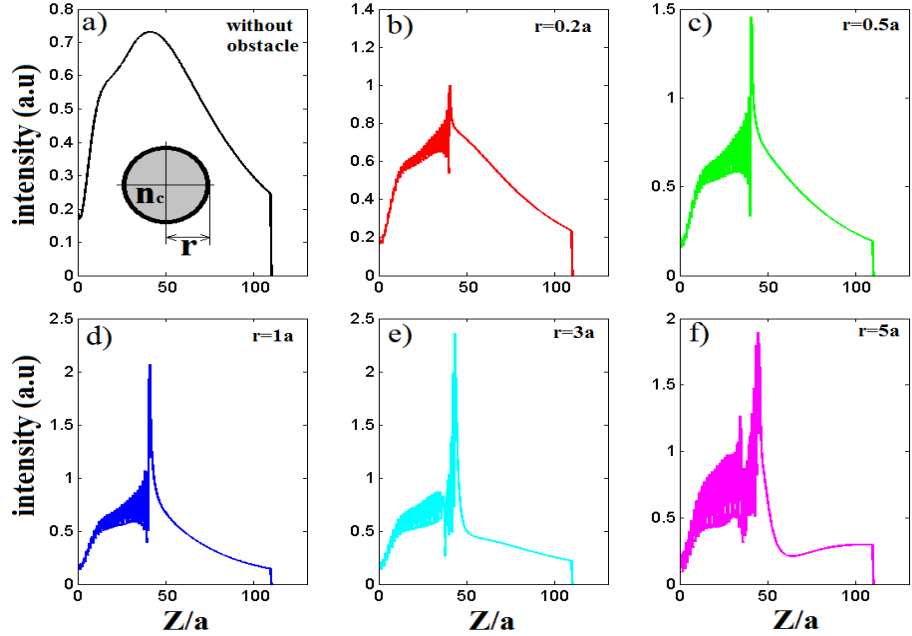

Figure 4.12. The on-axis intensity profiles of beams in Fig. 4.10 are shown for different circular shape obstacle sizes.

The self healing property of the generated limited diffraction beam is further examined in this section by using obstacles with different shapes and sizes. When we compare the self-healing performance of the limited diffraction beam under the presence of obstacles we observe various features. Fig. 4.10(a) - 4.10(f) presents variations of spatial intensity profiles for different radii of circular obstacles. The reference case without any obstacle is shown in Fig. 4.10(a). The remaining frames from (b) to (f) correspond to radii values of $0.20a$, $0.50a$, $1.0a$, $3.0a$, and $5.0a$. As the size of the circular object increases, the profiles of limited diffraction beam's transversal profile become more deformed. To indicate the degree of self-healing feature, we prepared cross sectional profiles of the beam in Fig. 4.11. The field distributions especially in frames (d), (e), and (f) are affected more than (a), (b), and (c). The axial intensity variations of the same cases are presented in Fig. 4.12. It can be shown that the back-reflected light increases as the size of the object on the path of the beam gets larger. In addition, monotonous decay of peak intensity shows different behavior in each case.

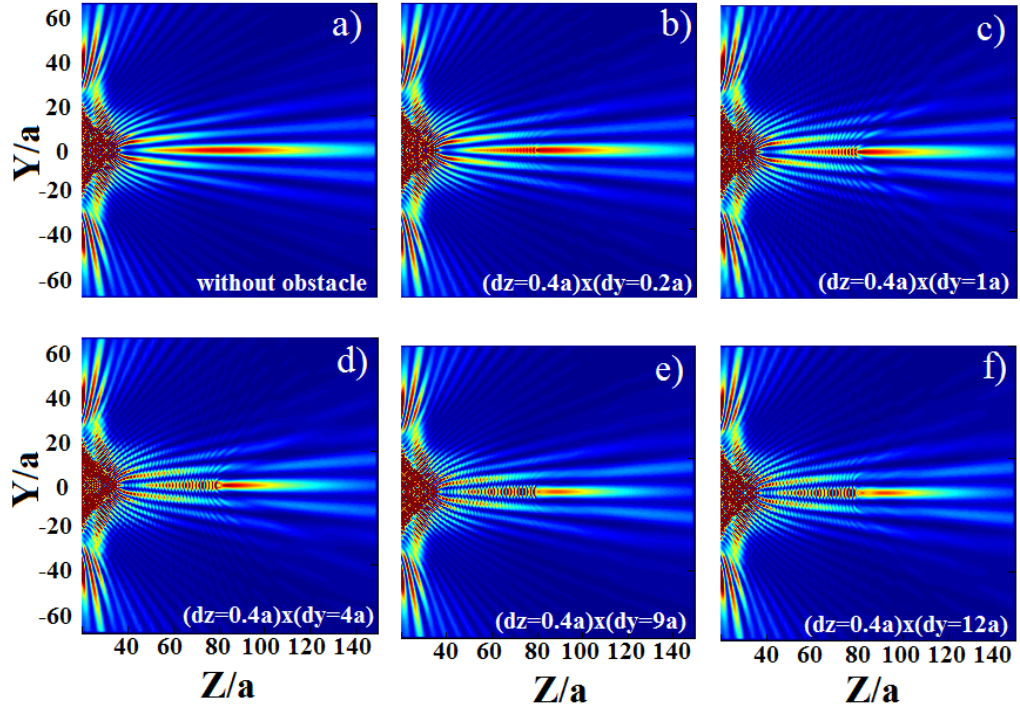

Figure 4.13. The spatial intensity profiles of the generated diffraction limited beams for different dimensions of rectangular shaped obstacle. Frame (a) corresponds to absence of obstacle and frames (b)-(f) correspond to rectangular obstacles whose dimensions are ($dz$, $dx$)={0.4$a$; 0.2$a$ -12$a$}.

Figs. 4.13(a)-(f) show spatial intensity profiles for rectangular shaped obstacles. The cross-sectional profiles are provided in Figs. 4.14(a)-(f). The reflected light increases as the size of the object increases. This trend is very similar to circular case as shown in Fig. 4.12. However, the decay characteristic of the maximum field behaves differently for rectangular case. The axial intensity distributions with respect to size of rectangular shaped obstacles are illustrated in Fig. 4.15. Generated limited diffraction beam is less affected in rectangular object case. This is due to the fact that the circular shape influences the phase front of the beam more than the obstacle that has planar front and back surfaces.

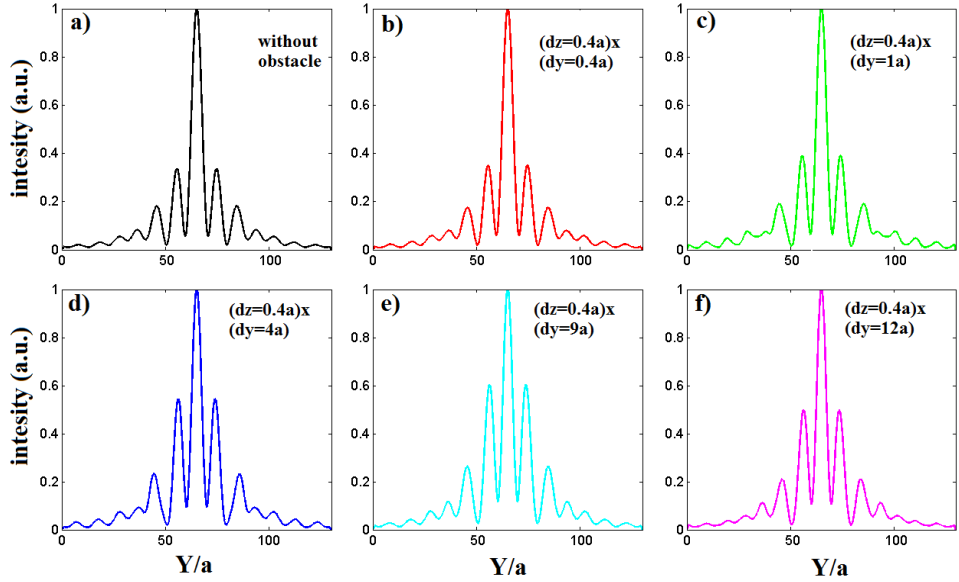

Figure 4.14. The cross-sectional intensity profiles of beams in Fig. 4.13 are shown at the distance $85a$ from the tip of the axicon PC.

Annular type PC gives additional flexibilities in terms of tuning the properties of the limited diffraction beam. For example, a careful adjustment of the refractive index of the inner air holes may give ways to manipulate the locations of the field's maximum amplitude. In addition to TM polarization, one may obtain a similar beam generation mechanism for TE polarization.

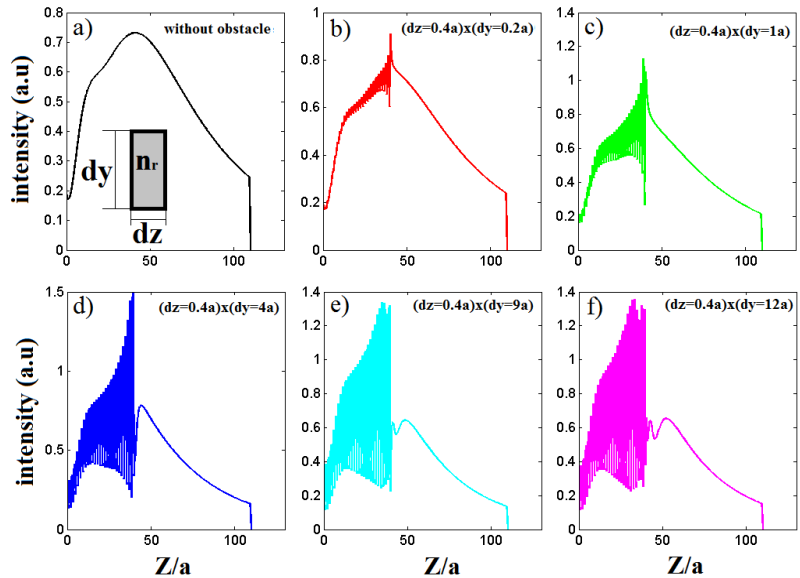

Figure 4.15. The on-axis intensity profiles of beams are shown for different rectangular shape obstacle sizes.

Using APC instead of common PC structures gives additional flexibilities on tuning the behavior of generated limited diffraction beam, as previously mentioned. Here we provide one example by showing that gradual change of the refractive index of the inner rods gives rise to longer propagation distance. The refractive indices from $n_2$ to $n_7$ are arranged to change gradually from 1.5 to 2.0 by a step size of 0.10 in Fig. 4.16(a). The beginning indices of the inner rods at the apex of the axicon are equal to $n_1=1.0$. The designed graded index axicon composed of APC acts as diverging (concave) lens. The light source follows certain paths that have different lengths and refracts with different output angles. Hence, multiple side lobes occur at the edges of the structure with different direction angles. Among those side lobes, the interference is observed, which results in formation of the central lobe along the propagation direction, as can be clearly inferred from the Fig. 4.16(b). In Fig. 4.16(c), the intensity profiles depending on the propagation distance, $P_d$ is observed. According to this graph, the generated limited diffraction beam can propagate up to $150a$ distance while keeping the intensity to be over 53% of maximum value and propagation distance reaches to $200a$ when the central intensity of the beam is higher than 36%.

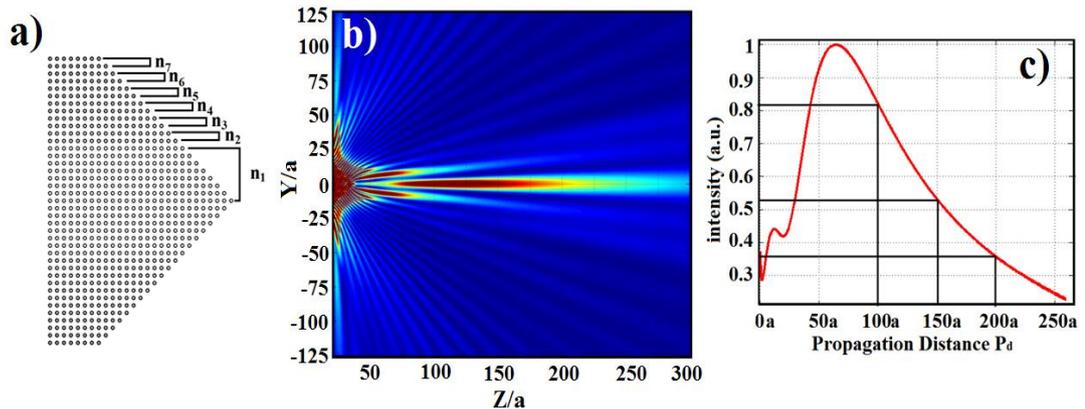

Figure 4.16. (a) The schematic presentation of 2D axicon-shape annular PC with graded indices of air holes. (b) The spatial intensity profiles of the generated diffraction limited beam. (c) The on-axis intensity profiles of the generated two-dimensional beam.

# 5. TWO-DIMENSIONAL COMPLEX PARITY-TIME SYMMETRIC PHOTONIC STRUCTURES*

## 5.1. Introduction

Parity-time (PT-) symmetric structures were initially proposed as exotic systems with unusual properties; despite their complex-valued potentials, the non-Hermitian Hamiltonians describing those systems can have real eigenvalues [219]. First regarded as a curiosity in quantum mechanics, such systems have recently been shown to have interesting and useful applications in classical wave systems, especially in optics. Indeed, PT-symmetric photonic systems have shown intriguing new features, such as PT phase transitions [220] and realization of unidirectional invisible media [221,222] or unidirectional waveguide transmitters [223,224]; some of these effects have already been experimentally realized [220,223,224].

PT symmetry requires that the complex potential, $U(\vec{r})=U^{\text{Re}}(\vec{r})+iU^{\text{Im}}(\vec{r})$, obey the symmetry requirement $U(\vec{r})=U^{*}(-\vec{r})$, which means that the real part of the potential is an even function, $U^{\text{Re}}(\vec{r})=U^{\text{Re}}(-\vec{r})$, whereas the imaginary part is odd, $U^{\text{Im}}(\vec{r})=-U^{\text{Im}}(-\vec{r})$. Although the imaginary part of the potential is generally difficult to obtain in nature, this is not the case in optics. The classical analog to the real part of the potential in optics is the refractive index, and the gain-loss is analogous to its imaginary part. Therefore, by combining the index and gain-loss modulations with the required symmetries, such optical systems become classical analogs of quantum systems described by PT-symmetric Hamiltonians [220–224].

To date, the pioneering works referenced above and recent extensive literature on optical PT symmetry covers mostly one-dimensional (1D) systems. On the other hand, recent works on systems with gain-loss modulations in two dimensions [225,226], and also on complex two-dimensional 2D crystals [227,228] where the gain-loss and index are simultaneously modulated, have shown the micro- and nanophotonics to be a platform for developing synthetic materials with novel beam propagation effects. However, none of these cases [225–228] can be attributed to PT-symmetric systems because they do not meet the requirements of PT symmetry.

In this chapter, we propose a 2D PT-symmetric complex photonic structure and show the new properties inherent of its 2D character. We explore the light propagation within it, both by realistic numerical calculations using the FDTD method and by analyzing the Bloch-like modes due to the complex modulation of the potential. We observe strong asymmetric clockwise–counterclockwise flows of light in the Bloch-like modes close to the crystallographic resonances or, equivalently, close to high-symmetry points. As a basic effect, we numerically show the measurable asymmetric transmission of a Gaussian light beam incident on a finite-sized structure resulting from asymmetric wave coupling.

## 5.2. Derivation of the 2D Honeycomb PT-symmetric Structure

To introduce the coupling effects in a 2D PT-symmetric photonic structure, we start from a 1D PT-symmetric optical system, the properties of which are summarized in Fig. 5.1. This is essentially a superposition of a 1D Bragg mirror [Fig. 5.1(a1)] and a balanced gain-loss modulation with the same periodicity but spatially displaced by a quarter-period [Fig. 5.1(a2)]. In the simplest case, we can consider the harmonic potential of the structure in the form: $n(x) = n[\cos(qx) + i\sin(qx)]$, more conveniently expressed as

$$n(x) = n\exp(iqx), \tag{5.1}$$

where $q$ is the reciprocal lattice vector of the modulation, and n is the amplitude of the complex index modulation. Clearly, such a modulation unidirectionally couples a wave with wavevector $k_B$ to $k_A=k_B+q$. In the right column of Fig. 5.1(a2), a left-propagating resonant wave, $k_B \approx -q/2$, is coupled to $k_A=k_B+q \approx q/2$ and is thus Bragg-reflected to the right. Alternatively, a harmonic Bragg reflector with real-valued potential,

$$n(x) = n\cos(qx) = \frac{n}{2}[\exp(iqx) + \exp(-iqx)], \tag{5.2}$$

symmetrically couples, at resonance, $\underline{k}_A \approx q/2$ with $k_B \approx -q/2$, as illustrated in the right column of Fig. 5.1(a1). Hence, the 1D PT-symmetric modulation given by Eq. (5.1) breaks the symmetry of left–right wave coupling and propagation, which becomes most pronounced at resonance. Note that this symmetry breaking is the main difference between the potentials in Eqs. (5.1) and (5.2), and is the reason for all the peculiarities shown by PT-symmetric systems, while reciprocity always holds [229].

Keeping this basic principle in mind, we consider the PT-symmetric complex crystal in 2D space. The simplest choice is the trivial extension of the 1D PT-symmetry to 2D, $n(\vec{r}) = n_x \exp(iq_x x) + n_y \exp(iq_y y)$, which simply factorizes the PT symmetries in both quadratures but does not lead to new 2D peculiarities. Therefore, we intend to build the nonfactorizable PT symmetries, *i.e.,* nonfactorizable unidirectional coupling between the plane wave components, assuming that it will introduce 2D peculiarities (in comparison with 1D PT-symmetric systems).

We chose a triangular lattice, as the simplest nontrivial case:

$$n(\vec{r}) = n_0 + \Delta n \sum_{j=AB,BC,CA} \exp(i\vec{q}_j \cdot \vec{r}), \tag{5.3}$$

which is generated by three vectors symmetrically rotated by angles of $2\pi/3$ with respect to one another, namely, $\vec{q}_{AB,CA}=(q/2,\pm q\sqrt{3}/2)$ and $\vec{q}_{BC}=(-q,0)$, as represented in the right column of Fig. 5.1(b2), where $n_0$ is the refractive index of the dielectric embedding medium, and $\Delta n$ determines the amplitude of the complex modulation. Note that considering only the real part of Eq. (5.3) leads to the corresponding dielectric PhC with 6-fold symmetry, as represented by Fig. 5.1(b1). At resonance, $|k_{A,B,C}|=q\sqrt{3}/3$, such a real structure (PhC case) reciprocally couples the plane wave components directed along the symmetry axes, as schematically shown in the inset of Fig. 5.1(b1). However, for the complex lattice described by Eq. (5.3), the coupling is analogous to that given by Eq. (5.1), being PT-symmetric in any direction Such a complex lattice exhibits a 3-fold symmetry, as shown in the inset of Fig. 5.1(b2). This can be expected to produce peculiarities in PT-symmetric systems. Next, to design a realistic 2D PT-symmetric structure, we replace the lower refractive index areas with low refractive index cylinders [central column in Fig. 5.1(b1)]. The right column of Fig. 5.1(b1) displays a 6-fold reciprocal space (Fourier transform) of the cylinder arrangement enabling symmetric coupling. However, when such cylinders alternatively exhibit gain and loss, as schematically represented in the central column of Fig. 5.1(b2), the complex distribution of the index contains the expected PT-symmetry. Indeed, the reciprocal space of the arrangement of cylinders [right column of Fig. 5.1(b2)] reproduces the three points in the configuration proposed in Eq. (5.3), leading to unidirectional coupling between wave components. Apart from the three points indicating the lattice vectors, $\vec{q}_{AB}, \vec{q}_{BC}$ and $\vec{q}_{CA}$, other higher-order harmonics of the complex distribution appear owing to the non-harmonic (stepwise) modulation of the potential.

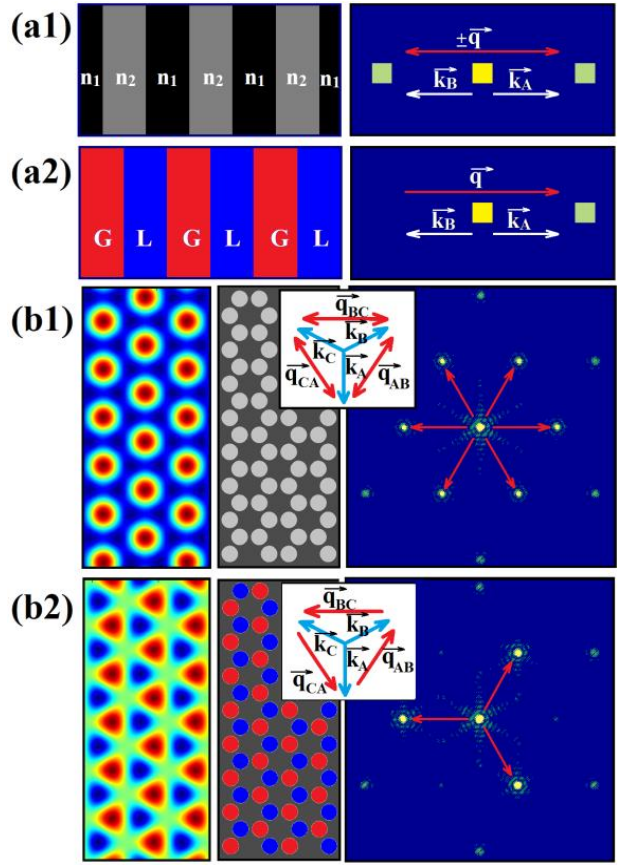

Figure 5.1. (a1) Left: 1D Bragg reflector. Right: Fourier transform (FT) of the structure, reciprocal lattice vectors and reciprocal coupling of wavevectors at resonance, $n_1 > n_2$. (a2) Left: Gain-Loss distribution (G/red, L/blue). Right: FT of the combined 1D PT-symmetric structure from (a1) and (a2), symmetric lattice vectors and asymmetric coupling at resonance. (b1) Left: Real part of Eq. (3), $n_0=1.1$, $\Delta n=0.1$. Center: Arrangement of cylinders. Right: FT of real cylinder's structure and lattice vectors. (b2) Left: Imaginary part of Eq. (3). Center: Honeycomb arrangement of gain-loss cylinders, $n=1.1\pm 0.1i$, $n_0=1.3$. Right: FT of full 2D PT-symmetric arrangement of cylinders. Insets in (b1) and (b2) show the symmetric and asymmetric coupling at resonance.

The triangular lattice is seemingly the simplest nontrivial case of a non-factorizable 2D PT-symmetric complex crystal. Further nontrivial cases could be realized for higher odd-fold rotational symmetry, which would also yield nontrivial 2D PT-symmetric quasi-crystals. Here we consider only this triangular case.

## 5.3. Asymmetric Chiral Excitation

We numerically check whether the proposed system displays the expected properties of complex PT-symmetric systems, in particular the asymmetric flow of light. Differently from 1D, the asymmetric coupling between wavevectors rotates the input by $\pm 2\pi/3$, depending on the input channel. In other words, the structure is expected to display a type of chiral non-reciprocity. This test is performed numerically using the well-established FDTD technique [230]. We consider two finite-size structures of the same symmetry containing the real and complex distributions shown in Figs. 5.2(a) and 5.2(b), respectively.

We first analyze the propagation of a short broadband pulse incident on the structure from the top in the vertical direction and calculate the transmitted intensity on two detectors, symmetrically located on both sides of the structure [T1 and T2 in Figs. 5.2(a) and 5.2(b)]. The resulting spectral transmission in the clockwise and counterclockwise directions, normalized to the incident pulse intensity, is represented in Figs. 5.2(c) and 5.2(d) for each structure. When comparing the transmissions, we clearly see the expected asymmetry arising precisely at resonant frequencies $a/\lambda \approx 0.3$, where "$a$" is the center-to-center distance between cylinders; note that $q=4\pi/3a$ [see Figs. 5.2(c) and 5.2(d)]. Although the T1 and T2 spectra coincide perfectly for all frequencies, for the PC case in Fig. 5.2(c), the counterclockwise (clockwise) transmission is enhanced (reduced) at resonance for the 2D PT-symmetric structure in Fig. 5.2(d). Note that, except for a higher-order resonance at $a/\lambda \approx 0.6$ (due to high-order mode coupling), the symmetry is still unbroken far from resonance, and both curves coincide well at other frequencies.

For the PC structure, an incident wave $k_A$ couples symmetrically to $k_B$ and $k_C$, as schematically shown in the inset of Fig. 5.1(b1). We can observe that the field distribution depicted in Fig. 5.2(e), obtained by numerical FDTD simulation, is perfectly symmetric. However, the asymmetric flow of light within the complex system enhances the transmission to the counterclockwise output channel, T2, in Fig. 5.2(f), whereas transmission to the T1 channel is suppressed. Figure 5.2(f) demonstrates at a glance the asymmetric coupling schematically represented in the inset of Fig. 5.1(b2); the incident wave $k_A$ is coupled to $k_B$ but not to $k_C$. Finally, we also find that the situation depicted in Fig. 5.2(e) is very similar to the field distribution from the 2D PT-symmetric structure far from resonance, where no symmetry breaking is predicted.

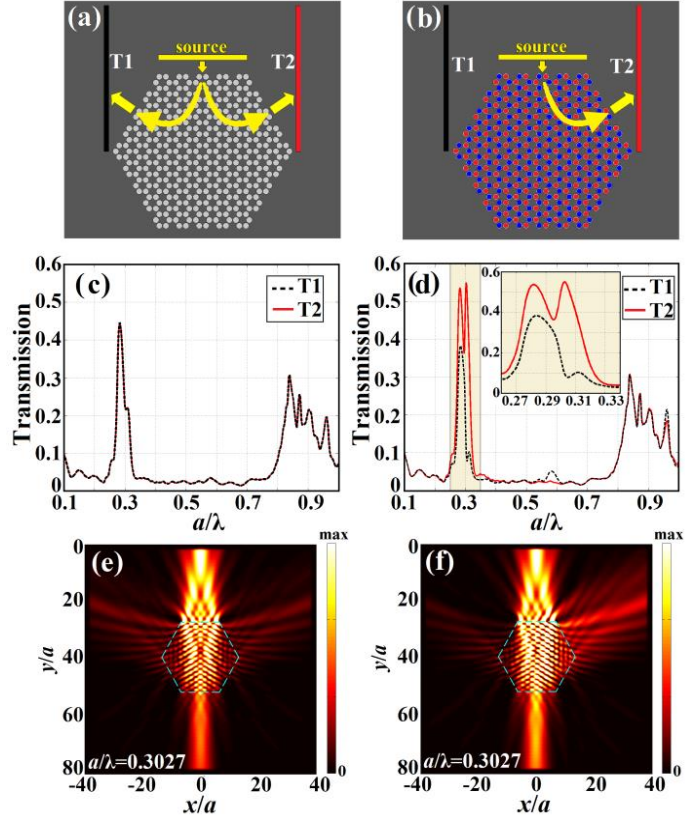

Figure 5.2. Schematic representation of: (a) 2D PC (real) and (b) 2D PT-symmetric (complex) structures, source and detectors (c) and (d) Clockwise and counterclockwise transmissions as a function of frequency, in $a/\lambda$ units, ("$a$" is the center-to-center distance between cylinders of radius, R = 0.45$a$.) for structures in (a) and (b), respectively. Inset in (d) is a magnified view within $a/\lambda$ = 0.25–0.35. (e) and (f) Normalized intensity distributions for an incident Gaussian beam (width 14a) on (a) and (b), respectively. See the supplementary material for a frequency scan of the distribution (f) showing the PT-phase transitions.

Finally, we note that whereas $k_A$ couples to $k_B$, -$k_A$ couples to –$k_C$. Thus, a -$k_A$ wave, incident from the base upward to the structure, would be transmitted clockwise instead of propagating counterclockwise within the structure owing to the nonreciprocal chirality of the system. Note that the closed set of lattice vectors ($q_{AB}$+$q_{BC}$+$q_{CA}$=0) enables the simultaneous resonance of two disjoint triads, namely ($k_A$, $k_B$, $k_C$) and (-$k_A$, -$k_C$, -$k_B$) in a circular chiral coupling. The counterclockwise chiral mode being excited by $k_A$, and the clockwise mode by -$k_A$; rending the chiral flow of light input dependent.

## 5.4. Chiral Bloch-like Modes Close to the PT-transition Point

For a PC, the Bloch modes are defined as localized electromagnetic states of the periodic media that are invariant in propagation. However, in a complex system described by a non-Hermitian Hamiltonian, complex Bloch-like modes may either amplify or decay in time. Below we calculate such Bloch-like modes analytically considering the simple case of a harmonic PT-symmetric complex crystal of triangular symmetry. We consider an incident plane wave with a polarization perpendicular to the plane of the crystal and a wavevector directed vertically, $\vec{k}=(0,-k)$, near resonance: $\vec{k}=\vec{k}_A+\Delta\vec{k}$. The small variations are considered to be in the same incident direction: $\Delta\vec{k}=(0,-\Delta k_y)$. Disregarding the second time derivatives, the wave equation can be written as;

$$-2i\omega \partial_t \vec{E} = \frac{c^2}{n(\vec{r})^2}\nabla^2\vec{E}+\omega^2\vec{E}. \tag{5.4}$$

We expand the electric field into the first three harmonics of the field, which are resonant in the lattice, namely: $\vec{k}_A=(0,-k_0)$, $\vec{k}_B=\vec{k}_A+\vec{q}_{AB}$, and $\vec{k}_C=\vec{k}_A-\vec{q}_{CA}$, and obtain, for the TM polarization:

$$E = \sum_{j=A,B,C} a_j \exp\left(i\left(\vec{k}_j+\Delta\vec{k}\right)\cdot\vec{r}\right). \tag{5.5}$$

Introducing the expansion in (5.5) into (5.4) yields coupled equations between their amplitudes, $a_A, a_B, a_C$:

$$-i\frac{n_0}{k_0 c}\partial_t \begin{pmatrix} a_A \\ a_B \\ a_C \end{pmatrix} = \begin{pmatrix} \vec{k}_A\cdot\Delta\vec{k} & \Delta n/n_0 & 0 \\ 0 & \vec{k}_B\cdot\Delta\vec{k} & \Delta n/n_0 \\ \Delta n/n_0 & 0 & \vec{k}_C\cdot\Delta\vec{k} \end{pmatrix} \begin{pmatrix} a_A \\ a_B \\ a_C \end{pmatrix}. \tag{5.6}$$

The dispersion diagrams, *i.e.,* the temporal eigenvalues and the associated Bloch-like modes, are obtained by diagonalization of the matrix in (5.6). Figures 5.3(a) and 5.3(b) display the real and imaginary parts, respectively, of the matrix eigenvalues for the three Bloch-like modes at the edge of the Brillouin zone, *i.e.,* at resonance between lattice vectors. The temporal evolution of the Bloch mode is defined by the matrix eigenvalues with a factor $i n_0/k_0 c$. As expected, sufficiently far from resonance, all the eigenvalues are real-valued (where the asymmetry of the coupling is not pronounced). Close to resonance, the PT phase transition occurs, and we obtain Bloch modes with complex eigenvalues, one with a negative imaginary part and hence amplified in time. Therefore, in an extended structure, after a finite propagation time, the field distribution is expected to exhibit the amplitude and phase corresponding to this amplified mode. Such amplitude and phase of the most amplified Bloch mode, as calculated analytically from Eq. (5.6), are depicted in Figs. 5.3(c) and 5.3(d), respectively.

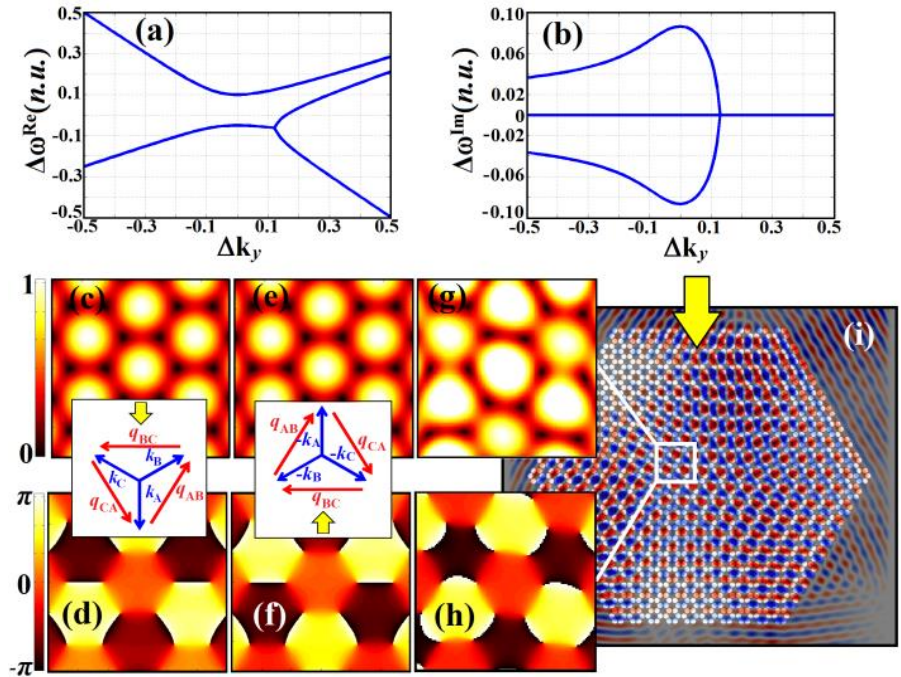

Figure 5.3. Calculated dispersion diagrams and Bloch modes. (a) Real and (b) imaginary parts of the matrix eigenvalues $\Delta\omega$, where $\Delta k_y$, on the horizontal axis, is the distance from resonance in $k_0$ units. (c) Amplitude and (d) phase of the amplified chiral Bloch mode for $\Delta k=0$, when illuminated from above and from below (e) and (f), respectively; the insets show the counterclockwise/clockwise asymmetric coupling. (g) Field intensity and (h) phase distribution within the hexagon for an incident Gaussian beam with carrier frequency $a/\lambda=0.303$, corresponding magnified $6a\times 6a$ region. (i) Field amplitude obtained directly from FDTD calculation, the arrow indicates the input channel. See the supplementary material for videos of amplitude and phase evolutions, and counterclockwise flow of light.

To check the analytic predictions, we analyze the field evolution after excitation by a relatively long Gaussian pulse with central frequency at resonance, and spectrum narrower than the width of the transmission resonance peak in Fig. 5.2(d). Within the structure (a larger version of the same honeycomb configuration), the incident radiation is redistributed among all the coupled harmonics approaching a stationary distribution of the growing Bloch-like mode, after a sufficiently long time. The analytically calculated amplitude and phase of the amplified chiral Bloch-like mode are shown in Figs. 5.3(c, d)/Figs. 5.3(e, f), when the structure is illuminated from above/below, respectively. The result presented in Fig. 5.3(i) is used to extract the amplitude and phase of the Bloch mode shown in Figs. 5.3(g) and 5.3(h), respectively. The results agree well with the analytically calculated amplified Bloch modes. The differences may be attributed mainly to the simplified model used (not accounting for the real shape of the scatter) and the interplay between higher-order harmonics, as well as to the finite size of the structure.

### 5.5. Implementation of Proposal

Finally, we propose a possible realization of the investigated 2D PT-symmetric complex structure, which could be implemented and measured in microphotonic devices. The configuration illustrated in Fig. 5.4(a) consists of a silicon slab with a honeycomb lattice of alternating p-n and n-p semiconductor junctions. Full three-dimensional (3D) FDTD numerical simulations were performed using the LUMERICAL software package [230]. The device is illuminated by a broadband pulse with a Gaussian profile, with a source 7 $\mu m$ width and 0.5 $\mu m$ height. Detectors T1 and T2 are symmetrically placed on either side of the structure as shown in Fig. 5.4(a) to record the transmission. The calculated normalized transmission spectra at T1 and T2 are depicted in Fig. 5.4(b). A measurable clockwise–counterclockwise asymmetry is observed in the transmission near resonance at the wavelength $\lambda=1.501\mu m$ (wavelength in a vacuum). The steady-state electric field distributions at the cross-sectional $xy$ plane ($z=0$) and $yz$ plane ($x=0$) are shown in Figs. 5.4(c) and 5.4(d), respectively. The electric field snapshot in Fig. 5.4(c) shows the asymmetric light transmission along the directions of T1 and T2 at the resonance frequency. Furthermore, the cross-sectional field distribution depicted in Fig. 5.4(d) proves the vertical confinement and guiding of the propagating beam inside the slab. As a result, the out-of-plane losses are almost negligible for this specific design.

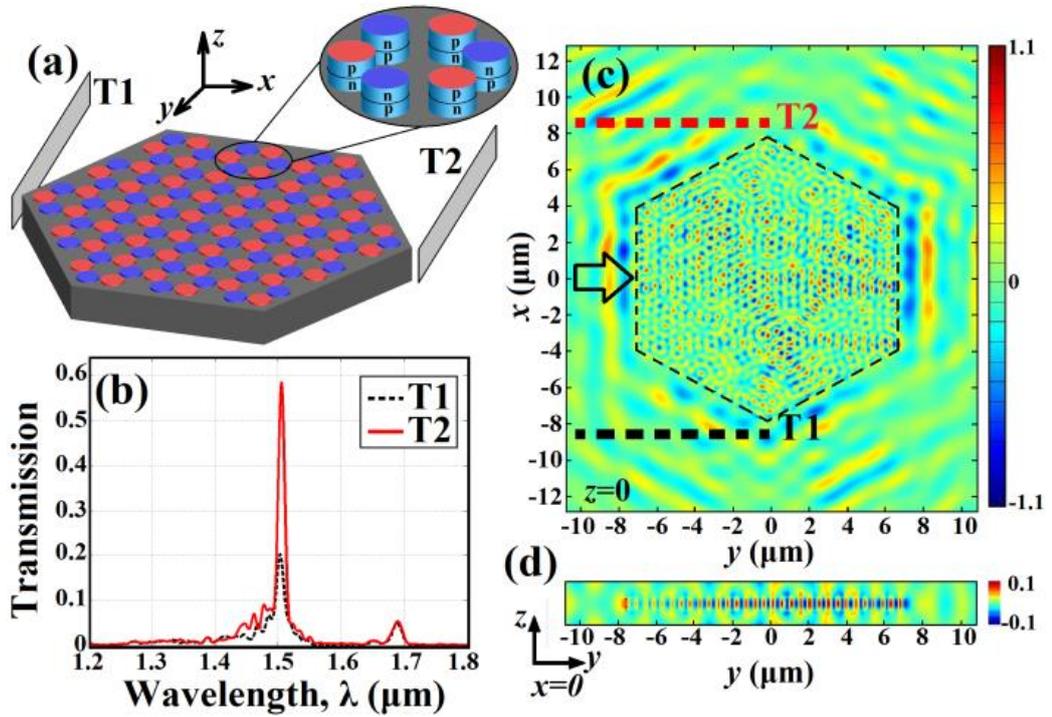

Figure 5.4. (a) Dielectric slab, $n=3.474$, 0.612 $\mu m$ high, with holes of radii 0.45$\mu m$ filled by *p-n*/*n-p* semiconductor junctions, $n=3.46\pm0.007i$; $a=1.0\mu m$, where red/blue circles indicate gain/loss areas (b) Clockwise–counterclockwise normalized transmission on detectors T1 and T2. Electric field distribution snapshots at cross-sectional planes (c) $z=0$ and (d) $x=0$. The black arrow in (c) indicates the input channel.

# 6. CONCLUSION

The aim of this thesis was to investigate and characterize new photonic designs for efficiently light manipulation, beam steering, beam splitting, wavelength de-multiplexing, light focusing, higher order mode converging, limited diffraction beam generation and asymmetric light transmission purposes. Moreover, we have provided analytical and numerical results in the design of all dielectric, low symmetric, graded index and gain-loss index modulated parity time photonic structures.

In Chapter 2, we investigated the effect of symmetry-reduction on the dispersive characteristics of all-dielectric PCs. Breaking the rotational symmetry of a PC unit-cell produces various anomalous optical characteristics such as complete PBG, tilted self-collimation, super-collimation, and wavelength selectivity. Besides, the low symmetric unit-cell provides additional parameters to control the dispersive features of Bloch modes. Next in section 2.1, the reduced symmetry of photonic crystals by introducing asymmetric unit cells in terms of crescent shape instead of circular ones improves the light manipulation capability via the appearance of anisotropic iso-frequency contours in the dispersion diagram. The optical characteristics of the structure were numerically investigated by means of finite-difference time-domain and plane wave expansion methods. The crescent-shaped photonic crystal demonstrates a high degree of control over the light propagation behavior in terms of focusing and self collimation of light beams. The routes of light beam can be tuned by altering the opening angle of the crescent shape. Engineering the placement of each crescent-shape cells may offer a platform for implementing various photonic functions including beam splitters and combiners, deflectors and routers without deploying any defects inside the periodic dielectric structure.

In section 2.2, we have proposed periodic all-dielectric structures with low symmetry for wavelength selective light manipulation. Contrary to common approaches that utilize highly symmetric photonic configurations, the proposed approach in this paper is highly different in terms of working principle (employment of self-collimation effect) and structural form of primitive PC cell (rotational $C_4$ symmetry).

As a potential application, the specific spectral property of wavelength division effect in a novel STAR-PC is analyzed and the spatial electric field distributions of different wavelengths are demonstrated. An efficient design of wavelength sensitive structure that works in the broad telecom window is properly modeled. The designed device operates within the wavelength interval of $1484.5\,\text{nm}-1621.5\,\text{nm}$ that corresponds to a bandwidth of $\Delta\omega/\omega_0 = 8.82\%$. By means of the proposed STAR-PC configuration, wavelength selective media can be designed without introducing defects or using complex materials. Moreover, the proposed method is based on engineering of the relevant dispersion curve of low symmetric PCs. The design does not require any physical perturbation of the structure or nonlinearities. Therefore, this innovative design based on self-collimation phenomenon may facilitate the construction of wavelength division structures operating with a fairly broad bandwidth and high coupling efficiency.

In chapter 3, we deal with inhomogeneous GRIN media and it optical characterizations. In section 3.1, inhomogeneous GRIN media are considered by considering them as optical lens possessing different light manipulation schemes such as focusing, de-focusing and collimation. The analytical investigation of continuous GRIN medium having HS profile is explored. In addition, the discrete version of continuous GRIN medium is designed by PCs and time domain analyses are conducted. The analytical approach utilizes Ray theory and computational tools are based on plane wave expansion and finite-difference time-domain methods. The detailed exploration of artificial GRIN structure reveals the significance of operating such medium not only at long wavelengths but also at short wavelength region. Meanwhile, strong focusing is possible if the operating frequency or structural length is appropriately adjusted. Conventional curved optical elements can be replaced with the current designs. Flat front and back surfaces, frequency selectivity of the structure and strong focusing capability make them invaluable element for variety of optical systems and devices.

In section 3.2, we have proposed A-GRIN PC designs to obtain mode conversion process transforming even mode to odd mode and vice versa. Numerical methods based on wave theory of light in inhomogeneous medium are used to design A-GRIN PC environment. Mode transformation process is deeply investigated by exploring both amplitude and phase information of the converted mode. Analytical approach based on Ray theory helps understanding beam trajectories in different GRIN media. The results produced by Exponential and Luneburg lens profiles are compared with the corresponding ideal cases. The proposed devices consist of PCs with a constant radius and a constant refractive index. The effective index variations are obtained by varying the lattice spacing in only transverse to the propagation direction. It is seen that Exponential profile's performance is superior to Luneburg lens profile in terms of creating higher-order mode. The numerical results are based on both frequency and time domain methods. Analytical derivations are carried out by using ray theory. The idea and implementations outlined in the present work may enable other mode conversions including third or even higher order modes.

In chapter 4, we have reported limited diffraction light propagation distance exceeding 92a (~50 $\mu$m ) by implementing 2D axicon shaped annular PC and optimizing the apex element location. In the numerical simulations, we observed that the shape of the generated two-dimensional beam's transversal intensity profile resembles to the zero-order Bessel function and the intensity on the central axis stays fairly constant for a certain propagation distance before diffraction shows its effect. We have explored the interaction of limited diffraction beam with obstacles of different sizes and shapes placed on the propagation path and demonstrated the self-healing ability of such beams. Furthermore, we show that using APC gives additional improvements on the properties of limited diffraction beam. Hence, enlargement of diffraction-resistive propagation distance is achieved by gradually changing the inner rods' refractive indexes.

Finally in chapter 5, we have proposed a simple 2D PT-symmetric photonic structure and analyze the propagation of light with it. As predicted, we see that close to resonance, the system exhibits a nonreciprocal chirality associated with asymmetric wave coupling between the plane wave components. Therefore, such a 2D PT-symmetric structure with a hexagonal shape asymmetrically transmits light beams incident on it. In addition, we analytically calculate the Bloch-like mode formations and find that indeed the more amplified mode agrees well with the complex field, and phase distributions in the structure at resonance. Following the proposed scheme, we design and numerically analyze, using full 3D FDTD simulations, a 2D PT-symmetric feasible configuration. The proposed 2D planar semiconductor structure could be produced by microfabrication and microstructuration of the electrodes to achieve the modulated gain-loss. It may be expected that new synthetic optical components could rely on such optical systems.

# REFERENCES

[1] L. Rayleigh, "On the maintenance of vibrations by forces of double frequency, and on the propagation of waves through a medium endowed with a periodic structure," Phil. Mag. S.5, 24(147), 145-59, 1887.

[2] E. Yablonovitch, "Inhibited spontaneous emission in solid-state physics and electronics," Phys. Rev. Lett. 58, 2059–2062, 1987.

[3] S. John, "Strong localization of photons in certain disordered dielectric superlattices", Phys. Rev. Lett. 58(23), 2486–2489, 1987.

[4] J. D. Joannopoulos, S. G. Johnson, J. N. Winn, and R. D. Meade, Photonic Crystals: Molding the Flow of Light (2nd ed.), *Princeton NJ: Princeton University Press*, 2008.

[5] Y. Akahane, T. Asano, B-S. Song, and S. Noda, "High-Q photonic nanocavity in a two-dimensional photonic crystal," Nature 425, 944-947, 2003.

[6] B. S. Song, S. Noda, T. Asano, and Y. Akahane, "Ultra-high-Q photonic double-heterostructure nanocavity," Nature Materials 4, 207 – 210, 2005.

[7] L. Wu, M. Mazilu, T. Karle, and T. F. Krauss, "Superprism phenomena in planar photonic crystals," IEEE J. Quantum Electron. 38, 915–918, 2002.

[8] H. Kosaka, T. Kawashima, A. Tomita, M. Notomi, T. Tamamura, T. Sato, and S. Kawakami, "Self-collimating phenomena in photonic crystals," Appl. Phys. Lett. 74, 1212–1214, 1999.

[9] E. Chow, A. Grot, L. W. Mirkarimi, M. Sigalas, and G. Girolami, "Ultracompact biochemical sensor built with two-dimensional photonic crystal microcavity," Opt. Lett. 29, 1093-1095, 2004.

[10] H. Kurt, M. N. Erim, and N. Erim, "Various photonic crystal bio-sensor configurations based on optical surface modes," Sensors and Actuators B: Chemical 165 (1), 68-75, 2012.

[11] T. F. Krauss, R. M. D. L. Rue, and S. Brand, "Two-dimensional photonic-bandgap structures operating at near infrared wavelengths," Nature 383(6602), 699–702, 1996.

[12] O. Painter, R.K. Lee, A. Scherer, A. Yariv, J.D. O'Brien, P.D. Dapkus, and L. Kim, "Two-dimensional photonic band-gap defect mode laser," Science, 284, 1819, 1999.

[13] S. Noda, A. Chutinan, and M. Imada, "Trapping and emission of photons by a single defect in a photonic bandgap structure," Nature 407, 606–610, 2000.

[14] H. Kurt, "Theoretical study of directional emission enhancement from photonic crystal waveguides with tapered exits," IEEE Photon. Technol. Lett. 20, 1682-1684, 2008.

[15] A. E. Akosman, M. Mutlu, H. Kurt, and E. Ozbay, "Compact wavelength de-multiplexer design using slow light regime of photonic crystal waveguides," Opt. Express, 19, 24129-24138, 2011.

[16] T. Gorishnyy, C. K. Ullal, M. Maldovan, G. Fytas, and E. L. Thomas, "Hypersonic Phononic Crystals," Phys. Rev. Lett. 94, 115501, 2005.

[17] A. Khelif, A. Choujaa, S. Benchabane, B. Djafari-Rouhani, and V. Laude, "Guiding and bending of acoustic Waves in Highly Confined Phononic Crystal Waveguides," Appl. Phys. Lett. 84(22), 4400-4402, 2004.

[18] S. Yang, J. H. Page, Z. Liu, M. L. Cowan, C. T. Chan, and P. Sheng, "Focusing of Sound in a 3D Phononic Crystal," Phys. Rev. Lett. 93(2) 024301 (2004).

[19] V. Romero-Garcia, R. Pico, A. Cebrecos, V. J. Sanchez-Morcillo, and K. Staliunas, "Enhancement of sound in chirped sonic crystals," Appl. Phys. Let. 102, 091906, 2013).

[20] K. Sakoda, Optical properties of photonic crystals, *Springer Verlag*, 2005.

[21] T. Krauss and R. De La Rue, "Photonic crystals in the optical regime--past, present and future," Progress in Quantum Electronics 23, 51-96 (1999).

[22] M. Plihal and A. A. Maradudin, "Photonic band structure of two-dimensional systems: The triangular lattice," Phys. Rev. B, vol. 44, no. 16, pp. 8565-8571, 1991.

[23] P. R. Villeneuve and M. Piché, "Photoinc band gaps in two-dimensional square and hexagonal lattices," Phys. Rev. B, vol. 46, no. 8, pp. 4969-4972, 1992.

[24] R. D. Meade, K. D. Brommer, A. M. Rappe, and J.D. Joannopoulos, "Existence of a photonic band gap in two dimensions," Appl. Phys. Lett., vol. 61, no. 4, pp. 495-497, 1992.

[25] K. M. Ho, C. T. Chan, and C. M. Soukoulis, "Existence of a photonic gap in periodic dielectric structures," Phys. Rev. Lett., vol. 65, no. 25, pp. 3152-3155, 1990.

[26] K. Yee, "Numerical solution ofinitial boundary value problems involving Maxwell's equations in isotropic media," Antennas and Propagation, IEEE Transactions on 14, pp. 302-307, 1966.

[27] A. Taflove and S. C. Hagness, Computational Electrodynamics, the FiniteDifference Time-Domain Method, 2nd ed. Norwood, MA, *Artech House*, 2000.

[28] D. S. Wiersma, "Disordered photonics," Nature Photonics, 7, 188 – 196, 2013.

[29] E. R. Martins, J.T. Li, Y. K. Liu, V. Depauw, Z. X. Chen, J. Y. Zhou, and T. F. Krauss, "Deterministic quasi-random nanostructures for photon control," Nature Communications, 4 (2665), 2013.

[30] M. Segev, Y. Silberberg, and D. N. Christodoulides, "Anderson localization of light," Nature Photon. 7, 197 – 204, 2013.

[31] B. Redding, S. F. Liew, R. Sarma, and H. Cao, "Compact spectrometer based on a disordered photonic chip," Nature Photonics, 7(9), 746-751, 2013.

[32] H. Cao, Y. G. Zhao, S. T. Ho, E. W. Seelig, Q. H. Wang, and R. P. H. Chang, "Random Laser Action in Semiconductor Powder," Phys. Rev. Lett. 82, 2278 – 2281, 1999.

[33] S. Gottardo, R. Sapienza, P. D. García, A. Blanco, D. S. Wiersma, and C. López, "Resonance-driven random lasing," Nature Photonics 2, 429 – 432, 2008.

[34] N. M. Lawandy, "Disordered media: Coherent random lasing," Nature Physics 6, 246 – 248, 2010.

[35] V. Roppo, D. Dumay, J. Trull, C. Cojocaru, S. M. Saltiel, K. Staliunas, R. Vilaseca, D. N. Neshev, W. Krolikowski, and Y. S. Kivshar, "Planar second-harmonic generation with noncollinear pumps in disordered media," Opt. Express 16, 14192–14199, 2008.

[36] Z. V. Vardeny, A. Nahata, and A. Agrawal, "Optics of photonic quasicrystals," Nature Photon. 7, 177 – 187, 2013.

[37] M. E. Zoorob, M. D. B. Charlton, G. J. Parker, J. J. Baumberg, and M. C. Netti, "Complete photonic bandgaps in 12-fold symmetric quasicrystals," Nature, 404 (6779), 740–743, 2000.

[38] N. D. Lai, J. H. Lin, Y. Y. Huang, and C. C. Hsu, "Fabrication of two- and three-dimensional quasi-periodic structures with 12-fold symmetry by interference technique," Opt. Express 14, 10746-10752, 2006.

[39] A. Della Villa, S. Enoch, G. Tayeb, V. Pierro, V. Galdi, and F. Capolino, "Band gap formation and multiple scattering in photonic quasicrystals with a Penrose-type Lattice," Phys. Rev. Lett. 94 183903, 2005.

[40] M. A. Kaliteevski, S. Brand, R. A. Abram, T. F. Krauss, R. M. De La Rue, and P. Millar, "Two-dimensional Penrose-tiled photonic quasicrystals: diffraction of light and fractal density of modes," Journal of Modern Optics 47 (11), 1771-1778, 2000.

[41] Y. A. Vlasov, M. I. Kaliteevski, and V. V. Nikolaev, "Different regimes of light localization in a disordered photonic crystal," Phys. Rev. B. 60, 1555–1562, 1999.

[42] M. Werchner, M. Schafer, M. Kira, S. W. Koch, J. Sweet, J. D. Olitzky, J. Hendrickson, B. C. Richards, G. Khitrova, H. M. Gibbs, A. N. Poddubny, E. L. Ivchenko, M. Voronov, and M. Wegener, "One dimensional resonant Fibonacci quasicrystals: noncanonical linear and canonical nonlinear effects," Opt. Express 17, 6813-6828, 2009.

[43] W. Gellermann, M. Kohmoto, B. Sutherland, and P. C.Taylor, "Localization of light waves in Fibonacci dielectric multilayers," Phys. Rev. Lett. 72, 633-636, 1994.

[44] Y. S. Chan, C. T. Chan, and Z. Y. Liu, "Photonic Band Gaps in Two Dimensional Photonic Quasicrystals," Phys. Rev. Lett. 80, 956-959, 1998.

[45] M. C. Rechtsman, H.-.C Jeong, P. M. Chaikin, S. Torquato, and P. J. Steinhardt, "Optimized Structures for Photonic Quasicrystals," Phys. Rev. Lett. 101, 073902, 2008.

[46] M. Florescu, S. Torquato, and P. J. Steinhardt, "Complete band gaps in two-dimensional photonic quasicrystals," Phys. Rev. B 80, 155112, 2009.

[47] W. Man, M. Megens, P. J. Steinhardt, and P. M. Chaikin, "Experimental measurement of the photonic properties of icosahedral quasicrystals," Nature 436(7053), 993–996, 2005.

[48] J. Hung Lin, W. L. Chang, H-Y. Lin, T-H. Chou, H-C. Kan, and C. C. Hsu, "Enhancing light extraction efficiency of polymer light-emitting diodes with a 12-fold photonic quasi crystal," Opt. Express 21, 22090-22097, 2013.

[49] E. Yablonovitch, "Photonic band-gap structures," J. Opt. Soc. Am. B 10, 283-295, 1993.

[50] S. G. Johnson, P. R. Villeneuve, S. Fan, and J. D. Joannopoulos, "Linear waveguides in photonic-crystal slabs," Phys. Rev. B 62, 8212–8222, 2000.

[51] H. Kurt, I. H. Giden, and K. Ustun, "Highly efficient and broadband light transmission in 90° nanophotonic wire waveguide bends," J. Opt. Soc. Am. B 28, 495-501, 2011.

[52] M. Lončar, J. Vučković, and A. Scherer, "Methods for controlling positions of guided modes of photonic-crystal waveguides," J. Opt. Soc. Am. B 18, 1362-1368, 2001.

[53] S. Foteinopoulou and C. M. Soukoulis, "Negative refraction and left-handed behavior in two-dimensional photonic crystals," Phys. Rev. B 67, 235107-235111, 2003.

[54] H. Kosaka, T. Kawashima, A. Tomita, M. Notomi, T. Tamamura, T. Sato, and S. Kawakami, "Photonic crystals for micro lightwave circuits using wavelength-dependent angular beam steering," Appl. Phys. Lett. 74, 1370-1372, 1999.

[55] D. Chigrin, S. Enoch, C. S. Torres, and G. Tayeb, "Self-guiding in two-dimensional photonic crystals," Opt. Express 11, 1203-1211, 2003.

[56] Z. Y. Li, B. Y. Gu, and G. Z. Yang, "Large absolute band gap in 2D anisotropic photonic crystals," Phys. Rev. Lett. 81, 2574-2577, 1998.

[57] H. Kurt and D. S. Citrin, "Annular photonic crystals," Opt. Express 13, 10316-10326, 2005.

[58] X. Zhu, Y. Zhang, D. Chandra, S. C. Cheng, J. M. Kikkawa, and S. Yang, "Two-dimensional photonic crystals with anisotropic unit cells imprinted from poly (dimethylsiloxane) membranes under elastic deformation," Appl. Phys. Lett. 93, 161911-161913, 2008.

[59] H. F. Ho, Y. F. Chau, H. Y. Yeh, and F. L. Wu, "Complete bandgap arising from the effects of hollow, veins, and intersecting veins in a square lattice of square dielectric rods photonic crystal," Appl. Phys. Lett. 98, 263115-263117, 2011.

[60] B. Rezaei, T. F. Khalkhali, A. S. Vala, and M. Kalafi, "Absolute band gap properties in two-dimensional photonic crystals composed of air rings in anisotropic tellurium background," Opt. Commun. 282, 2861-2869, 2009.

[61] Y. Zhang, L. Kong, Z. Feng, and Z. Zheng, "PBG structures of novel two-dimensional annular photonic crystals with triangular lattice," Optoelectronics Lett. 6, 281-283, 2010.

[62] J. Hou, D. Gao, H. Wu, and Z. Zhou, "Polarization insensitive self-collimation waveguide in square lattice annular photonic crystals," Opt. Commun. 282, 3172-3176, 2009.

[63] H. Wu, L. Y. Jiang, W. Jia, and X. Y. Li, "Imaging properties of an annular photonic crystal slab for both TM-polarization and TE-polarization," J. Opt. 13, 095103-095111, 2011.

[64] H. Kurt and D. S. Citrin, "Graded index photonic crystals," Opt. Express 15, 1240-1253, 2007.

[65] E. Centeno, D. Cassagne, and J. P. Albert, "Mirage and superbending effect in two-dimensional graded photonic crystals," Phys. Rev. B 73, 235119-235119, 2006.

[66] H. Kurt and D. S. Citrin, "A novel optical coupler design with graded-index photonic crystals," IEEE Photon. Technol. Lett. 19, 1532-1534, 2007.

[67] C. Tan, T. Niemi, C. Peng, and M. Pessa, "Focusing effect of a graded index photonic crystal lens," Opt. Commun. 284, 3140-3143, 2011.

[68] H. Kurt, E. Colak, O. Cakmak, H. Caglayan, and E. Ozbay, "The focusing effect of graded index photonic crystals," Appl. Phys. Lett. 93, 171108-171110, 2008.

[69] B. Vasić and R. Gajić, "Self-focusing media using graded photonic crystals: Focusing, Fourier transforming and imaging, directive emission, and directional cloaking," J. Appl. Phys. 110, 053103-053110, 2011.

[70] M. Lu, B. K. Juluri, S-C. S. Lin, B. Kiraly, T. Gao, and T. J. Huang, "Beam Aperture Modification and Beam Deflection Using Gradient-Index Photonic Crystals," J. Appl. Phys. 108, 103505-103509, 2010.

[71] B. Vasic, G. Isic, R. Gajic, and K. Hingerl, "Controlling electromagnetic fields with graded photonic crystals in metamaterial regime," Opt. Express 18, 20321-20333, 2010.

[72] I. Khromova and L. Melnikov, "Anisotropic photonic crystals: generalized plane wave method and dispersion symmetry properties," Opt. Commun. 281, 5458-5466, 2008.

[73] H. Xie and Y. Y. Lu, "Modeling two-dimensional anisotropic photonic crystals by Dirichlet-to-Neumann maps," J. Opt. Soc. Am. A 26, 1606-1614, 2009.

[74] B. Rezaei and M. Kalafi, "Tunable full band gap in two-dimensional anisotropic photonic crystals infiltrated with liquid crystals," Opt. Commun. 282, 1584-1588, 2009.

[75] S. W. Leonard, J. P. Mondia, H. M. van Driel, O. Toader, S. John, K. Busch, A. Birner, U. Gösele, and V. Lehmann, "Tunable two-dimensional photonic crystals using liquid-crystal infiltration," Phys. Rev. B 61, 2389-2392, 2000.

[76] C. S. Kee, K. Kim, and H. Lim, "Tuning of anisotropic optical properties of two-dimensional dielectric photonic crystals," Physica B: Condensed Matter 338, 153-158, 2003.

[77] T. Trifonov, L. F. Marsal, A. Rodríguez, J. Pallarès, and R. Alcubilla, "Effects of symmetry reduction in two dimensional square and triangular lattices," Phys. Rev. B 69, 235112, 2004.

[78] R. P. Zaccaria, P. Verma, S. Kawaguchi, S. Shoji, and S. Kawata, "Manipulating full photonic band gaps in two dimensional birefringent photonic crystals," Opt. Express 16, 14812-14820, 2008.

[79] F. Guan, Z. Lin, and J. Zi, "Opening up complete photonic bandgaps by tuning the orientation of birefringent dielectric spheres in three-dimensional photonic crystals," J. Phys.: Condens. Matter 17, 343-349, 2005.

[80] A. I. Cabuz, E. Centeno, and D. Cassagne, "Superprism effect in bidimensional rectangular photonic crystals," Appl. Phys. Lett. 84, 2031-2033, 2004.

[81] Y. Xu, X. J. Chen, S. Lan, Q. Guo, W. Hu, and L. J. Wu, "The all-angle self-collimating phenomenon in photonic crystals with rectangular symmetry," J. Opt. A: Pure Appl. Opt. 10, 1-5, 2008.

[82] Y. Ogawa, Y. Omura, and Y. Iida, "Study on Self-Collimated Light-Focusing Device Using the 2-D Photonic Crystal With a Parallelogram Lattice," J. Lightwave Technol. 23, 4374-4381, 2005.

[83] D. Gao, Z. Zhou, and D. S. Citrin, "Self-collimated waveguide bends and partial bandgap reflection of photonic crystals with parallelogram lattice," J. Opt. Soc. Am. A 25, 791-795, 2008.

[84] P. Yeh, "Electromagnetic propagation in birefringent layered media," J. Opt. Soc. Am. 69, 742-756, 1979.

[85] S. Johnson and J. Joannopoulos, "Block-iterative frequency-domain methods for Maxwell's equations in a planewave basis," Opt. Express 8, 173–190, 2001.

[86] D. E. Aspnes, "Local-Field Effects and Effective-Medium Theory: A Microscopic Perspective," Am. J. Phys. 50, 704-709, 1982.

[87] J. P. Berenger, "A perfectly matched layer for the absorption of electromagnetic waves," J. Comput. Phys. 114, 185–200, 1994.

[88] A. Yariv and P. Yeh, Optical Waves in Crystals: Propagation and Control of Laser Radiation, *John Wiley & Sons Press*, 1983.

[89] G. Si, A. J. Danner, S. Lang Teo, E. J. Teo, J. Teng, and A. A. Bettiol, "Photonic crystal structures with ultrahigh aspect ratio in lithium niobate fabricated by focused ion beam milling," J. Vac. Sci. Technol. B 29, 021205-021209, 2011.

[90] J. Feng, Y. Chen, J. Blair, H. Kurt, R. Hao, D. S. Citrin, C. J. Summers, and Z. Zhou, "Fabrication of annular photonic crystals by atomic layer deposition and sacrificial etching," J. Vac. Sci. Technol. B 27, 568-572, 2009.

[91] R. R. Panepucci, H. B. Kim, R. V. Almeida, and M. D. Jones, "Photonic crystals in polymers by direct electron-beam lithography presenting a photonic band gap," J. Vac. Sci. Technol. B 22, 3348 – 3351, 2004.

[92] P. Borel, A. Harpøth, L. Frandsen, M. Kristensen, P. Shi, J. Jensen, and O. Sigmund, "Topology optimization and fabrication of photonic crystal structures," Opt. Express 12, 1996-2001, 2004.

[93] H. Kosaka, T. Kawashima, A. Tomita, M. Notomi, T. Tamamura, T. Sato, and S. Kawakami, "Superprism phenomena in photonic crystals," Physical Review B 58, 10096–10099, 1998.

[94] H. Kurt, M. Turduev, and I. H. Giden, "Crescent shaped dielectric periodic structure for light manipulation," Optics Express 20 (2012) 7184-7194.

[95] M. Turduev, I. Giden, and H. Kurt, "Modified annular photonic crystals with enhanced dispersion relations: polarization insensitive self-collimation and nanophotonic wire waveguide designs," Journal of Optical Society of America B 29, 1589-1598, 2012.

[96] I. H. Giden and H. Kurt, "Modified annular photonic crystals for enhanced band gap properties and iso-frequency contour engineering," Appl. Opt. 51, 1287-1296, 2012.

[97] K. Ren and X. Ren, "Controlling light transport by using a graded photonic crystal," Appl. Opt. 50, 2152-2157, 2011.

[98] A. Cicek and B. Ulug, "Polarization-independent waveguiding with annular photonic crystals," Opt. Express 17, 18381-18386, 2009.

[99] H. Ishio, J. Minowa, and K. Nosu, "Review and status of wavelength-division-multiplexing technology and its applications," IEEE J. of Lightw. Technol. 2, 448–463, 1984.

[100] M. Gerken and D. A. B. Miller, "Photonic nanostructures for wavelength division multiplexing," Proc. SPIE 5597 82-96, 2004.

[101] P. Dumon, et al., "Compact wavelength router based on a silicon-on-insulator arrayed waveguide grating pigtailed to a fiber array," Opt. Express 14, 664–669, 2006.

[102] G. E. Keiser, "A review of WDM technology and applications," Optical Fiber Technology 5, 3-39, 1999.

[103] M. Gerken and D. A. B. Miller, "Multilayer thin-film structures with high spatial dispersion," Appl. Opt. 42, 1330-1345, 2003.

[104] A. Sharkawy, S. Shi, and D. W. Prather, "Multichannel wavelength division ultiplexing with photonic crystals," Appl. Opt. 40, 2247-2252, 2001.

[105] H. A. B. Salameh and M. I. Irshid, "Wavelength-division demultiplexing using graded-index planar structures," J. of Lightw. Technolo. 24, 2401-2408, 2006.

[106] V. K. Do, et al., "Wavelength Demultiplexer Based on a Two-Dimensional Graded Photonic Crystal," IEEE Photon. Technol. Lett. 23, 1094-1096, 2011.

[107] D. Yilmaz, I. H. Giden, M. Turduev and H. Kurt, "Design of wavelength selective medium by GRIN PC," Journal of Quantum Electronics 49, 477-484, 2013.

[108] T. Matsumoto, S. Fujita, and T. Baba, "Wavelength demultiplexer consisting of Photonic crystal superprism and superlens," Optics Express 13, 10768-10776, 2005.

[109] B. Momeni, et al., "Compact wavelength demultiplexing using focusing negative index photonic crystal superprisms," Optics Express 14, 2413-2422, 2006.

[110] T. Niemi, L. H. Frandsen, K. K. Hede, A. Harpoth, P. I. Borel, and M. T. Kristensen, "Wavelength-division demultiplexing using photonic crystal waveguides," IEEE Photonics Technology Letters 18, 226-228, 2006.

[111] M. Djavid and F. Monifi and A. Ghaffari, and M. S. Abrishamian, "Heterostructure wavelength division demultiplexers using photonic crystal ring resonators," Optics Communications 281, 4028–4032, 2008.

[112] H. Benisty, C. Cambournac, F. V. Laere, and D. V. Thourhout, "Photonic-Crystal Demultiplexer With Improved Crosstalk by Second-Order Cavity Filtering," Journal of Lightwave Technology 28, 1201-1208, 2010.

[113] B. E. Nelson, M. Gerken, D. A. B. Miller, R. Piestun, C. Lin, and J. S. Harris, "Use of a dielectric stack as a one-dimensional photonic crystal for wavelength demultiplexing by beam shifting," Optics Letters 25, 1502-1504, 2000.

[114] A. Lupu, E. Cassan, S. Laval, L. El Melhaoui, P. Lyan, and J. M. Fedeli, "Experimental evidence for superprism phenomena in SOI photonic crystals," Optics Express 12 5690-5696, 2004.

[115] L. Wu, M. Mazilu, and T. F. Krauss, "Beam steering in planar-photonic crystals: from superprism to supercollimator," Journal of Lightwave Technology 21, 561-566, 2003.

[116] J. Witzens, T. Baehr-Jones, and A. Scherer, "Hybrid superprism with low insertion losses and suppressed cross-talk," Physical Review E 71, 026604-1-9, 2005.

[117] T. Baba, and T. Matsumoto, "Resolution of photonic crystal superprism," Applied Physics Letters 81, 2325-2327, 2002.

[118] B. Momeni and A. Adibi, "Optimization of photonic crystal demultiplexers based on the superprism effect," Applied Physics B 77, 555-560, 2003.

[119] A. F. Oskooi, D. Roundy, M. Ibanescu, P. Bermel, J. D. Joannopoulos, and S. G. Johnson, "MEEP: A flexible free-software package for electromagnetic simulations by the FDTD method," Computer Physics Communications 181, 687–702, 2010.

[120] S. H. Moosavi Mehr and S. Khorasani, "Influence of Asymmetry on the Band Structure of Photonic Crystals," Proc. SPIE 7609, 76091G, 2010.

[121] R. E. Hamam, M. Ibanescu, S. G. Johnson, J. D. Joannopoulos, M. Soljačić, "Broadband super-collimation in a hybrid photonic crystal structure," Opt. Express 17, 8109–8118, 2009.

[122] K. Staliunas and R. Herrero, "Nondiffractive propagation of light in photonic crystals," Physical Review E 73, 016601, 2006.

[123] Y. Loiko, K. Staliunas, R. Herrero, C. Cojocaru, J. Trull, V. Sirutkaitis, D. Faccio, and T. Pertsch, "Towards observation of sub-diffractive pulse propagation in photonic crystals," Optics Communications 279, 377–383, 2007.

[124] Y. Loiko, C. Serrat, R. Herrero, K. Staliunas, "Quantitative analysis of subdiffractive light propagation in photonic crystals," Optics Communications 269, 128-136, 2007.

[125] W. Smigaj and B. Gralak, "Semianalytical design of antireflection gratings for photonic crystals," Physical Review B 85, 035114-035127, 2012.

[126] M. Gerken and D. A. B. Miller, "Wavelength demultiplexer using the spatial dispersion of multilayer thin-film structures," IEEE Photonics Technology Letters 15, 1097-1099, 2003.

[127] J. Witzens, M. Loncar, and A. Scherer, "Self-Collimation in Planar Photonic Crystals," IEEE Journal of Selected Topics in Quantum Electronics 8, 1246-1257, 2002.

[128] R.K. Luneburg: Mathematical Theory of Optics *University of California Press*, *Berkeley*, 1964 and Mimeographed Lectured Notes on Mathematical Theory of Optics, *Brown University Providence,* 1944.

[129] E.W. Marchand: Gradient Index Optics, *Academic Press*, New York, 1978.

[130] C. Gomez-Reino, M. V. Perez, and C. Bao, Gradient-Index Optics: Fundamentals and Applications, *Springer*, 2002.

[131] J. Maxwell: *Cambridge and Dublin*, Math. J. 8, 88,1854.

[132] R.W. Wood: Physical Optics, *McMillan*, New York, 1905.

[133] A.L. Mikaelian: SU Academy Reports 81, 569 *in Russian*, 1951.

[134] S.-Y. Lin, V. M. Hietala, L. Wang, and E. D. Jones, "Highly dispersive photonic band-gap prism," Opt. Lett. 21(21), 1771–1773, 1996.

[135] D. W. Prather, S. Y. Shi, D. M. Pustai, C. H. Chen, S. Venkataraman, A. Sharkawy, G. J. Schneider, and J. Murakowski, "Dispersion-based optical routing in photonic crystals," Opt. Lett. 29(1), 50-52, 2004.

[136] C. Luo, S. G. Johnson, J. D. Joannopoulos, and J. B. Pendry, "All-angle negative refraction without negative effective index," Phys. Rev. B 65(20), 201104-1-201104-4, 2002.

[137] D. T. Moore, "Gradient-index optics: A review," Appl. Opt. 19(7), 1035–1038, 1980.

[138] M. J. Adams, An introduction to optical waveguides, 14, *Chichester Wiley*, 1984.

[139] J. C Palais, Fiber optic communications, *Prentice Hall*, 1988.

[140] B. Saleh and M. C. Teich, Fundamental of Photonics, *Wiley-Interscience*, 1991.

[141] A. Sharma, A. K. Ghatak, "A variational analysis of single mode graded-index fibers," Opt. Commun. 36(1), 22-24, 1981.

[142] M. Feit and J. Fleck, Jr., "Light propagation in graded-index optical fibers," Appl. Opt. 17(24), 3990-3998, 1978.

[143] Cregan, R. F., B. J. Mangan, J. C. Knight, T. A. Birks, P. St J. Russell, P. J. Roberts, and D. C. Allan. "Single-mode photonic band gap guidance of light in air," Science 285(5433), 1537-1539, 1999.

[144] G. Yabre, "Comprehensive Theory of Dispersion in Graded-Index Optical Fibers," J. Lightwave Technol. 18(2), 166-177, 2000.

[145] E. Centeno and D. Cassagne, "Graded photonic crystals," Opt. Lett. 30(17), 2278–2280, 2005.

[146] H. T. Chien, C. Lee, H. K. Chiu, K. C. Hsu, C. C. Chen, J. A. Ho, and C. Cho, "The comparison between the graded photonic crystal coupler and various coupler," IEEE J. Lightwave Technol. 27(14), 2570-2574, 2009.

[147] F. Gaufillet and É. Akmansoy, "Graded photonic crystals for graded index lens," Opt. Commun. 285(10), 2638–2641, 2012.

[148] H. W. Wang and L. W. Chen, "High transmission efficiency of arbitrary waveguide bends formed by graded index photonic crystals," J. Opt. Soc. Am. B 28(9), 2098-2104, 2011.

[149] E. Akmansoy, E. Centeno, K. Vynck, D. Cassagne, and J.-M. Lourtioz, "Graded photonic crystals curve the flow of light: an experimental demonstration by the mirage effect," Appl. Phys. Lett. 92(13), 133501-133503, 2008.

[150] H. Kurt, B. Oner, M. Turduev, and I. Giden, "Modified Maxwell fish-eye approach for efficient coupler design by graded photonic crystals," Opt. Express 20, 22018-22033, 2012.

[151] B. Oner, M. Turduev, and H. Kurt, "High-efficiency beam bending using graded photonic crystals," Opt. Lett. 38, 1688-1690, 2013.

[152] M. Turduev, B. Oner, I. Giden, and H. Kurt, "Mode transformation using graded photonic crystals with axial asymmetry," J. Opt. Soc. Am. B 30, 1569-1579, 2013.

[153] D. Luo, G. Alagappan, X.W. Sun, Z. Raszewski, J.P. Ning, "Superbending effect in two-dimensional graded photonic crystals," Opt. Commun. 282(2), 329-332, 2009.

[154] E. Cassan, D. Khanh-Van, C. Caer, D. Marris-Morini, L. Vivien, "Short-Wavelength Light Propagation in Graded Photonic Crystals," J. Lightwave Technol. 29(13), 1937-1943, 2011.

[155] A. G. Nalimov and V. V. Kotlyar, "Hyperbolic secant slit lens for subwavelength focusing of light," Opt. Lett. 38, 2702-2704, 2013.

[156] L. Maigyte, V. Purlys, J. Trull, M. Peckus, C. Cojocaru, D. Gailevičius, M. Malinauskas, and K. Staliunas, "Flat lensing in the visible frequency range by woodpile photonic crystals," Opt. Lett. 38, 2376-2378, 2013.

[157] N. Kumar, R. Herrero, M. Botey, and K. Staliunas, "Flat lensing by periodic loss-modulated materials," J. Opt. Soc. Am. B 30, 2684-2688, 2013.

[158] N. Kumar, L. Maigyte, M. Botey, R. Herrero, and K. Staliunas, "Beam shaping in two-dimensional metallic photonic crystals," J. Opt. Soc. Am. B 31, 686-690, 2014.

[159] D. Bertilone and C. Pask, "Exact Ray Paths in a Graded- Index Taper," Appl. Opt. 26(7), 1189-1194, 1987.

[160] A. W. Snyder and J. D. Love, "Optical Waveguide Theory,"Chapman & Hall, New York, 1983.

[161] A. Fletcher, T. Murphy, and A. Young, "Solutions of Two Optical Problems," Proc R. Soc. Lond. A 223(1153), 216-225, 1954.

[162] H. Kurt and D. S. Citrin, "Photonic-crystal heterostructure waveguides," IEEE J. Quant. Electron. 43, 78-84, 2007.

[163] W. D. Zhou, J. Sabarinathan, P. Bhattacharya, B. Kochman, E. Berg, P.C. Yu, and S. Pang, "Characteristics of a photonic bandgap single defect microcavity electroluminescent device," IEEE J. Quant. Electron. 37, 1153-1160, 2001.

[164] M. Yano, F. Yamagishi, and T. Tsuda, "Optical MEMS for photonic switching-compact and stable optical cross connect switches for simple, fast, and flexible wavelength applications in recent photonic networks," IEEE J. Select. Top. Quant. Electron. 11, 383-394, 2005.

[165] M. Shah, J. D. Craw, and S. Wang, "Optical waveguide mode conversion experiments," Appl. Phys. Lett. 20, 66-69, 1972.

[166] F. Auracher, "Mode order converter," Opt. Commun. 11, 187-195, 1974.

[167] Y. Huang, G. Xu, and S. Ho, "An ultra compact optical mode order converter," IEEE Photon. Technol. Lett. 18, 2281-2283, 2006.

[168] B. Lee and S. Shin. "Mode order converter in a multimode waveguide," Opt. Lett. 28, 1660-1662, 2003.

[169] M. Yang, H. Chen, K. Webb, S. Minin, S. Chuang, and G. Cueva, "Demonstration of mode conversion in an irregular waveguide," Opt. Lett. 31, 383-385, 2006.

[170] J. Castro, D. Geraghty, S. Honkanen, C. Greiner, D. Iazikov, and T. Mossberg, "Demonstration of mode conversion using anti-symmetric waveguide Bragg gratings," Opt. Express 13, 4180-4184, 2005.

[171] J. Castillo, J. Castro, R. Kostuk, and D. Geraghty, "Study of Multichannel Parallel Anti-Symmetric Waveguide Bragg Gratings for Telecom Applications," IEEE Photon. Technol. Lett. 19, 85-87, 2007.

[172] J. Kurz, J. Huang, X. Xie, T. Saida, and M. Fejer, "Mode multiplexing in optical frequency mixers," Opt. Lett. 29, 551-553, 2004.

[173] M. Pruessner, J. Khurgin, T. Stievater, W. Rabinovich, R. Bass, J. Boos, and V. Urick, "Demonstration of a mode-conversion cavity add-drop filter," Opt. Lett. 36, 2230-2232, 2011.

[174] J. Tan, M. Lu, A. Stein, and W. Jiang, "High-purity transmission of a slow light odd mode in a photonic crystal waveguide," Opt. Lett. 37, 3189-3191, 2012.

[175] V. Delaubert, M. Lassen, D. Pulford, H. Bachor, and C. Harb, "Spatial mode discrimination using second harmonic generation," Opt. Express 15, 5815-5826, 2007.

[176] B. Desiatov, I. Goykhman, and U. Levy, "Nanoscale mode selector in silicon waveguide for on chip nanofocusing applications," Nano Lett. 9, 3381–3386, 2009.

[177] J. Leuthold, J. Eckner, E. Gamper, P.A. Besse, and H. Melchior, "Multimode interference couplers for the conversion and combining of zero- and first-order modes," IEEE J. Lightwave Technol. 16, 1228-1239, 1998.

[178] M. Wienold, A. Tahraoui, L. Schrottke, R. Sharma, X. Lü, K. Biermann, R. Hey, and H. Grahn, "Lateral distributed-feedback gratings for single-mode, high-power terahertz quantum-cascade lasers," Opt. Express 20, 11207-11217, 2012.

[179] T. Swietlik, G. Franssen, R. Czernecki, M. Leszczynski, C. Skierbiszewski, I. Grzegory, T. Suski, P. Perlin, C. Lauterbach, and U. T. Schwarz, "Mode Dynamics of High Power (InAl)GaN Based Laser Diodes Grown on Bulk GaN Substrate," J. Appl. Phys. 101, 083109, 2007.

[180] S. Blaaberg, P.M. Petersen, and B. Tromborg, "Structure, Stability, and Spectra of Lateral Modes of a Broad-Area Semiconductor Laser," IEEE J. Quant. Electron. 43, 959-973, 2007.

[181] C.D. Poole, J.M. Wiesenfeld, D. J. DiGiovanni, and A. M. Vengsarkar, "Optical fiber-based dispersion compensation using higher order modes near cutoff," IEEE J. Lightwave Technol. 12, 1746-1758, 1994.

[182] M. Notomi, "Theory of light propagation in strongly modulated photonic crystals: Refractionlike behavior in the vicinity of the photonic band gap," Phys. Rev. B 62, 10696, 2000.

[183] J. E. Durnin, J. J. Miceli, and J. H. Eberly, "Diffraction-free beams," Phys. Rev. Lett. 58, 1499-1501, 1987.

[184] J. H. McLeod, "The axicon: A new type of optical element," J. Opt. Soc. Am. 44, 592–597, 1954.

[185] D. McGloin and K. Dholakia, "Bessel beams: diffraction in a new light," Contemp. Phys. 46, 15-28, 2005.

[186] A. Onae, T. Kurosawa, Y. Miki, and E. Sakuma, "Nearly diffraction-free $CO_2$ laser beam," J. Appl. Phys. 72, 4529-4532, 1992.

[187] L. Niggel, T. Lanz, and M. Maier, "Properties of Bessel beams generated by periodic grating of circular symmetry," J. Opt. Soc. Am. A 14, 27–33, 1997.

[188] R. M. Herman and T. A. Wiggins, "Production and uses of diffractionless beams," J. Opt. Soc. Am. A 8, 932–942, 1991.

[189] J. Arlt, K. Dholakia, J. Soneson, and E. M. Wright, "Optical dipole traps and atomic waveguides based on Bessel light beams," Phys. Rev. A 63, 063602, 2001.

[190] R. Grunwald, U. Grieber, F. Tschirschwitz, E. T. J. Nibbering, T. Elsaesser, V. Kebbel, H.-J. Hartmann, and W. Jueptner, "Generation of femtosecond Bessel beams with microaxicon arrays," Opt. Lett. 25, 981–983, 2000.

[191] A. Vasara, J. Turunen, and A. T. Friberg, "Realization of general nondiffracting beams with computer generated holograms," J. Opt. Soc. Am. A 6, 1748–1754, 1989.

[192] T. Grosjean, F. Baida, and D. Courjon, "Conical optics: the solution to confine light," Appl. Opt. 46, 1994–2000, 2007.

[193] K. Uehara and H. Kikuchi, "Generation of nearly diffraction-free laser beams," Appl. Phys. B 48, 125–129, 1989.

[194] G. Rui, Y. Lu, P. wang, H. Ming, and Q. Zhan, "Generation of enhanced evanescent Bessel beam using band-edge resonance," J. Appl. Phys. 108, 074304, 2010.

[195] H. Kurt, "Limited-diffraction light propagation with axicon-shape photonic crystals," J. Opt. Soc. Am. B 26, 981-986, 2009.

[196] M. Abramowitz and I. A. Stegun, "Handbook of Mathematical Functions: with Formulas, Graphs, and Mathematical Tables," New-York: Dover, 362, 1964.

[197] V. Garces-Chavez, D. McGolin, H. Melville, W. Sibbett, and K. Dholakia, "Simultaneous micromanipulation in multiple planes using a self reconstructing light beam," Nature 419, 145-147, 2002.

[198] J. Broky, G. A. Siviloglou, A. Dogariu, and D. N. Christodoulides, "Self-Healing properties of optical Airy beams," Opt. Exp. 16, 12880-12890, 2008.

[199] Z. Bouchal, J.Wagner, and M. Chlup, "Self-reconstruction of distorted nondiffracting beam," Opt. Comm. 151, 207-211, 1998.

[200] C. A. McQueen, J. Arlt, and K. Dholakia, "An experiment to study a "nondiffracting" light beam," Am. J. Phys. 67, 912-915, 1999.

[201] Z. Ding, H. Ren, Y. Zhao, J. S. Nelson, and Z. Chen, "High-resolution optical coherence tomography over a large depth range with an axicon lens," Opt. Lett. 27, 243-245, 2002.

[202] P. Polesana, D. Faccio, P. Di Trapani, A. Dubietis, A. Piskarkas, A. Couairon, and M. A. Porras, "High localization, focal depth and contrast by means of nonlinear Bessel Beams," Opt. Exp. 13, 6160-6167, 2005.

[203] Y. Matsuoka, Y. Kizuka, and T. Inoue, "The characteristic of laser micro drilling using a Bessel Beam," Appl. Phys. A: Mater. Sci. Proc. A 84, 423-430, 2006.

[204] W. M. Saj, "Light focusing on a stack of metal-insulator-metal waveguides sharp edge," Opt. Exp. 17, 13615-13623, 2009.

[205] B. P. S. Ahluwalia, X. –C. Yuan, S.H. Tao, W.C. Cheong, L. S. Zhang, and H. Wang, "Micromanipulation of high and low indices microparticles using a microfabricated double axicon," J. Appl. Phys. 99, 113104, 2006.

[206] S. K. Morrison and Y. S. Kivshar, "Engineering of directional emission from photonic-crystal waveguides," Appl. Phys. Lett. 86, 081110, 2005.

[207] D. Tang, L. Chen, and W. Ding, "Efficient beaming from photonic crystal waveguides via self-collimation effect," Appl. Phys. Lett. 89, 131120, 2006.

[208] I. Bulu, H. Caglayan, and E. Ozbay, "Beaming of light and enhanced transmission via surface modes of photonic crystals," Opt. Lett. 30, 3078-3080, 2005.

[209] H. Kurt, "The directional emission sensitivity of photonic crystal waveguides to air hole removal," Appl. Phys. B 95, 341–344, 2009.

[210] M. S. Kumar, S. Menabde, S. Yu, and N. Park, "Directional emission from photonic crystal waveguide terminations using particle swarm optimization," J. Opt. Soc. Am. B 27, 343-349, 2010.

[211] H. Kurt, "Theoretical study of directional emission enhancement from photonic crystal waveguides with tapered exits," IEEE Photon. Tech. Lett. 20, 1682-1684, 2008.

[212] M. Xin, L. Zhang, C. Eng Png, J. H. Teng, and A. J. Danner, "Asymmetric open cavities for beam steering and switching from line-defect photonic crystals," J. Opt. Soc. Am. B 27, 1153-1157, 2010.

[213] Z.-H. Chen, Z.-Y. Yu, Y.-M. Liu, P.-F. Lu, and Y. Fu, "Multiple beam splitting to free space from a V groove in a photonic crystal waveguide," Appl. Phys. B 102, 857-861, 2011.

[214] A. Vasara, J. Turunen, and A. T. Friberg, "Holographic generation of diffraction-free beams," Appl. Opt. 27, 3959–3962, 1988.

[215] T. Grosjean, F. Baida, and D. Courjon, "Conical optics: the solution to confine light," Appl. Opt. 46, 1994-2000, 2007.

[216] O. Brzobohaty, T. Cizmar, and P. Zemanek, "High quality quasi-Bessel beam generated by round-tip axicon," Opt. Exp. 17, 12688–700, 2008.

[217] J. Arlt and K. Dholakia, "Generation of high-order Bessel beams by use of an axicon," Opt. Commun. 177, 297-301, 2000.

[218] A. Sharkawy, D. Pustai, S. Shi, D. W. Prather, S. McBridge, and P. Zazucchi, "Modulating dispersion properties of low index photonic crystal structures using microfluidics," Opt. Exp. 13, 2814-2821, 2005.

[219] C. M. Bender and S. Boettcher, "Real spectra in non-Hermitian Hamiltonian having PT symmetry," Phys. Rev. Lett. 80, 5243–5246, 1998.

[220] A. Guo, et al., "Observation of PT-symmetry breaking in complex optical potentials," Phys. Rev. Lett. 103, 93902, 2009.

[221] S. Longhi, "Invisibility in PT-symmetric complex crystals," J. Phys. A, Math. Theor. 44, 485302, 2011.

[222] Zi. Lin, H. Ramezani, T. Eichelkraut, T. Kottos, H. Cao, and D.N. Christodoulides, "Unidirectional Invisibility Induced by PT-Symmetric Periodic Structures," Phys. Rev. Lett. 106, 213901, 2011.

[223] C. E. Ruter, K. G. Makris, R. El-Ganainy, D. N. Christodoulides, M. Segev, and D. Kip, "Observation of parity–time symmetry in optics," Nat. Phys. 6, 192–195, 2010.

[224] L. Feng, Y.-L. Xu, W. S. Fegadolli, M.-H. Lu, J. E. Oliveira, V. R. Almeida, Y.-F. Chen, and A. Scherer, "Experimental demonstration of a unidirectional reflectionless parity-time metamaterial at optical frequencies," Nat. Mater. 12, 108–113, 2012.

[225] K. Staliunas, R, Herrero, and R. Vilaseca, "Subdiffraction and spatial filtering due to periodic spatial modulation of the gain-loss profile," Phys. Rev. A 80, 013821, 2009.

[226] M. Botey, R. Herrero, and K. Staliunas, "Light in materials with periodic gain-loss modulation on a wavelength scale", Phys. Rev. A 82, 013828, 2010.

[227] R. Herrero, M. Botey, M. Radziunas, and K. Staliunas "Beam shaping in spatially modulated broad-area semiconductor amplifiers." Opt. Lett. 37, 5253-5255, 2012.

[228] M. Radziunas, M. Botey R. Herrero, and K. Staliunas, "Intrinsic beam shaping mechanism in spatially modulated broad area semiconductor amplifiers," Appl. Phys. Lett. 103, 132101, 2013.

[229] S. Fan, R. Baets, A. Petrov, Z. Yu, J.D. Joannopoulos, W. Freude, A. Melloni, M. Popović, M. Vanwolleghem, D. Jalas, M. Eich, M. Krause, H. Renner, E. Brinkmeyer, and C. R. Doerr, "Comment on Nonreciprocal Light Propagation in a Silicon Photonic Circuit," Science 335, 38, 2012.

[230] Lumerical FDTD Solutions, Inc.http://www.lumerical.com

# APPENDIX A

The transverse HS refractive index profile of the half GRIN medium along the *y*-axis given by

$$n(y) = n_0 \sec h(\alpha y), \tag{A1}$$

The ray propagation equation for describing light behaviour within the planar graded index structure can be obtained from the expression of a differential arc length along a ray connecting two points within the medium. Ray trajectory calculation method is illustrated in Fig. 3.1.1. The reciprocal slope of the path at the initial point ($x_0$, $y_0$) and end point ($x$, $y$) are also given in the same figure. The differential arc length can be calculated as follows:

$$ds = \sqrt{(dx)^2 + (dy)^2}; \quad \frac{dx}{ds} = \frac{1}{\sqrt{1+(\dot{y})^2}}, \tag{A2}$$

In a transverse gradient the third optical direction is invariant along any ray within the medium Ref. [13]-[Expressions (1.60), (6.3)]. In this case, maximum ray penetration depth is a constant value and expressed as follows [13]:

$$I_0 = n(y)\frac{dx}{ds} = \frac{n(y)}{\sqrt{1+(\dot{y})^2}} = const. \tag{A3}$$

In this regard, substitution of Eq. A1 into Eq. A3 provides

$$x = \int_{y_0}^{y} \frac{I_0}{\sqrt{n(y)^2 - I_0^2}} dy = \int_{y_0}^{y} \frac{\cosh(\alpha y)}{\sqrt{A^2 - \sinh^2(\alpha y)}} dy, \tag{A4}$$

where *A* is a constant value and equals to $\sqrt{\frac{n_0^2 - I_0^2}{I_0^2}}$.

To find the ray path, we use a transformation of the variables as defined below

$$u = \sinh(\alpha y), \tag{A5}$$

Under this change of variables and performing integration with respect to new variable *u*, the ray trajectory between two points in the HS index continuous GRIN structure becomes

$$x = \int_{y_0}^{y} \frac{\cosh(\alpha y)}{\sqrt{A^2 - \sinh^2(\alpha y)}} dy = |y(x) \Rightarrow u(x)|$$

$$= \frac{1}{\alpha} \int_{u_0}^{u} \frac{1}{\sqrt{A - u^2}} du = \frac{1}{\alpha} \left( \sin^{-1}\left(\frac{u}{A}\right) - \sin^{-1}\left(\frac{u_0}{A}\right) \right),$$ (A6)

where $u_0$ is the value of $u$ at $x = 0$ that is $u_0 = \sinh(\alpha y_0)$.

Eq. A6 can be written as

$$\alpha x + \sin^{-1}\left(\frac{u_0}{A}\right) = \sin^{-1}\left(\frac{u}{A}\right),$$ (A7)

After denoting $\tau = \alpha x + \sin^{-1}(u_0/A)$ and taking sine of the both sides of Eq. A7 gives

$$A \sin(\tau) = u(x).$$ (A8)

Here $u(x)$ indicates the ray position at each point within the medium and ray slope $\dot{u}(x)$ in the new $u$-$x$ coordinate system as follows:

$$\dot{u}(x) = A\cos(\tau)\dot{\tau}(x) = A\cos(\tau)\left(\alpha x + \sin^{-1}\left(\frac{u_0}{A}\right)\right)' = A\alpha \cos(\tau).$$ (A9)

After constituting the parameter $\tau$ to Eq. A9 and performing trigonometric manipulation gives us formula of a ray slope as follows:

$$\dot{u}(x) = A\alpha \cos(\alpha x + \sin^{-1}(\frac{u_0}{A})) = A\alpha\left(\cos(\alpha x)\cos(\sin^{-1}(\frac{u_0}{A})) - \frac{u_0}{A}\sin(\alpha x)\right).$$ (A10)

Integration of the slope information can give us the ray position,

$$u(x) = \int \dot{u}(x) dx = A\cos\left(\sin^{-1}\left(\frac{u_0}{A}\right)\right)\sin(\alpha x) + u_0 \cos(\alpha x).$$ (A11)

Equation A11 is a mathematical expression of ray position at each $x$ point within the continuous HS GRIN medium. To represent ray position and ray slope information using the initial conditions we should calculate $\tau$ and $\dot{u}$ at $x = 0$ position, that is

$$\tau_0 = \tau_{x=0} = \sin^{-1}\left(\frac{u_0}{A}\right); \quad \dot{u}_0 = A\alpha \cos(\tau_0); \quad \cos(\tau_0) = \frac{\dot{u}_0}{A\alpha}.$$ (A12)

Then Eq. A10 and Eq. A11 become

$$\dot{u}(x) = \dot{u}_0 \cos(\alpha x) - \alpha u_0 \sin(\alpha x), \tag{A13}$$

$$u(x) = \frac{\dot{u}_0}{\alpha} \sin(\alpha x) + u_0 \cos(\alpha x). \tag{A14}$$

Introducing position and slope of the axial and field rays we can obtain (2-by-2) matrices to describe light propagation in a continuous HS GRIN medium, *i.e.*,

$$\begin{pmatrix} u(x) \\ \dot{u}(x) \end{pmatrix} = \begin{pmatrix} u_0 \cos(\alpha x) + \frac{\dot{u}_0}{\alpha} \sin(\alpha x) \\ -\alpha u_0 \sin(\alpha x) + \dot{u}_0 \cos(\alpha x) \end{pmatrix} = \begin{pmatrix} \cos(\alpha x) & \frac{\sin(\alpha x)}{\alpha} \\ -\alpha u_0 \sin(\alpha x) & \cos(\alpha x) \end{pmatrix} \begin{pmatrix} u_0 \\ \dot{u}_0 \end{pmatrix}. \tag{A15}$$

Note that, expression A15 is analogous of the ray-transfer (ABCD) matrices used in geometrical optics for analyzing light propagation through any geometrical system. Taking into account the expressions A15, A5, A9 and A10 we can determine the ray position and slope information in a Cartesian coordinates, that is

$$y(x) = \frac{1}{\alpha} \sinh^{-1}(u(x)) = \frac{1}{\alpha} \sinh^{-1}\left( u_0 \cos(\alpha x) + \dot{u}_0 \frac{\sin(\alpha x)}{\alpha} \right), \tag{A16}$$

$$\dot{y}(x) = \frac{\dot{u}(x)}{\alpha \cosh[\alpha y(x)]} = \frac{-\alpha u_0 \sin(\alpha x) + \dot{u}_0 \cos(\alpha x)}{\alpha \cosh\left\{ \sinh^{-1}\left[ u_0 \cos(\alpha x) + \dot{u}_0 \frac{\sin(\alpha x)}{\alpha} \right] \right\}}. \tag{A17}$$

The last two expressions are enough to describe the ray trajectories of light in GRIN medium with a HS profile.

As can be seen in Fig. 3.1.1(b) when the light ray enters the free space after exiting the HS GRIN structure it refracts obeying Snell's law and travels in a straight line. As illustrated in Fig. 3.1.1(b), the output ray intersects with the optical axis (OA) and the distance between that point and end face of the structure is defined as the back focal length $\Delta F$. Incident angle $\theta_1$ of the light ray at $(x_e, y_e)$ position and the refracted angle $\theta_2$ are given as an inset in same figure. Applying Snell's law to exiting ray can be formulated as follows:

$$n(y_e) \sin \theta_1 = n_{air} \sin \theta_2, \tag{A18}$$

where $n(y_e) = n_0 \operatorname{sech}(\alpha y_e)$, $n_{air} = 1$ and $\theta_2 = arctg(\dot{y}(d))$

Putting all parameters into Eq. A18 gives

$$n_0 \sec h(\alpha y_e)\sin(arctg(\dot{y}(d))) = \sin\theta_2. \tag{A19}$$

We know that: $\sin(arctg(x)) = x/\sqrt{x^2+1}$, hence, an angle of refraction $\theta_2$ can be found as follows:

$$\theta_2 = \sin^{-1}\left\{n_0 \sec h(\alpha y_e)\frac{\dot{y}(d)}{\sqrt{\dot{y}(d)^2+1}}\right\}, \tag{A20}$$

From Fig. 3.1.1(b), the tangent of refracted angle $\theta_2$ can be determined as follows

$$\tan\theta_2 = \frac{y_e}{\Delta F}. \tag{A21}$$

Then $\Delta F$ can be expressed as follows:

$$\Delta F = \frac{y_e}{\tan\left(\sin^{-1}\left\{n_0 \sec h(\alpha y_e)\dfrac{\dot{y}(d)}{\sqrt{\dot{y}(d)^2+1}}\right\}\right)}. \tag{A22}$$

The final formula of $\Delta F$ can be determined after using an expression $\tan(\sin^{-1}(x)) = x/\sqrt{1-x^2}$:

$$\Delta F = \frac{y_e\sqrt{\dot{y}(d)^2[1-n_0^2 \sec h^2(\alpha y_e)]+1}}{n_0 \sec h(\alpha y_e)\dot{y}(d)}. \tag{A23}$$

One can see from Eq. A23 that there is a close link between this equation and Eq. 3.1.7 that express conditions of three special cases, *i.e.,* collimation, focusing, and defocusing.

**APPENDIX B**

The transverse Exponential refractive index profile of the half GRIN medium along the *y*-axis given by

$$n_{\exp}(y) = n_0 \exp(-\alpha y) \quad y \geq 0, \tag{B1}$$

The ray propagation equation for describing light behavior within the planar graded index structure can be obtained from the expression of a differential arc length along a ray connecting two points within the medium. Ray trajectory calculation method is illustrated in Fig. 3.2.6. The reciprocal slope of the path at the initial point ($x_0$, $y_0$) and end point ($x_0$, $y_0$) are also given in the same figure. The differential arc length can be calculated as follows:

$$ds = \sqrt{(dx)^2 + (dy)^2}; \quad \frac{dx}{ds} = \frac{1}{\sqrt{1+(\dot{y})^2}}, \tag{B2}$$

In a transverse gradient the third optical direction is invariant along any ray within the medium Ref. [42]-[expressions (1.60), (6.3)]. In this case, maximum ray penetration depth is a constant value and expressed as follows [44]:

$$I_0 = n(y)\frac{dx}{ds} = \frac{n(y)}{\sqrt{1+(\dot{y})^2}} = const. \tag{B3}$$

In this regard, Eq. B3 can be rewritten as

$$x = \int_{y_o}^{y} \frac{I_0}{\sqrt{n(y)^2 - I_0^2}} dy = \int_{y_o}^{y} \frac{\frac{I_0}{n(y)}}{\sqrt{1-\left(\frac{I_0}{n(y)}\right)^2}} dy. \tag{B4}$$

To find the ray path, we use a transformation of the variables as defined below

$$u = \frac{I_0}{n(y)} = \frac{I_0}{n_0} e^{\alpha y} = A e^{\alpha y}; \quad A = \frac{I_0}{n_0}, \tag{B5}$$

Under this change of variables and performing integration with respect to new variable *u*, the ray trajectory between two points in the half Exponential index continuous GRIN structure becomes

$$x = \int_{y_0}^{y} \frac{Ae^{\alpha y}}{\sqrt{1-(Ae^{\alpha y})^2}} dy = |y(x) \Rightarrow u(x)| = \quad (B6)$$

$$\frac{1}{\alpha} \int_{u_0}^{u} \frac{1}{\sqrt{1-u^2}} du = \frac{1}{\alpha}(\sin^{-1}(u) - \sin^{-1}(u_0)),$$

where $u_0$ is the value of $u$ at $x = 0$ that is $u_0 = Ae^{\alpha y_0}$.

Eq. B6 can be written as

$$\alpha x + \sin^{-1}(u_0) = \sin^{-1}(u), \quad (B7)$$

After denoting $\tau = \alpha x + \sin^{-1}(u_0)$ and taking sine of the both sides of Eq. (B7) gives

$$\sin(\tau) = u(x). \quad (B8)$$

Here $u(x)$ indicates the ray position at each point within the medium and ray slope $\dot{u}(x)$ in the new *u-x* coordinate system as follows:

$$\dot{u}(x) = \cos(\tau)\dot{\tau}(x) = \cos(\tau)(\alpha x + \sin^{-1}(u_0))' = \cos(\tau)\alpha. \quad (B9)$$

After constituting the parameter $\tau$ to Eq. B9 and performing trigonometric manipulation gives us formula of a ray slope as follows:

$$\dot{u}(x) = \alpha \cos(\alpha x + \sin^{-1}(u_0)) = \alpha(\cos(\alpha x)\cos(\sin^{-1}(u_0)) - u_0 \sin(\alpha x)) \quad (B10)$$

Integration of the slope information can give us the ray position

$$u(x) = \int \dot{u}(x)dx = \cos(\sin^{-1}(u_0))\sin(\alpha x) + u_0 \cos(\alpha x). \quad (B11)$$

Expression (B11) is a mathematical expression of ray position at each *x* point within the continuous GRIN medium. To represent ray position and ray slope information using the initial conditions we should calculate $\tau$ and $\dot{u}$ at $x = 0$ position, that is

$$\tau_0 = \tau(x=0) = \sin^{-1}(u_0); \quad \dot{u}_0 = \alpha \cos(\tau_0); \quad \cos(\tau_0) = \frac{\dot{u}_0}{\alpha}. \quad (B12)$$

Then Eq. B10 and Eq. B11 become

$$u(x) = \frac{\dot{u}_0}{\alpha}\sin(\alpha x) + u_0 \cos(\alpha x), \tag{B13}$$

$$\dot{u}(x) = \dot{u}_0 \cos(\alpha x) - \alpha u_0 \sin(\alpha x), \tag{B14}$$

Introducing position and slope of the axial and field rays we can obtain (2x2) matrices to describe light propagation in a continuous GRIN medium, *i.e.*

$$\begin{pmatrix} u(x) \\ \dot{u}(x) \end{pmatrix} = \begin{pmatrix} u_0 \cos(\alpha x) + \frac{\dot{u}_0}{\alpha}\sin(\alpha x) \\ -\alpha u_0 \sin(\alpha x) + \dot{u}_0 \cos(\alpha x) \end{pmatrix} = \begin{pmatrix} H_f & H_a \\ \dot{H}_f & \dot{H}_a \end{pmatrix} \begin{pmatrix} u_0 \\ \dot{u}_0 \end{pmatrix}, \tag{B15}$$

where axial and field rays are defined as follows

$$\begin{aligned} H_f &= \cos(\alpha x); H_a = \frac{\sin(\alpha x)}{\alpha}; \\ \dot{H}_f &= -\alpha \sin(\alpha x); \dot{H}_a = \cos(\alpha x). \end{aligned} \tag{B16}$$

Note that, expression B15 is analogous of the ABCD law used in geometrical optics for analyzing light propagation thorough any geometrical system. Taking into account the expressions B15, B5, B9 and B10 we can determine the ray position and slope information in a Cartesian coordinates, that is

$$y(x) = \frac{1}{\alpha}\ln\left(\frac{u(x)}{A}\right) = \frac{1}{\alpha}\ln\left(\frac{\dot{u}_0 \sin(\alpha x) + u_0 \alpha \cos(\alpha x)}{A\alpha}\right), \tag{B17}$$

$$\dot{y}(x) = \frac{\dot{u}(x)}{\alpha u(x)} = \frac{1}{\alpha}\frac{\dot{H}_f u_0 + \dot{H}_a \dot{u}_0}{H_f u_0 + H_a \dot{u}_0} = \left(\frac{-\alpha u_0 \sin(\alpha x) + \dot{u}_0 \cos(\alpha x)}{\alpha u_0 \cos(\alpha x) + \dot{u}_0 \sin(\alpha x)}\right). \tag{B18}$$

The last two expressions are enough to describe the ray trajectories of light in GRIN medium with an Exponential profile.

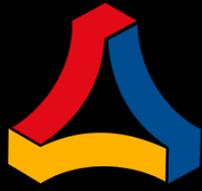
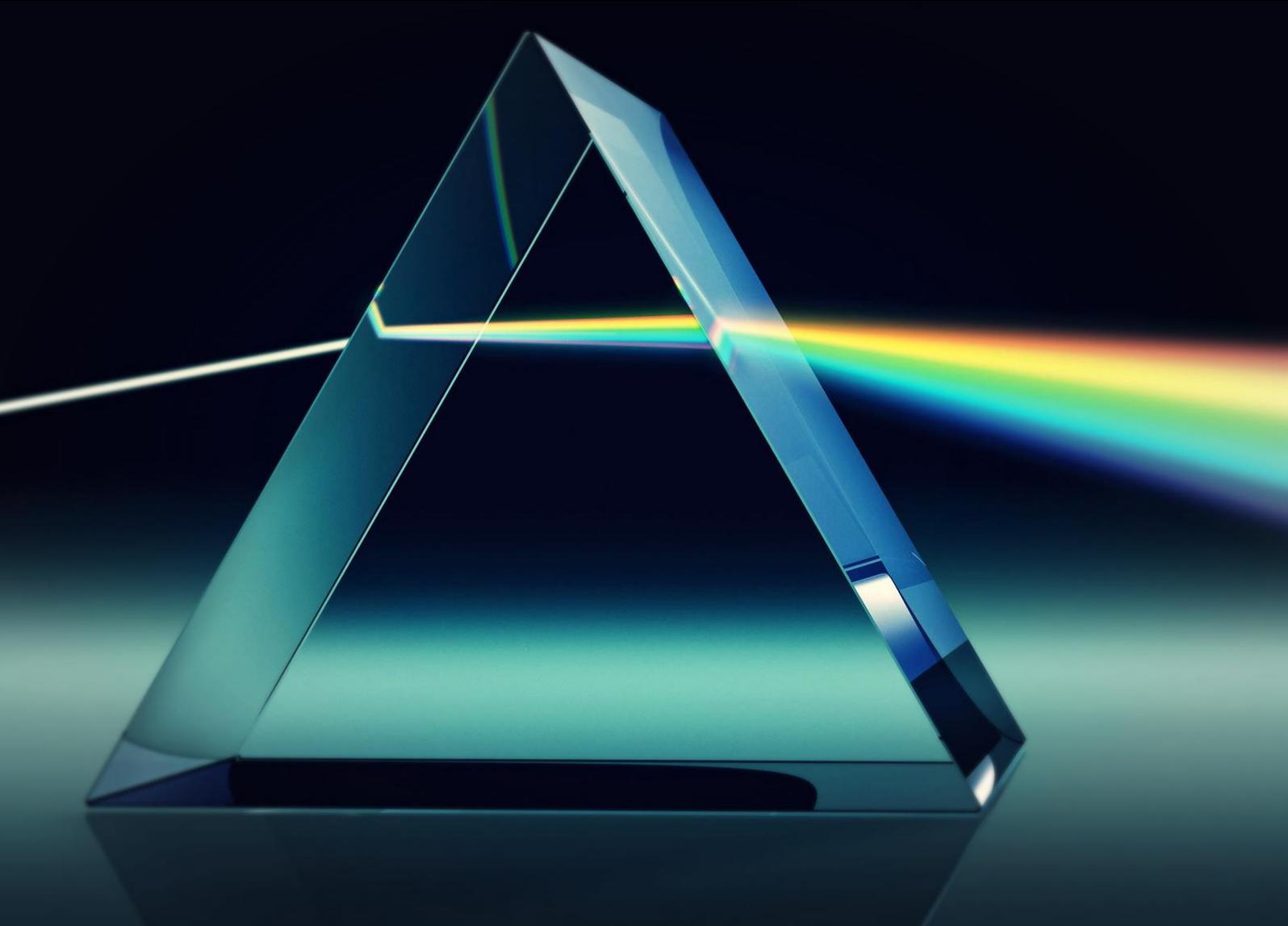